\documentclass[12pt]{article}
\input psfig.sty

\usepackage{amsmath,amssymb,graphicx} 
\usepackage{cite}

\newcommand{\drawsquare}[2]{\hbox{%
\rule{#2pt}{#1pt}\hskip-#2pt
\rule{#1pt}{#2pt}\hskip-#1pt
\rule[#1pt]{#1pt}{#2pt}}\rule[#1pt]{#2pt}{#2pt}\hskip-#2pt
\rule{#2pt}{#1pt}}

\newcommand{\Yfund}{\raisebox{-.5pt}{\drawsquare{6.5}{0.4}}}

\newcommand{\beq}{\begin{eqnarray}}
\newcommand{\eeq}{\end{eqnarray}}

\newcommand{\centeron}[2]{{\setbox0=\hbox{#1}\setbox1=\hbox{#2}\ifdim
                           \wd1>\wd0\kern.5\wd1\kern-.5\wd0\fi \copy0
                           \kern-.5\wd0\kern-.5\wd1\copy1\ifdim\wd0>\wd1
                           \kern.5\wd0\kern-.5\wd1\fi}}
\newcommand{\ltap}{\>\centeron{\raise.35ex\hbox{$<$}}
                   {\lower.65ex\hbox{$\sim$}}\>}
\newcommand{\gtap}{\>\centeron{\raise.35ex\hbox{$>$}}
                   {\lower.65ex\hbox{$\sim$}}\>}

\newcommand\ZZ{\hbox{\zfont Z\kern-.4emZ}}
\font\zfont = cmss10 

\begin{document}
\begin{titlepage}
\begin{flushright}
{\tt hep-ph/0404096} \\
\end{flushright}

\vspace*{0.8cm}
\begin{center}
\vspace*{0.5cm}
{\LARGE \bf 
TASI Lectures on Extra Dimensions and \\
\vspace*{0.2cm} Branes\footnote{Lectures at 
the Theoretical Advanced Study Institute 2002, 
University of Colorado, Boulder, CO June 3-28, 2002.}} \\
\vspace*{1.5cm}

\mbox{\bf
{Csaba Cs\'aki}} \\

\vspace*{0.4cm}

{\it Institute of High Energy Phenomenology,\\
Newman Laboratory of Elementary Particle Physics, \\
Cornell University, Ithaca, NY 14853} \\
{\tt  csaki@lepp.cornell.edu}
\end{center}

\vspace*{1cm}

\begin{abstract}
\vskip 3pt
\noindent This is a pedagogical introduction into theories with branes and
extra dimensions. We first discuss the construction of such models from an 
effective field theory point of view, and then discuss large extra dimensions
and some of their phenomenological consequences. Various possible phenomena
(split fermions, mediation of supersymmetry breaking and orbifold breaking
of symmetries) are discussed next. The second half of this review is entirely
devoted to warped extra dimensions, including the construction of the 
Randall-Sundrum solution, intersecting branes, radius stabilization, KK 
phenomenology and bulk gauge bosons.

\end{abstract}

\end{titlepage}

\newpage

\tableofcontents
\section{Introduction}
\label{sec:intro}
\setcounter{equation}{0}
\setcounter{footnote}{0}

Theories with extra dimensions have recently attracted enormous attention.
These lectures attempt to give an introduction to these new models. 
First we attempt to motivate theories with branes in extra dimensions,
and then explain how to write down an effective theory of branes, and what 
kind of 4D excitations one would expect to see in such models and how 
these modes couple to the standard model (SM) 
fields. The first lecture is closed with a 
discussion of the main phenomenological implications of models with large 
extra dimensions. In the second lecture we discuss various possible
scenarios for theories with flat extra dimensions 
that could be relevant to model building. First we discuss split fermions,
which could give a new way of explaining the fermion mass hierarchy problem,
and perhaps also explain proton stability. Next we discuss mediation of 
supersymmetry breaking via a flat extra dimension, and finally we close this
lecture with a discussion of symmetry breaking via orbifolds. These topics 
were chosen such that besides getting to know some of the most important 
directions in model building the reader will also be introduced to most of 
the relevant techniques used in this field.

The second half of this review deals exclusively with warped extra dimensions:
the Randall-Sundrum model and its variations. In the third lecture we first 
show in detail how
to obtain the Randall-Sundrum solution, and how gravity would behave in such
theories. Then localization of gravity to brane intersections is discussed,
followed by a possible solution to the hierarchy problem in infinite extra dimensions. The final lecture
discusses various issues in warped extra dimensions, including the graviton
KK spectrum, radius stabilization and radion physics, quasi-localization
of fields, bulk gauge fields and the AdS/CFT correspondence.

Inevitably, many important topics have been left out from this review.
These include the cosmology of extra dimensional models, running and 
unification in AdS space, universal extra dimensions and KK dark 
matter, black hole physics, brane induced gravity, electroweak symmetry 
breaking with extra dimensions, and the list goes on. The purpose of these 
lectures is not to cover every imoprtant topic, but rather to provide the 
necessary tools for the reader to be able to delve deeper into some of the 
topics currently under investigation. I have tried to include a representative
list of references for the topics covered, and also some list for the 
topics left out from the review. Clearly, there are many hundreds of references
that are relevant to the topics covered here, and I apologize to everyone
whose work was not quoted here. Other recent reviews of the subject can be found in~\cite{otherreviews}.

\section{Large Extra Dimensions}
\label{sec:LED}
\setcounter{equation}{0}
\setcounter{footnote}{0}
Extra dimensions were  first introduced in the 1920's by
Kaluza and Klein~\cite{KK}, who were trying to unify electromagnetism with gravity, by 
assuming that the photon field originates from fifth component ($g_{\mu 5}$)
of a five dimensional metric tensor. The early 1980's lead to
a revitalization of these ideas partly due to the realization that a 
consistent string theory will necessarily include extra dimensions. 
In order for such theories with extra dimensions to not flatly contradict 
with our observed four space-time dimensions, we need to be able to hide the
existence of the extra dimensions in all observations that have been made
to date. The most plausible way of achieving this is by assuming that the
reason why we have not observed the extra dimensions yet is that contrary to 
the ordinary four space-time dimensions which are very large (or infinite), 
these  hypothetical extra dimensions are finite, that is they are compactified.
Then one would need to be able to
probe length scales corresponding to the size of the extra dimensions to be
able to detect them. If the size of the extra dimensions is small, then one
would need extremely large energies to be able to see the consequences of the
extra dimensions. Thus by making the size of the extra dimensions very small,
one can effectively hide these dimensions. So the most important question
that one needs to ask is how large could the size of the extra dimensions
be without getting into conflict with observations. This is the first point
we will address below. We will see that answering this question will naturally
lead us towards theories with fields localized to branes. In order to be
able to examine such theories, we will consider how to write down an 
effective Lagrangian for a theory with a brane, and find out how the different
modes in such theories would couple to the particles of the SM 
of particle physics. We close this section by explaining how to calculate
various processes including the Kaluza-Klein modes of the graviton, and 
by briefly sketching ideas about black-hole production in theories with large
extra dimensions. The very basics of Kaluza-Klein theories have been 
explained in the lectures by Keith Dienes~\cite{Keith}. Therefore I will assume that the
reader is familiar with the concept of Kaluza-Klein decomposition of a higher
dimensional field. Otherwise, the lectures should be self-contained.

\subsection{Matching the higher dimensional theory to the 4D effective
theory}
The first question that we would like to answer is how large 
the extra dimensions could possibly be without us having them noticed
until now. For this we need to understand how the effectively
four dimensional world that we observe would be arising from the higher
dimensional theory. In more formal terms, this procedure is called matching
the effective theory to the fundamental higher dimensional theory. Thus 
what we would like to find out is how the observed gauge and gravitational 
couplings (which should be thought of as effective low-energy couplings)
would be related to the ``fundamental parameters'' of the higher dimensional
theory. 

Let us call the fundamental (higher dimensional) Planck scale of the theory
$M_*$, assume that there are $n$ extra dimensions, and that the radii of
the extra dimensions are given by $r$. In order to be able to perform the 
matching, we need to first write down the action for the higher dimensional
gravitational theory, including the dimensionful constants, since these are 
the ones we are trying to match. For this it is very useful to examine the mass 
dimensions of the various quantities that will appear. The infinitesimal
distance is related to the coordinates and the metric tensor by 
\begin{equation}
ds^2= g_{MN} dx^M dx^N
\end{equation}
We will always be using the $(+,-,-,\ldots ,-)$ sign convention for the metric.
Assuming that the coordinates carry proper dimensions (that is they are NOT
angular variables) the metric tensor is dimensionless, $[g]=0$. Since we can
calculate the Christoffel symbols  as
\begin{equation}
\Gamma_{MN}^A \sim g^{AB} \partial_M g_{NB}
\end{equation} 
we get that the Christoffel symbols carry dimension one, $[\Gamma ]=1$.
Since $R_{MN} \sim \Gamma^2$, the Ricci tensor will carry dimension
two, $[R_{MN}]=2$, and similarly the curvature scalar $[R]=2$. 
The main point is that all of this is independent of the total number of
dimensions, since these were based on local equations. In order to 
generalize the Einstein-Hilbert action to more than four dimensions, we simply
assume that the action will take the same form as in four dimensions:
\begin{equation}
S_{4+n} \sim \int d^{4+n}x \sqrt{g^{(4+n)}} R^{(4+n)}.
\end{equation}
In order to make the action dimensionless, we need to multiply by the 
appropriate power of the fundamental Planck scale $M_*$. Since $d^{4+n} x$
carries dimension $-n-4$, and $R^{(4+n)}$ carries dimension $2$, 
this has to be the power $n+2$, thus we take
\begin{equation}
\label{extradaction}
S_{4+n} = -M_*^{n+2} \int d^{4+n}x \sqrt{g^{(4+n)}} R^{(4+n)}.
\end{equation}
What we need to find out is how the usual 
four dimensional action 
\begin{equation}
\label{4daction}
S_{4} = -M_{Pl}^{2} \int d^{4}x \sqrt{g^{(4)}} R^{(4)}.
\end{equation}
is contained in this higher dimensional expression. Here $M_{Pl}$ is the observed
4D Planck scale $\sim 10^{18}$ GeV.
For this we need to make
some assumption about the geometry of the space-time. We will for now
assume, that spacetime is flat, and that the $n$ extra dimensions are compact.
So the metric is given by
\begin{equation}
\label{metric}
ds^2 = (\eta_{\mu\nu}+h_{\mu\nu}) dx^\mu dx^\nu -r^2 d \Omega_{(n)}^2,
\end{equation}
where $x_\mu$ is a four dimensional coordinate, $ d \Omega_{(n)}^2$ 
corresponds to the line element of the flat extra dimensional space in some parametrization, $\eta_{\mu\nu}$
is the flat (Minkowski) 4D metric, and $h_{\mu\nu}$ is the 4D fluctuation
of the metric around its minimum. The reason why we have only put in 
4D fluctuations is that our goal is to find out how the usual 4D action is
contained in the higher dimensional one. For this the first thing to find
out is how the 4D graviton is contained in the higher dimensional metric, this 
is precisely what is given in (\ref{metric}). This does not mean that there 
wouldn't be additional terms (and in fact there will be as we will see very 
soon). From this we can now calculate the expressions that appear in 
(\ref{extradaction}):
\begin{equation}
\sqrt{g^{(4+n)}} = r^n \sqrt{g^{(4)}}, \ \ R^{(4+n)}= R^{(4)},
\end{equation}
where these latter quantities are to be calculated from $h$. Therefore we get 
\begin{equation}
\label{matchingaction}
S_{4+n} = -M_*^{n+2} \int d^{4+n}x \sqrt{g^{(4+n)}} R^{(4+n)}=
-M_*^{n+2} \int d \Omega_{(n)} r^n \int d^{4}x \sqrt{g^{(4)}} R^{(4)}.
\end{equation} 
The factor $\int d \Omega_{(n)} r^n$ is nothing but the volume of the 
extra dimensional space which we denote by $V_{(n)}$. For toroidal 
compactification it would simply be  given by $V_{(n)} =(2\pi r)^n$. 
Comparing (\ref{matchingaction}) with (\ref{4daction}) we find the 
matching relation for the gravitational couplings that we have looked for:
\begin{equation}
M_{Pl}^2= M_*^{n+2} V_{(n)}=M_*^{n+2} (2\pi r)^n.
\label{grmatching}
\end{equation}
Let us now repeat the same matching procedure for the gauge couplings. 
Assume that the gauge fields live in the extra dimensions, and use 
 a normalization where the gauge fields are not canonically normalized:
\begin{equation}
\label{gaugeaction}
S^{(4+n)}=-\int d^{4+n}x \frac{1}{4 g_*^2} F_{MN}F^{MN} \sqrt{g^{(4+n)}}.
\end{equation}
$M,N$ denote indices that range from 1 to $4+n$,
and $g_*$ denotes the higher dimensional (``fundamental'') gauge coupling. 
Clearly, the four dimensional
part of the field strength $F_{\mu\nu}$ is 
included in the full higher dimensional $F_{MN}$. Again performing the 
integral over the extra dimension we find:
\begin{equation}
\label{gaugeact2}
S^{(4)}=-\int d^{4}x \frac{V_{(n)}}{4 g_*^2} F_{\mu\nu}F^{\mu\nu} 
\sqrt{g^{(4)}}.
\end{equation}
Thus the matching of the gauge couplings is given by
\begin{equation}
\label{gaugematching}
\frac{1}{g_{eff}^2}=\frac{V_{(n)}}{g_*^2}.
\end{equation}
Note, that it is clear from this equation, that the coupling constant of a 
higher dimensional gauge theory is not dimensionless, but rather it has 
dimension $[g_*]=-n/2$. As a consequence it is not a renormalizable theory,
but can be thought of as the low-energy effective theory of some more 
fundamental theory at even higher energies.

Now let us try to understand the consequences of (\ref{grmatching}) and 
(\ref{gaugematching}). Since the gauge coupling is dimensionful in extra 
dimensions, one needs to ask what should be its natural size. The simplest 
assumption is that the same physics that sets the strength of gravitational
couplings would also set the gauge coupling, and thus
\begin{equation}
g_* \sim \frac{1}{M_*^{\frac{n}{2}}}.
\end{equation}
Then we would have the following two equations:
\begin{eqnarray}
&& \frac{1}{g_4^2}= V_{(n)} M_*^n \sim r^n M_*^n \nonumber \\
&& M_{Pl}^2= V_{(n)} M_*^{n+2} \sim r^n M_*^{n+2},
\end{eqnarray}
from which it follows that 
\begin{equation}
r\sim \frac{1}{M_{Pl}} g_4^{\frac{n+2}{n}}.
\end{equation}
This would imply that in a ``natural'' higher dimensional theory
$r\sim 1/M_{Pl}$! In this case there would be no hope of 
finding out about the existence of these tiny extra dimensions in the
foreseeable future. This is what the prevailing view has been until the 90's
about extra dimensions. However, we should note that these arguments crucially
depended on the ASSUMPTION that every field propagates in all dimensions.
The purpose of these lectures is to understand, what kind of physical phenomena
one could expect if some or all the fields of the standard model were localized
in the extra dimensions to a ``brane''. This possibility has been first raised 
in~\cite{RubShap,Visser} (see also \cite{Akama,DvaliShifman}). 

Before jumping into the detailed description of theories with branes,
we would like to first understand what the restrictions on the size of 
extra dimensions would be, if contrary to the previous assumption 
the SM fields were localized to 4 dimensions, and only gravity (or other yet 
unobserved fields) were to propagate into the extra dimension. In this case,
new physics will only appear in the gravitational sector, and only 
when distances as short as the size of the extra dimension are actually reached. 
However, it is very hard to test gravity at very short distances. The
reason is that gravity is a much weaker interaction than all the other forces.
Over large distances gravity is dominant because there is only one
type of gravitational charge, so it cannot be screened. However, as one starts
going to shorter distances, inter-molecular van der Waals forces and
eventually bare electromagnetic forces will be dominant, which will
completely overwhelm the gravitational forces. This is the reason why 
the Newton-law of gravitational interactions 
has only been tested down to about a fraction of a millimeter using 
essentially Cavendish-type experiments~\cite{gravityexp}. Therefore, the real bound on the
size of an extra dimension is 
\begin{equation}
r \leq 0.1\ {\rm mm}
\end{equation}
if only gravity propagates in the extra dimension. How would a large value
close to the experimental bound affect the fundamental Planck scale $M_*$?
Since we have the relation $M_{Pl}^2 \sim M_*^{n+2} r^n$, if $r> 1/M_{Pl}$,
the fundamental Planck scale $M_*$ will be lowered from $M_{Pl}$. How low
could it possibly go down? If $M_* < 1$ TeV, that would imply that quantum
gravity should have already played a role in the collider experiments that 
have been performed until now. Since we have not seen a hint of that, 
one has to impose that $M_* \geq 1$ TeV. So the lowest possible 
value (and thus the largest possible size of the extra dimensions) would be 
for $M_* \sim 1$ TeV. Such models are called theories with ``Large extra
dimensions'', proposed by Arkani-Hamed, Dimopoulos and Dvali~\cite{ADD}, see also
\cite{AADD}. For earlier papers where the possibility of lowering the fundamental
Planck scale has been mentioned see~\cite{antoniadis,Lykken}.
Let us check, how large a radius one would need, if in fact 
$M_*$ was of the order of a TeV. Reversing the expression 
$M_{Pl}^2 \sim M_*^{n+2} r^n$ we would now get
\begin{equation}
\frac{1}{r}=M_* \left( \frac{M_*}{M_{Pl}}\right)^\frac{2}{n} = 
(1 {\rm TeV}) 10^{-\frac{32}{n}},
\end{equation}
where we have used $M_* \sim 10^3$ GeV and $M_{Pl} \sim 10^{19}$ GeV.
To convert into conventional length scales one should keep the 
conversion factor
\begin{equation}
1 {\rm GeV}^{-1} =2 \cdot 10^{-14} {\rm cm}
\end{equation}
in mind. Using this we finally get
\begin{equation}
r\sim 2\cdot 10^{-17} 10^{\frac{32}{n}} \ {\rm cm}.
\end{equation}
For $n=1$ this would give the absurdly large value of $r=2\cdot10^{15}$ cm,
which is grater than the astronomical unit of $1.5 \times 10^{13}$ cm. This is clearly
not possible: there can't be one flat large extra dimension if one would like
to lower $M_*$ all the way to the TeV scale. However, already for two extra
dimensions one would get a much smaller number $r\sim 2$ mm. This is 
just borderline excluded by the latest gravitational experiments performed
in Seattle~\cite{Eotwash}. Conversely, one can set a bound on the size of two large extra 
dimensions from the Seattle experiments, which gave $r\leq 0.2$ mm$=10^{12}$
1/GeV. This results in $M_* \geq 3$ TeV. We will see that for two extra 
dimensions there are in fact more stringent bounds than the 
direct bound from gravitational measurements.

For $n>2$ the size of the extra dimensions is less than $10^{-6}$ cm, which
is unlikely to be tested directly via gravitational measurements any time soon.
Thus for $n>2$ $M_* \sim 1$ TeV is indeed a possibility that one has to 
carefully investigate. If $M_*$ was really of order the TeV scale, there 
would no longer be a large hierarchy between the  fundamental Planck scale
$M_*$ and the scale of weak interactions $M_w$, thus this would resolve the 
hierarchy problem. In this case gravity would appear weaker than the other
forces at long distances because it would get diluted by the large volume
of the extra dimensions. However, this would only be an apparent hierarchy 
between the strength of the forces, as soon as one got below scales of order
$r$ one would start seeing the fundamental gravitational force, and the 
hierarchy would disappear. However, as soon as one postulates the equality 
of the strength of the weak and gravitational interactions one needs to 
ask why this is not the scale that sets the size of the extra dimensions 
themselves. Thus by postulating a very large radius for the extra dimensions
one would merely translate the hierarchy problem of the scales of interactions 
into the problem of why the size of the extra dimension is so large compared 
to its natural value.

\subsection{What is a brane and how to write an effective theory for it?}

Above we have seen that theories where certain particles (especially the 
light SM particles) are localized to four dimensions, while other particles
could propagate in more dimensions could be very interesting. In these lectures
we would like to study theories of this sort. We will refer to 
the surface along which some of the particles are localized as 
``branes'', which stands for a membrane that could have more spatial 
dimensions  than the usual 2 dimensional membrane. A p-brane will mean that
the brane has  p spatial dimensions, so a 2-brane is just the usual
membrane, a 1-brane is just a string, while the most important object for
us in these lectures will be a 3-brane, which has 3 spatial dimensions just 
like our observed world, which could be embedded into more dimensions.

What is really a brane? In field theory, it is best to think of it as a 
topological defect (like a soliton), which could have fields localized to 
its surface. For example, as we will see in the next lecture in detail,
domain walls localize fermions to the location of the domain wall. 
String theories also contain objects called D-branes (short for Dirichlet 
banes). These are surfaces where open string can end on. 
These open strings will give rise to all kinds of fields localized 
to the brane, including gauge fields. In the supergravity 
approximation these D-branes will also appear as solitons of the 
supergravity equations of motion. 

In our approach, we will use a low-energy 
effective field theory description. We will (usually) not care very much 
where these branes come from, but simply assume that there is some consistent
high-energy theory that would give rise to these objects. Therefore,
our theory will be valid only up to some cutoff scale, above which
the dynamics that actually generates the brane has to be taken into account. For our discussion we will follow
the description and notation of Sundrum in~\cite{Raman}.

To describe the branes, let us first set up some notation. The 3-brane 
will be assumed to be described by a flat four-dimensional space-time 
equivalent to $R^4$, while the extra dimensions by $R^n$ or if compactified by 
$T^n$ (an $n$ dimensional torus). The coordinates in the bulk 
(bulk stands for all of spacetime) are denoted by
$X^M$, $M=0,1,\ldots , 3+n$, the coordinates on the brane are denoted by
$x_\mu$, $\mu =0,1,2,3$, while the coordinates along the extra dimensions
only are denoted by $x_m$, $m=4,\ldots , 3+n$. For now we will 
concentrate on the bosonic degrees of freedom. What are these degrees of 
freedom that we need to discuss in a low-energy effective theory? Since we 
need to discuss the physics of higher dimensional gravity, this should include
the metric in the $4+n$ dimensions $G_{MN}(X)$, and also the position 
of the brane in the extra dimensions $Y^M(x)$. Note, that the metric is a 
function of the bulk coordinate $X^M$, while the position of the brane is 
a function of the coordinate {\it along} the brane $x_\mu$. In addition,
we would like to take into account the fields that are localized to live along 
the brane. These could be some scalar fields $\Phi (x)$, gauge fields 
$A_\mu (x)$ or fermions $\Psi_L (x)$. These fields are also functions of the
coordinate along the brane $x_\mu$. 
\begin{figure}[h!t]
\centerline{\includegraphics[width=0.5\hsize]{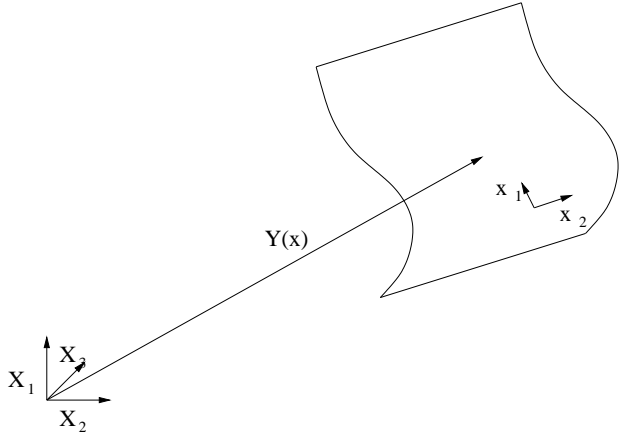}}
\caption{Parametrization of the position of the brane in the bulk.} 
\label{brane}
\end{figure}

The effective theory that we are trying to build up should describe small 
fluctuations of the the fields around the vacuum state. So we have to specify 
what we actually mean by the vacuum. We will assume that we have a flat brane 
embedded into flat space. The corresponding choice of vacua is then given by
\begin{eqnarray}
\label{vacuum}
&& G_{MN}(X)=\eta_{MN} \nonumber \\
&& Y^M(x)=\delta^M_\mu x^\mu.
\end{eqnarray} 
Here and everywhere in the lectures we will use the metric convention
$\eta_{MN}={\rm diag} (+,-,$ $-, \ldots ,-)$. The bulk action will then 
just be the usual higher dimensional Einstein-Hilbert action as discussed
in the previous section, plus perhaps a term from a 
bulk cosmological constant $\Lambda$:
\begin{eqnarray}
S_{bulk}=-\int d^{4+n}X \sqrt{|G|} (M_*^{n+2} R^{(4+n)}+\Lambda).
\end{eqnarray}
In order to be able to write down an effective action for the brane localized
fields one has to first find the induced metric on the brane. 
This is the metric that should be used to contract Lorentz indices of the 
brane field. To get the induced metric, we need to find what the distance
between two point on the branes is, $x$ and $x+dx$:
\begin{equation}
ds^2=G_{MN} dY(x)^M dY(x)^N=G_{MN} \frac{\partial Y^M}{\partial x^\mu} dx^\mu
\frac{\partial Y^N}{\partial x^\nu} dx^\nu.
\end{equation}
From this the induced metric can be easily read off:
\begin{equation}
g_{\mu\nu}= G_{MN} (Y(x)) \partial_\mu Y^M \partial_\nu Y^N.
\label{induced}
\end{equation}
This is the general expression for arbitrary background metric and for an 
arbitrary brane. For the flat vacuum that we have chosen one can easily see 
from (\ref{induced}) that the induced metric will fluctuate around 
the flat 4D Minkowski metric $\eta_{\mu\nu}$.

After this we can discuss the basic principle for writing down the
brane induced part of the action: it has to be invariant both under
the general coordinate transformation of the bulk coordinates 
$X$ {\it and} under the general coordinate transformations of $x$. It is 
clear that the  invariance under the general coordinate transformation of the 
bulk coordinates just corresponds to the usual general covariance of a 
higher dimensional gravitational theory. The additional requirement 
that the action also be invariant under the coordinate transformations of 
the brane coordinate $x$ is an expression of the fact that $x$ is just one 
possible parametrization of the surface (brane), which itself can not 
have a physical significance, and any different choice of parametrization
has to give the same physics. This will ensure the usual 4D Lorentz
invariance of the brane induced action. Thus there are two separate coordinate
transformations that the action has to be invariant under, that corresponding
to $X$ and to $x$. Practically this means that one has to contract 
bulk indices with bulk indices, and brane indices with brane indices. For
example $\partial_\mu Y^M$ would be a vector under both the bulk and 
the brane coordinate invariance, and both indices have to be contracted
to form a scalar that can appear in the action, while the induced metric
$g_{\mu\nu}$ would be a tensor under the brane reparametrization, but a scalar
under bulk coordinate invariance, etc. Thus the general form of the 
brane action would be of the form
\begin{equation}
S_{brane}=\int d^4 x \sqrt{|g|} \left[ -f^4-R^{(4)} +\frac{g^{\mu\nu}}{2} 
D_\mu \Phi D_\nu \Phi -V(\Phi) -\frac{g^{\mu\nu}g^{\rho\sigma}}{4} F_{\mu\rho}
F_{\nu\sigma}+\ldots \right].
\end{equation}
Here the possible constant piece $f$ corresponds to the energy density of the brane,
called the brane-tension. This brane tension has to be small (in the units of the fundamental
Planck scale) in order to be able to neglect its back-reaction on the gravitational background.
In Chapter 4 we will investigate warped backgrounds where this back-reaction will be taken into account. 
As discussed above, in addition to the usual bulk coordinate invariance which 
everyone is familiar with, there is also a 4D reparametrization invariance, 
which is corresponding to the fact that a different parametrization of the
surface describing the brane would yield the same physics, $x\to x'(x)$ 
is an invariance of the Lagrangian. Thus one needs an additional gauge fixing 
condition, which will eliminate the non-physical components from $Y^M$.
There are four coordinates, so one needs four conditions, which can be picked
as
\begin{equation}
Y^\mu (x)=x^\mu,
\end{equation}
which is a complete gauge fixing. Thus out of the $4+n$ components of 
$Y^M$ only the components along the extra dimension $Y^m (x)$, $m=4,5,\ldots 
,3+n$ correspond to physical degrees of freedom. These $n$ physical fields 
correspond to the position of the brane within the bulk.

Let us now discuss how to normalize the fields in order to end up with
canonically normalized 4D actions. The bulk metric
$G_{MN}$ is dimensionless, and we expand it in fluctuations around the 
background which we assume to be flat space:
\begin{equation}
G_{MN}=\eta_{MN}+\frac{1}{2 M_*^{\frac{n}{2}+1}} h_{MN}.
\end{equation}
This way the graviton fluctuation $h_{MN}$ has dimension $\frac{n}{2}+1$
which is the right one for a bosonic field in $4+n$ dimensions.
The prefactor was chosen such that the kinetic term reproduces the 
canonically normalized kinetic term when expanding the Einstein-Hilbert
action in $h$. 

To get a canonically normalized field for the coordinates describing the
position of the brane $Y^m(x)$ we expand the leading term in the 
brane-action which is just the brane tension:
\begin{equation}
\int d^4x \sqrt{|g|} \left[ -f^4 +\ldots \right],
\end{equation}
where for the induced metric the leading dependence on $Y$ is
\begin{equation}
g_{\mu\nu} =G_{MN} \partial_\mu Y^m \partial_\nu Y_m = 
\eta_{\mu\nu}+\partial_\mu Y^m \partial_\nu Y_m +\ldots,
\end{equation}
and expanding the determinant of the metric in powers of $Y$ we
obtain
\begin{equation}
\det g=-\partial_\mu Y^m\partial^\mu Y_m,
\end{equation}
from which the leading term in the action is
\begin{equation}
S=\int d^4x f^4 \partial_\mu Y^m \partial^\mu Y_m.
\end{equation}
Thus the canonically normalized field will be 
\begin{equation}
Z^m \equiv f^2 Y^m.
\end{equation}
Note, that it is the brane tension which sets the size of the kinetic term 
for the $Y^m$ fields. In particular, if the tension is negative, then one
would have a field with negative kinetic energy, which is thus a 
physical ghost. This shows that a brane with negative tension is likely 
unstable, the brane wants to crumble, unless somehow these modes with negative
kinetic energy are projected out, by not allowing the brane to move.
This possibility will be encountered when we consider branes at orbifold
fixed points. 

\subsection{Coupling of SM fields to the various graviton components}

In the previous section we have discussed how one should construct 
an action for fields localized on branes coupled to gravity. Next we would
like to explicitly construct the generic interaction Lagrangians
between the matter on the brane (which we will simply call the SM matter)
and the various graviton modes. For our discussion we will follow the work of
Giudice, Rattazzi and Wells~\cite{GRW}. We have seen above that the SM fields 
feel only the induced metric
\begin{equation}
g_{\mu\nu} (x)=G_{MN}(X) \partial_\mu Y^M \partial_\nu Y^N.
\end {equation}
It is quite clear from the previous subsection how to deal with the $Y^m$ 
fields, so we will concentrate on the modes of the bulk graviton,
and for now set the fluctuations of the brane to zero, that is set 
$Y^M=\delta^M_\mu x^\mu$, and thus $Y^m=0$. In this case
\begin{equation}
g_{\mu\nu} (x)=G_{\mu\nu} (x_\mu,x^m=0).
\end{equation}
The action is given by 
\begin{equation}
S=\int d^4x {\cal L}_{SM} \sqrt{g} (g_{\mu\nu}, \Phi ,\Psi ,A,\ldots )
\end{equation}
The definition of the energy-momentum tensor is given by
\begin{equation}
\sqrt{g} T^{\mu\nu} =\frac{\delta S_{SM}}{\delta g_{\mu\nu}}.
\end{equation}
From this it is clear that at linear order the interaction between the 
SM matter and the graviton field is given by (expanding in the fluctuation
around flat space again $g_{\mu\nu}=\eta_{\mu\nu}+\frac{h_{\mu\nu} 
(x_\mu )}{M_*^{\frac{n}{2}+1}}$):
\begin{equation}
\label{gravcoupling}
S_{int}=\int d^4x T^{\mu\nu} \frac{h_{\mu\nu} (x_\mu )}{M_*^{\frac{n}{2}+1}}.
\end{equation}
Thus generically the graviton couples linearly to the 
energy-momentum tensor of the matter (this is in fact the definition of 
the stress-energy tensor, but it is a quantity that we know quite well
and is easy to find). Note, that in the above expression what appears
is the graviton field at the position of the brane, which is not a mass
eigenstate field from the 4D point of view, but rather a superposition 
of all KK modes. Using the lecture by K. Dienes we write the graviton field
in the KK expansion as 
\begin{equation}
h_{MN} (x,y)=\sum_{k_1=-\infty}^{\infty} \ldots \sum_{k_n=-\infty}^{\infty}
\frac{h^{\vec{k}}_{MN} (x)}{\sqrt{V_n}} e^{i\frac{\vec{k}\cdot \vec{y}}{R}},
\end{equation}
where we have denoted the coordinates along the extra dimension by
$y^m$ (until now they were simply denoted by $x^m$), 
and assumed a toroidal compactification with volume
$V_n=(2\pi R)^n$. Plugging the KK expansion back into (\ref{gravcoupling})
we find the coupling of the SM fields to the individual KK modes to be
\begin{equation}
\sum_{\vec{k}} \int d^4 x T^{\mu\nu} \frac{1}{M_*^{\frac{n}{2}+1}}
\frac{h_{\mu\nu}^{\vec{k}}}{\sqrt{V_n}}=\sum_{\vec{k}}\int d^4 x
\frac{1}{M_{Pl}}T^{\mu\nu}h_{\mu\nu}^{\vec{k}},
\end{equation}
where we have again used the relation between the fundamental Planck scale 
and the observed one. Thus we can see that an individual 
KK mode couples with strength $1/M_{Pl}$ to the SM fields. However, since there
are many of them, the total coupling in terms of the field at the brane sums
up to a coupling proportional to $1/M_*$, as we have seen in 
(\ref{gravcoupling}).

Next let us discuss what the different modes contained in the bulk graviton field 
are. Clearly, the graviton is a $D$ by $D$ symmetric tensor, where
$D=n+4$ is the total number of dimensions. Therefore this tensor has in principle
$D(D+1)/2$ components. However, we know that general relativity has a large 
gauge symmetry -- $D$ dimensional general coordinate invariance. Therefore
we can impose $D$ separate conditions to fix the gauge, for example using the
harmonic gauge 
\begin{equation}
\partial_M h^M_N =\frac{1}{2} \partial_N h^M_M.
\end{equation}
However, just like in the ordinary Lorentz gauge for gauge theories, this is not yet a complete
gauge fixing. Gauge transformations which satisfy the equation 
$\Box \epsilon_M=0$ are still allowed (where the gauge transformation is $h_{MN}\to h_{MN}+\partial_M \epsilon_N+
\partial_N \epsilon_M$), and this means that another $D$ conditions
can be imposed. This means that generically a graviton has 
$D(D+1)/2-2D=D(D-3)/2$ independent degrees of freedom. For $D=4$ this gives the usual
2 helicity states for a massless spin two particle, however in $D=5$ we get 5 components,
in $D=6$ we get 9 components, etc. This means that from the 4D point of view a higher dimensional
graviton will contain particles other than just the ordinary 4D graviton. This is quite clear,
since the higher dimensional graviton has more components, and thus will have to contain 
more fields. The question is what these fields are and how many degrees of freedom they contain.
We will go through these modes carefully in the following.

\begin{itemize}
\item The 4D graviton and its KK modes
\end{itemize}

These live in the upper left 4 by 4 block of the bulk graviton which is given by a 
$4+n$ by $4+n$ matrix:
\begin{equation}
\left( \begin{array}{ccc|ccccc} & & & & & & & \\ & G_{\mu\nu}^{\vec{k}} & & & & & & \\ & & & & & & &\\ 
\hline & & & & & & & \\ & & & & & & & \\ & & & & & & & \\ & & & & & & & \\ & & & & & & &\\ \end{array}
\right)
\end{equation}
These modes are labeled by the vector $\vec{k}$, which is an $n$-component vector,
specifying the KK numbers along the various extra dimensions. These modes are generically 
massive, except for the zero mode. In the limit of no sources these modes satisfy the 4D equation
\begin{equation}
(\Yfund +\hat{k}^2) G_{\mu\nu}^{\vec{k}}=0
\end{equation}
which as usual would imply 10 components, however the gauge conditions
\begin{equation}
\partial^\mu G_{\mu\nu}^{\vec{k}}=0, \ \ \ G_{\mu}^{\mu \vec{k}}= 0
\end{equation}
eliminate 5 components, and we are left generically with 5 degrees of freedom, which is just the 
right number for a {\it massive} graviton in four dimensions. The reason is that a massive graviton
contains a normal 4D massless graviton with two components, but also ``eats'' a massless gauge 
field and a massless scalar, as in the usual Higgs mechanism. Thus $5=2+2+1$.
 
\begin{itemize}
\item 4D vectors  and their  KK modes
\end{itemize}

The off-diagonal blocks of the bulk graviton form vectors under the four dimensional Lorentz
group. This is clear, since they come from the $G_{\mu j}$ components of the graviton:
\begin{equation}
\left( \begin{array}{cccc|ccccc} & & & & & & & \\ & & & & & V_{\mu j}^{\vec{k}} & & \\ & & & & & & &\\ 
\hline & & & & & & & \\ & & & & & & & \\ & V_{\mu j}^{\vec{k}}& & & & & & \\ & & & & & & & \\ & & & & & & &\\ \end{array}
\right)
\end{equation}
Naively one could think there there would be $n$ such four dimensional vectors, however we have seen
before that the 4D graviton has to eat one 4D vector to form a full massive KK tower, thus 
there are only $n-1$ massive KK towers describing spin 1 particles in 4D. The missing of the 
last tower is expressed by the constraint
\begin{equation}
\hat{k}^j V_{\mu j}^{\vec{k}} =0.
\end{equation}
The usual Lorentz gauge condition can also be imposed
\begin{equation}
\partial^\mu V_{\mu j}^{\vec{k}}=0.
\end{equation}
Each of these massive vectors absorbed a scalar via the Higgs mechanism. 

\begin{itemize}
\item 4D scalars and their  KK modes
\end{itemize}

The remaining lower right $n$ by $n$ block of the graviton matrix clearly corresponds to 
4D scalar fields: 
\begin{equation}
\left( \begin{array}{ccc|ccccc} & & & & & & & \\ & & & & & & & \\ & & & & & & &\\ 
\hline & & & & & & & \\ & & & & & & & \\ & & & & & S^{\vec{k}}_{ij}& & \\ & & & & & & & \\ & & & & & & 
&\\ \end{array}
\right)
\end{equation}
Originally there are $n(n+1)/2$ scalars, however as we discussed before the graviton eats
one scalar, and the remaining $n-1$ vectors eat one scalar each. In addition, there is a special
scalar mode, whose zero mode sets the overall size of the internal manifold, and therefore
this special scalar is usually called the {\it radion}. Thus there are $n(n+1)/2-n-1=(n^2-n-2)/2$ scalars
left. The equations that express the fact that $n$ fields were eaten are 
\begin{equation}
\hat{k}^jS^{\vec{k}}_{jk}=0,
\end{equation}
while the fact that we usually separate out the radion as a special field not included among the 
scalars is expressed as the additional condition
\begin{equation}
S^{\vec{k} j}_j =0.
\end{equation}
The radion is then given by $h^{\vec{k} j}_j$.

Let us now count the total number of degrees of freedom taken into account:
$5$ (graviton) $+3(n-1)$ (vectors) $+(n^2-n-2)/2$ (scalars) $+1$ (radion) 
$=(4+n)(1+n)/2=D(D-3)/2$. Thus we can see that all the modes of the graviton have been accounted for.

The explicit expressions for the canonically normalized 4D fields are given in unitary gauge by
(using the notation $\kappa=\sqrt{\frac{3(n-1)}{n+2}}$):
\begin{eqnarray}
&{\rm radion} & H^{\vec{k}}=\frac{1}{\kappa} h^{\vec{k} j}_j \nonumber \\
&{\rm scalars} & S^{\vec{k}}_{ij} = h^{\vec{k}}_{ij}-\frac{\kappa}{n-1} 
(\eta_{ij}+\frac{\hat{k}_i\hat{k}_j}{\hat{k}^2}) H^{\vec{k}} \nonumber \\
&{\rm vectors} & V_{\mu j}^{\vec{k}} =\frac{i}{\sqrt{2}}h^{\vec{k}}_{\mu j} \nonumber \\
&{\rm gravitons} & G_{\mu\nu}^{\vec{k}}=h^{\vec{k}}_{\mu \nu}+\frac{\kappa}{3}(\eta_{\mu\nu}
+\frac{\partial_\mu\partial_\nu}{\hat{k}^2}) H^{\vec{k}}.
\end{eqnarray}
The equation of motion in the presence of sources is given for the above fields by
\begin{equation}
(\Yfund +\hat{k}^2) \left( \begin{array}{c} G_{\mu\nu}^{\vec{k}} \\ V_{\mu j}^{\vec{k}} \\S^{\vec{k}}_{ij} \\
H^{\vec{k}} \end{array} \right) = \left( \begin{array}{c} 
\frac{1}{M_{Pl}}[-T_{\mu\nu}+(\eta_{\mu\nu}
+\frac{\partial_\mu\partial_\nu}{\hat{k}^2}) T^\mu_\mu/3 ] \\ 0 \\ 0 \\
\frac{\kappa}{3 M_{Pl}} T^\mu_\mu \end{array} \right).
\end{equation}
We can see why the radion is special: besides setting the overall size of the compact dimension
it also is the only field besides the 4D graviton that couples to brane sources. It couples to the 
trace of the energy-momentum tensor. All the other fields are not important when one tries to 
calculate the coupling of the matter fields on the brane to the modes of the bulk graviton.

For example, if we take QED on the brane, the Lagrangian is given by
\begin{equation}
{\cal L}=\sqrt{g} (i \bar{\Psi} \gamma^a D_a \Psi -\frac{1}{4} F_{\mu\nu}F^{\mu\nu}),
\end{equation}
where the covariant derivative  for fermions is given by
\begin{equation}
D_a= e^M_a (\partial_\mu-ieQ A_\mu +\frac{1}{2} \sigma^{bc}e_b^\nu\partial_\mu e_{\nu c} ),
\end{equation}
where $e^M_a$ is the vielbein, and $\sigma^{bc}$ the spin-connection (we will not go into 
detail in these lectures on how to couple fermions to gravity, for those interested in more
details we refer to \cite{Raman}).
From this the energy-momentum tensor is given by 
\begin{equation}
T_{\mu\nu} =\frac{i}{4}\bar{\Psi}(\gamma_\mu\partial_\nu +\gamma_\nu\partial_\mu )\Psi
-\frac{i}{4} (\partial_\mu \bar{\Psi}\gamma_\nu+\partial_\nu \bar{\Psi}\gamma_\mu ) \Psi+
\frac{1}{2} eQ\bar{\Psi} (\gamma_\mu A_\nu+\gamma_\nu A_\mu )\Psi +F_{\mu\lambda}F^{\lambda}_\nu
+\frac{1}{4} \eta_{\mu\nu}F^{\lambda\rho}F_{\lambda\rho}.
\end{equation}
Typical Feynman diagrams involving the graviton which we would get from this $T$ when coupled to gravity on the 
brane are given by:

\centerline{\includegraphics[width=0.15\hsize]{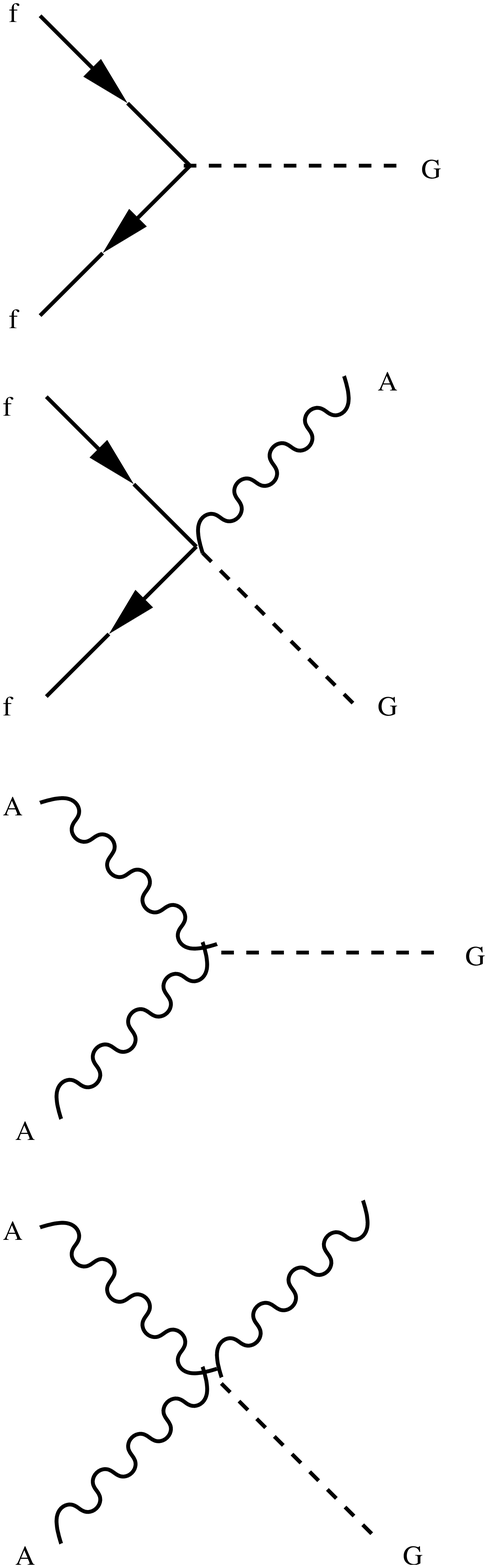}}

The last one appears only in non-Abelian gauge theories. There are similar vertices involving the radion field.

\subsection{Phenomenology with large extra dimensions}
We have discussed before that the mass splitting of the KK modes in large extra dimensional
theories is extremely small, 
\begin{equation}
\Delta m \sim \frac{1}{r} = M_* \left(\frac{M_*}{M_{Pl}}\right)^\frac{2}{n} = 
\left(\frac{M_*}{{\rm TeV}}\right)^{\frac{n+2}{2}} 10^{\frac{12n-31}{n}} \ \ {\rm eV}.
\end{equation}
This implies that for a typical particle physics process with high energies there
is an enormous number of KK modes available. This suggests that since the splitting of 
the KK modes is very small, it is useful to turn the sum over KK modes into an integral.
If $N$ denotes the number of KK modes whose momentum along the extra dimension is less than
$k$, then clearly
\begin{equation}
dN = S_{n-1} k^{n-1}dk,
\end{equation}
where $S_n= (2\pi)^{\frac{n}{2}}/ \Gamma (n/2)$ is the surface of an $n$ dimensional sphere 
with unit radius. The actual mass of a given KK mode is given by
$m=\frac{|k|}{R}$, therefore we get that
\begin{equation}
dN=S_{n-1}m^{n-1}R^{n-1} dm R = S_{n-1} \frac{M_{Pl}^2}{M_*^{n+2}} m^{n-1}dm.
\end{equation}
In order to calculate an inclusive cross-section for the production of graviton modes,
what one needs to do is to first calculate the cross-section for the production 
of an individual mode with mass $m$, $d\sigma_m/dt$, and then using the above formula
get 
\begin{equation}
\frac{d^2\sigma}{dt dm}=S_{n-1} \frac{M_{Pl}^2}{M_*^{n+2}} m^{n-1} \frac{d\sigma_m}{dt}.
\end{equation}
Since the cross section for an individual KK mode is proportional to $1/M_{Pl}^2$,
the inclusive cross section will have  a behavior of the form
\begin{equation}
\frac{d^2\sigma}{dt dm} \sim S_{n-1} \frac{m^{n-1}}{M_*^{n+2}} .
\end{equation}

With this we have basically covered all elements of calculating processes with large
extra dimensions. In the following we will briefly list some of the most interesting
features/constraints on these models. Clearly, not everything will be covered here,
for those interested in further details we refer to the original papers. A nice general 
overview of the phenomenology of large extra dimensions is given in~\cite{ADDphen}. For collider signals
see~\cite{GRW,HLZ,Hewett,otherADDphen}. For the running of couplings and unification in extra dimensions
see~\cite{DDG}. For consequences in electroweak precision physics see~\cite{ewprecision}. For neutrino
physics with large extra dimensions see~\cite{ADDneutrino}. For topics related to inflation with flat 
extra dimensions see~\cite{ADDinflation}. Issues related to radius stabilization for large extra dimensions
is discussed in~\cite{ADDradion}. Connections to string theory model building can be found in for 
example in~\cite{shiutye,pomarolquiros,stringapproach}. More detailed description of supernova cooling into extra dimensions
is in~\cite{SN}, while of black hole production in~\cite{BH,otherBH}.

\begin{itemize}
\item Graviton production in colliders~\cite{ADDphen,GRW,MPP}
\end{itemize}

Some of the most interesting processes in theories with large extra dimensions
involve the production of a single graviton mode at the LHC or at a linear collider.
Some of the typical Feynman diagrams for such a process are given by~\cite{ADDphen,GRW,MPP,HLZ}:

\vspace*{0.5cm}

\centerline{\includegraphics[width=0.25\hsize]{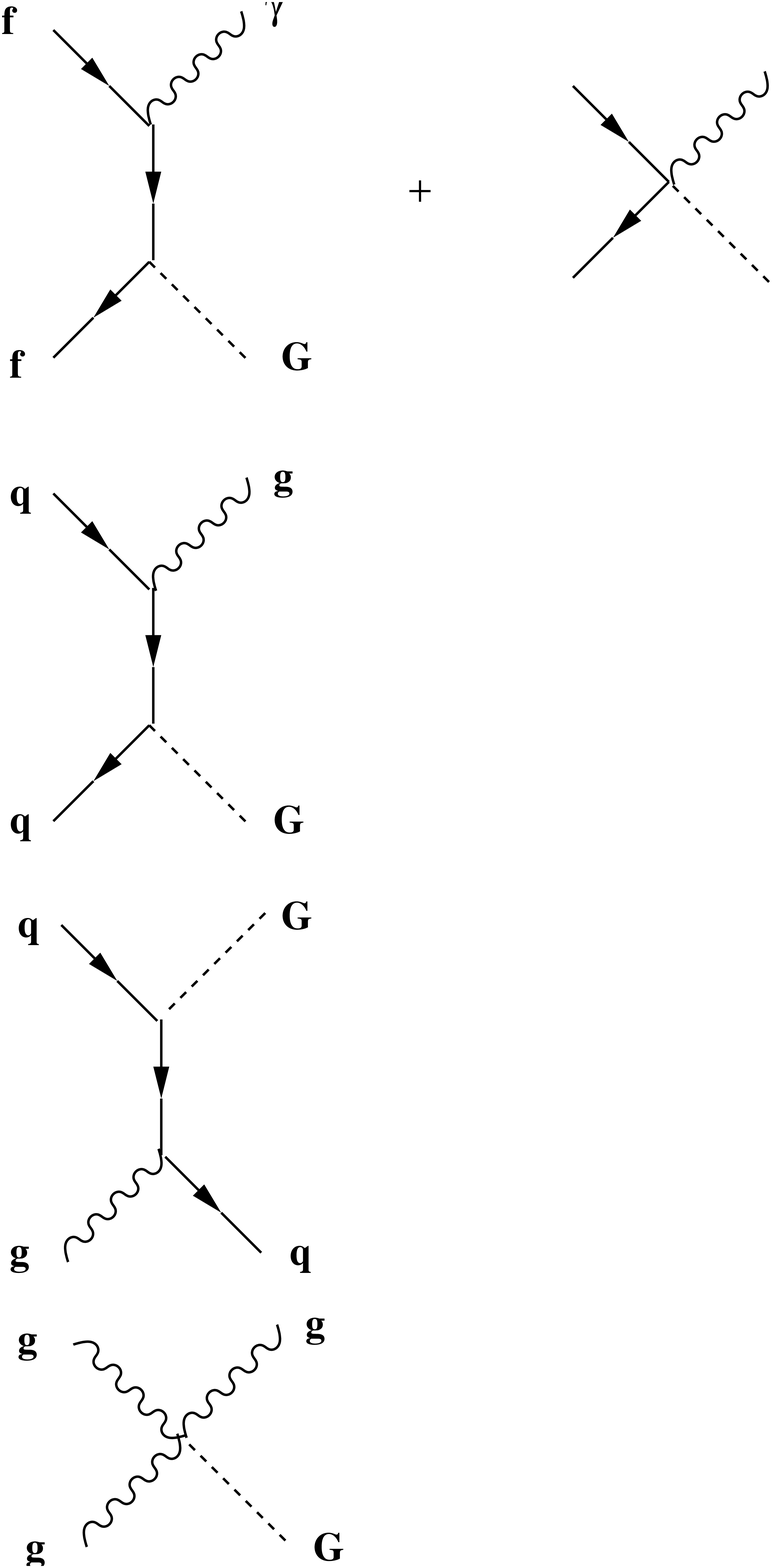}}

Note, that the lifetime of an individual graviton mode is of the order
$\Gamma \sim \frac{m^3}{M_{Pl}^2}$, which means that each graviton produces is extremely 
long lived, and once produced will not decay again within the detector. Therefore, 
it is like a stable particle, which is very weakly interacting since the interaction of individual KK 
modes is suppressed by the 4D Planck mass, 
and thus takes away undetected energy and momentum. Thus in the above 
diagrams wherever we see a graviton, what one really observes is missing energy. This would
lead to the spectacular events when a single photon recoils against missing energy in the linear 
collider, or a single jet against missing energy at the LHC. These are processes with 
small SM backgrounds
(only coming from Z production with initial photon radiation, followed by the Z decaying 
to neutrinos), and also usually quite different from the canonical 
signals of supersymmetry, where one would usually 
have two jets, or two photons produced in the presence of missing energy. 

\begin{itemize}
\item Virtual graviton exchange~\cite{Hewett,GRW,HLZ}
\end{itemize}

Besides the direct production of gravitons, another interesting consequence of 
large extra dimensions is that the exchange of virtual gravitons can lead to enhancement 
of certain cross-sections above the SM values. For example, in a linear collider 
the $e^+e^-\to f\bar{f}$ process would also get contributions from the diagram
\vspace*{1cm}

\centerline{\includegraphics[width=0.25\hsize]{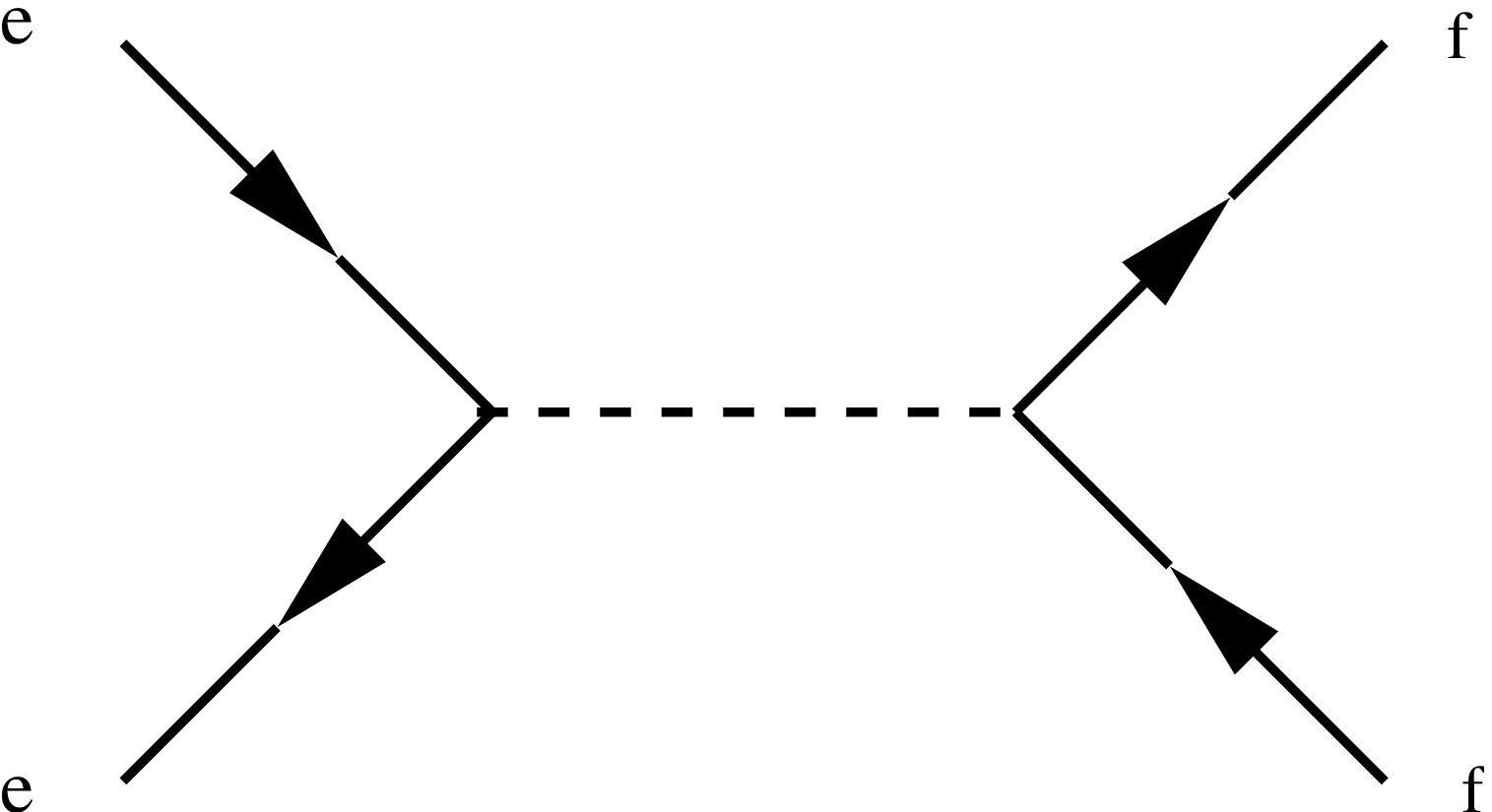}}

The scattering amplitude for this process can be calculated using our previous rules:
\begin{equation}
{\cal A} \sim \frac{1}{M_{Pl}^2} \sum_k \left[ T_{\mu\nu} \frac{P^{\mu\nu\alpha\beta}}{s-m^2}
T_{\alpha\beta} +\frac{\kappa^2}{3} \frac{T^\mu_\mu T^\nu_\nu}{s-m^2}\right]
\equiv S(s) \tau,
\end{equation}
where $S(s)=\frac{1}{M_{Pl}^2}\sum_k \frac{1}{s-m^2}$ and $\tau =T^{\mu\nu}T_{\mu\nu}-
\frac{1}{n+2} T^\mu_\mu T^\nu_\nu$. Note, that for $n\geq 2$ the above sum is UV divergent, 
which implies that the result will be UV sensitive.
For more details see \cite{Hewett,GRW,HLZ}.

\begin{itemize}
\item Supernova cooling~\cite{ADDphen,SN}
\end{itemize}

Some of the strongest constraints on the large extra dimension scenarios come 
from astrophysics, in particular from the fact that just as axions could lead to a too fast
cooling of supernovae, since gravitons are also weakly coupled particles they could also 
transport a significant fraction of the energy within a supernovae. These processes have been
discussed in detail in Refs. \cite{SN}, here we will just briefly mention the essence of
the calculation. The production of axions in supernovae is proportional to the axion
decay constant $1/f_a^2$. The production of gravitons as we have seen is roughly
proportional to $1/M_{Pl}^2 (T/\delta m)^n \sim T^n/M_{*}^{n+2}$, where $T$ is a typical 
temperature within the supernova. This means that the bounds 
obtained for the axion cooling calculation can be applied using the substitution
$1/f_a^2\to  T^n/M_{*}^{n+2}$. For a supernova $T\sim 30$ MeV, and the usual axion bound
$f_a \geq 10^9$ GeV implies a bound of order $M_*\geq 10 - 100$ TeV for $n=2$. For $n>2$ one does
not get a significant bound on $M_*$ from this process.

\begin{itemize}
\item Cooling into the bulk~\cite{ADDphen}
\end{itemize}

Another strong constraint on the cosmological history of models with large extra dimensions comes from
the fact that at large temperatures emissions of gravitons into the bulk would
be a very likely process. This would empty our brane from energy density, and 
move all the energy into the bulk in the form of gravitons. To find out at which temperature
this would cease to be a problem, one has to compare the cooling rates of the brane energy density
via the ordinary Hubble expansions and the cooling via the graviton emission.
The two cooling rates are given by
\begin{equation}
\frac{d\rho}{dt}_{expansion}\sim -3 H\rho \sim -3 \frac{T^2}{M_{Pl}^2}\rho,
\end{equation}
\begin{equation}
\frac{d\rho}{dt}_{evaporation}\sim \frac{T^n}{M_*^{n+2}}.
\end{equation}
These two are equal at the so called ``normalcy temperature'' $T_*$, below which 
the universe would expand as a normal 4D universe. By equating the above two rates we get
\begin{equation} 
T_* \sim \left( \frac{M_*^{n+2}}{M_{Pl}}\right)^{\frac{1}{n+1}} =10^{\frac{6n-9}{n+1}}
\ \ {\rm MeV}.
\end{equation}
This suggests, that after inflation the reheat temperature of the universe should be such,
that one ends up {\it below} the normalcy temperature, otherwise one would overpopulate the
bulk with gravitons, and overclose the universe. This is in fact a very stringent 
constraint on these models, since for example for $n=2$, $T_* \sim 10$ MeV, so there is just barely
enough space to reheat above the temperature of nucleosynthesis. However, this makes
baryogenesis a tremendously difficult problem in these models.

\begin{itemize}
\item Black hole production at colliders~\cite{BH}
\end{itemize}

One of the most amazing predictions of theories with large extra dimensions would be 
that since the scale of quantum gravity is lowered to the TeV scale, one could
actually form black holes from particle collisions at the LHC. Black holes
are formed when the mass of an object is within the horizon size corresponding to the mass
of the object. For example, the horizon size corresponding to the mass of the Earth is 8 mm's, so since the radius 
of the Earth is 6000 km's most of the mass is outside the horizon of the Earth and it is not a black hole. 

What would be
the characteristic size of the horizon in such models? This usually can be read off from the
Schwarzschild solution which in 4D is given by
\begin{equation}
ds^2 = (1-\frac{GM}{r}) dt^2-\frac{dr^2}{(1-\frac{GM}{r})} +r^2 d^2 \Omega,
\end{equation}
and the horizon is at the distance where the factor multiplying $dt^2$ vanishes:
$r_H^{4D}= GM$. In $4+n$ dimensions in the Schwarzschild solution the prefactor is 
replaced by $1-\frac{GM}{r} \to 1- \frac{M}{M_*^{2+n} r^{1+n}}$, from which the 
horizon size is given by 
\begin{equation}
r_H \sim \left( \frac{M}{M_*}\right)^\frac{1}{1+n} \frac{1}{M_*}.
\end{equation}
The exact solution gives a similar expression except for a numerical prefactor in the above 
equation. Thus we know roughly what the horizon size would be, and a black hole will form if
the impact parameter in the collision is smaller, than this horizon size. Then the particles
that collided will form a black hole with mass $M_{BH}=\sqrt{s}$, and the cross section as 
we have seen is roughly the geometric cross section corresponding to the 
horizon size of a given collision energy 

\begin{equation}
\sigma \sim \pi r_H^2 \sim \frac{1}{M_{Pl}^2} 
(\frac{M_{BH}}{M_*})^{\frac{2}{n+1}}
\end{equation}
The cross section would thus be of order $1/$TeV$^2\sim 400$ pb, and the LHC would produce
about $10^7$ black holes per year! These black holes would not be stable, but decay via 
Hawking radiation. This has the features that every particle would be produced with an equal
probability in a spherical distribution. In the SM there are 60 particles, out of which there
are 6 leptons, and one photon. Thus about 10 percent of the time the black hole would decay into 
leptons, 2 percent of time into photons, and 5 percent into neutrinos, which would be observed as
missing energy. These would be very specific signatures of black hole production at the
LHC. For black hole production in cosmic rays see~\cite{BHcosmic}.

\section{Various Models with Flat Extra Dimensions}
\label{sec:flat}
\setcounter{equation}{0}

In the previous section we have discussed theories with large extra 
dimensions: the motivation, basic idea, some calculational tools and 
some of the most interesting consequences. In this section we will
consider some topics that are related to flat extra dimensions (that is 
theories where the gravitational background along the extra dimension is
flat, as compared to the warped extra dimensional scenario discussed
in the following two lectures). 

These models do not necessarily assume that the size of the extra dimension is as large
as in the large extra dimension scenario discussed previously. The first model, theories
with split fermions will still be closely related to large extra dimensions, while the other 
two examples: mediation of supersymmetry breaking via extra dimensions and symmetry breaking 
via orbifold compactifications will be models of their own, usually in a supersymmetric context,
and thus in those models we will not assume the presence of large extra dimensions at all.

\subsection{Split fermions, proton decay and flavor hierarchy}

If there are indeed large extra dimensions, and the scale of gravity is of order $M_* \sim $ TeV, then
one has to confront the following issue. It is usually a well-accepted fact that quantum gravity generically
breaks all global symmetries (but not the gauge symmetries), and therefore if one is assuming
that a theory has a global symmetry, one can only do this up to symmetry breaking
operators suppressed by the scale of quantum gravity. However, this would generically cause problems with
proton decay.
In the SM baryon number is an accidental global symmetry of the Lagrangian (this means that every 
renormalizable operator consistent with gauge invariance also conserves baryon number, without having 
to explicitly require that). However, one can easily write down non-renormalizable operators that do 
violate baryon number, and would give rise to proton decay. However, in the SM if there is no new physics
up to some high scale (the GUT or the Planck scales) then these operators are expected to be suppressed 
by this large scale, and proton decay could be sufficiently suppressed.

In principle, any new physics beyond the
SM could give rise to new baryon number violating 
operators, suppressed by the scale of new physics. For example, in the 
supersymmetric standard model the exchange of superpartners could in principle lead to some baryon number
violating operators, which would be disastrous, since they would only be suppressed by the scale of the mass of
the superpartners. Therefore, in the MSSM one has to make sure that all the 
interactions with the superpartners also conserve baryon number, which can be achieved by 
{\it imposing} the so-called R-parity. Then again the remaining baryon number violating operators 
will be suppressed by the next (very high) scale of physics. However, in the large extra dimensional
case the situation is slightly worse. The reason is that since we are assuming that the scale of quantum
gravity itself is of order $M_*$ we can not really rely on a global symmetry to forbid the unwanted 
baryon number violating operators. For example the operator 
\begin{equation}
\frac{1}{M_*^2} QQQL
\end{equation}
would cause very rapid proton decay. 

A very nice way out of this problem that relies specifically on the existence of the extra dimensions
was proposed by Arkani-Hamed and Schmaltz~\cite{AS}. 
Their idea is to make use of the extra dimensions in a way that
can explain why the dangerous operators are very suppressed without having to worry about quantum gravity
spoiling the suppression. This will be done
by localizing the SM fermions at slightly different  points along the extra 
dimension. This localization of fermions at different points is sometimes called the ``split fermion'' scenario.
The reason why this could be interesting for the suppression of proton decay is that if the fermions are
split, and their wave functions have a relatively narrow width compared to the distance of splitting,
then operators in the effective 4D theory that involve fermions localized at different points along the 
extra dimensions could have a very large suppression due to the small overlap of the fermion wave 
functions. This way one can generate large suppression factors without the use of any symmetry, 
and this could be used both to highly suppress baryon number violating operators, and also
to generate the observed fermion mass hierarchy of the SM fermions. In this section we will follow the 
discussion of~\cite{AS} of split fermions.

Until now we have not discussed the mechanism of localization of various fields. 
If we want to pursue the program of localizing fermions at slightly different points, we will have to
go into the detail of the localization mechanism of fermions in extra dimensions. The discussion below
will thus also give an example of how one should be thinking of a ``brane'' from field theory.

We will discuss the simplest
case, namely a single extra dimension. For this, we have to first understand how fermions in 5D are 
different from fermions in 4D. Fermions are forming representations of the 5D Lorentz group, which is
different from the 4D Lorentz group, since it is larger. In particular, the 5D Clifford algebra
{\it contains} the $\gamma_5$ Dirac matrix as well. While in 4D the smallest representation of the 
Lorentz group is a two-component Weyl fermion, and one gets four component Dirac fermions only if 
one requires parity invariance, in 5D due to the fact the Clifford algebra contains $\gamma_5$ 
the smallest irreducible representation is the four component Dirac fermion. Since a Dirac fermion
contains two two-component fermions with opposite chirality, whenever one talks about higher 
dimensional fermions one has to start out with an intrinsically {\it non-chiral} set of 4D fermions.
Since the SM is a chiral gauge theory, non-chirality of the higher dimensional fermions has to 
be overcome somehow. We will see that the localization mechanism and (in the next two subsection) orbifold
projections can achieve this. First we will concentrate on describing a viable localization mechanism
that produces chiral fermions localized at different points along the extra dimension.

We will use the following representation for the $\gamma$ matrices:
\begin{equation}
\gamma^i=\left( \begin{array}{cc} 0 & \sigma^i \\ \bar{\sigma}^i & 0 \end{array} \right)
\ , i=0,\ldots ,3 \ \ \ \gamma^5 =-i\left( 
\begin{array}{cc} 1 &  \\ & -1 \end{array} \right).
\end{equation}
The two type of 5D Lorentz invariants that can be formed from two four-component 5D spinors $\Psi_1$
and $\Psi_2$ are the usual $\bar{\Psi}_1 \Psi_2$ which corresponds to the usual 4D Dirac mass term,
and $\Psi_1^T C_5 \Psi_2$, which corresponds to the Majorana mass term, and where $C_5$ is the 5D 
charge conjugation matrix $C_5=\gamma^0\gamma^2\gamma^5$. 

To describe the localization mechanism for fermions, we consider a 5D spinor in the background of a scalar
field $\Phi$ which forms a domain wall. This domain wall is a physical example of brane that has a finite width
(a ``fat brane''). The background configuration of this scalar is denoted by $\Phi (y)$. 
The action for a 5D spinor in this background is given by
\begin{equation}
S=\int d^4x dy \bar{\Psi} \left[ i \gamma_\mu \partial_4^\mu  +i \gamma_5\partial_y +\Phi (y)
\right] \Psi .
\end{equation}
Note, that we have added a Yukawa type coupling between the scalar field and the fermion, which will be
essential for the localization of the fermions on the domain wall. From this the 5D Dirac equation is 
given by
\begin{equation}
\left[ i \gamma_\mu \partial_4^\mu  +i \gamma^5\partial_y +\Phi (y)
\right] \Psi =0
\end{equation}
We will look for solutions to this which are left- or right-handed 4D modes:
\begin{equation}
i \gamma^5 \Psi_L = \Psi_L, \ \ \ i \gamma^5 \Psi_R = -\Psi_R.
\end{equation}
To find the 4D eigenmodes of this system, we look for the eigensystem of the $y$-dependent piece
of the above equation. To simplify the equation (as usual when solving a Dirac-type equation) we also
multiply by the conjugate of the differential operator to get the equation:
\begin{equation}
\left[ -i \gamma^5\partial_y +\Phi (y)\right] \left[ i \gamma^5\partial_y +\Phi (y)\right] \Psi_{L,R}^{(n)}=
\mu_n^2 \Psi_{L,R},
\end{equation}
which gives
\begin{equation}
\left( -\partial_y^2 +\Phi (y)^2\pm \dot{\Phi} (y)\right) \Psi_{L,R} =
\mu_n^2 \Psi_{L,R}.
\end{equation} 
If $\Phi (y)$ had a linear profile, this would exactly give a harmonic oscillator equation. However, even for
the generic background one can define creation and annihilation operators as for the usual harmonic oscillator
using the definition
\begin{equation}
a=\partial_y +\Phi (y), \ \ a^\dagger =-\partial_y +\Phi (y).
\end{equation}
With these operators one can turn the above Schr\"odinger-like problem into a SUSY quantum mechanics 
problem, which means that one can define the operators $Q$ and $Q^\dagger$ such, that $\left\{ Q,Q^\dagger 
\right\} =H$, where $H$ is the Hamiltonian of the system. For the case at hand $Q=a \gamma^0 P_L$, 
$Q^\dagger = a^\dagger \gamma^0 P_R$ will give the right anti-commutation relation 
(here $P_{L,R}$ are the left and right-handed projectors. The SUSY quantum
mechanics-like Schr\"odinger problems have very special properties. For example, the eigenvalues of the $L$ and
$R$ modes always come in pairs, {\it except} possibly for the zero modes. The pairing of eigenmodes 
just corresponds to the expected vectorlike behavior of the bulk fermions: for every L mode there is an R-mode,
since these are massive modes that is exactly what one would expect.  
The fact, that the zero modes need not be paired is the 
most important part of the statement, since these zero modes are the most interesting for us: they are the 
ones that could give chiral 4D fermions. Therefore let us examine the equation for the zero mode.
Since $\left\{ Q,Q^\dagger  \right\} =H$, the solution to $H\Psi=0$ are the solutions to $Q\Psi =0$ or
$Q^\dagger \Psi =0$, that is
\begin{equation}
\left[ \pm \partial_y +\Phi (y)\right] \Psi_{L,R}(y)=0.
\end{equation}
From this we find the left and right-handed zero modes to be
\begin{eqnarray}
&& \Psi_L \sim e^{-\int_0^y \Phi(y') dy'}, \nonumber \\
&& \Psi_R \sim e^{+\int_0^y \Phi(y') dy'}.
\end{eqnarray}
Since we have an exponential with two different signs for the exponent, clearly both of these
solutions can not be normalizable at the same time. Therefore we get {\it chiral} zero modes localized
to the domain wall.  For example, if the profile of the domain wall is linear, $\Phi (y) \sim 2 \mu^2 y$,
then the solution for the left-handed zero mode is:
\begin{equation}
\Psi_L (y) = \frac{\mu^\frac{1}{2}}{(\pi/2)^\frac{1}{4}} e^{-\mu^2y^2}
\end{equation}
This would be a zero mode localized at $y=0$, which is exactly the point at which the domain wall profile
switches sign, that is where $\Phi (y)=0$. Thus we have described a dynamical mechanism to localize a chiral 
fermion to a domain wall. These are sometimes called domain-wall fermions, and are extensively used
for example in trying to put chiral fermions on a lattice~\cite{Kaplan}.

Let us now  consider the case when we have many fermions in the bulk, $\Psi_i$, and their action
is given by
\begin{equation}
S=\int d^4x dy \bar{\Psi_i} \left[ i \gamma_\mu \partial_4^\mu  +i \gamma_5\partial_y +\lambda \Phi (y)-m
\right]_{ij} \Psi_j .
\end{equation}
There will now be chiral fermion zero modes centered at the zeroes of 
\begin{equation}
(\lambda \Phi (y) -m)){ij}.
\end{equation}
If we again imagine that the domain wall is linear in the region where all these fermions will be localized,
then the center of the fermion wave functions will be at 
\begin{equation}
y^i=\frac{m^i}{2\mu^2},
\end{equation}
where the $m_i$'s are the eigenvalues of the $m_{ij}$ matrix. Thus the different fermions will be 
localized at different positions!

Let us now try to write down what the Yukawa couplings between such fermion zero modes would be in the presence
of a bulk Higgs field that connects these fermions. 
The 5D action would be given by 
\begin{equation}
S=\int d^5 x\bar{L} [ i \gamma^M \partial_M +\Phi (y)]L+\int d^5 x\bar{E^c} [ i \gamma^M \partial_M +\Phi (y)
-m]E^c+\int d^5 x \kappa HL^T C_5 E^c,
\end{equation}
where the last term is the bulk Yukawa coupling written such that it is both gauge invariant and 5D Lorentz
invariant. From this the effective 4D Yukawa coupling for the zero modes will be of the form 
\begin{equation}
\int d^4 x \kappa Hl e^c \int dy \Psi_L(y) \Psi_{e^c}(y-r),
\end{equation}
where $l,e^c$ are the zero modes of the bulk fields, and $\Psi_L(y), \Psi_{e^c}(y-r)$ are the wave functions
of these zero modes. These wave functions are assumed  to be Gaussians centered around different points ($y=0$ and $y=r$)
in the linear domain wall approximation, and so this integral will just be a convolution of two 
Gaussians, which is also a Gaussian:
\begin{equation}
\int dy \Phi_L(y) \Phi_{e^c}(y)=\frac{\sqrt{2}\mu}{\sqrt{\pi}} \int dy e^{-\mu^2 y^2} e^{-\mu^2 (y-r)^2}=
e^{-\frac{\mu^2r^2}{2}}.
\end{equation}
Thus the effective 4D Yukawa coupling between the zero modes is 
\begin{equation}
\kappa e^{-\frac{\mu^2 r^2}{2}},
\end{equation}
which could be exponentially small even if the bulk Yukawa coupling $\kappa$ is of order one. The exponential
suppression factor depends on the relative position of the localized zero modes. Thus split fermions could
naturally generate the fermion mass hierarchy observed in the SM. Basically, this method translates the 
issue of fermion masses into the geography of fermion localization along the extra dimension.

To close this subsection, we get back to our original motivation of proton stability. Imagine, that the
leptons and the quarks are localized on opposite ends of a fat brane. A dangerous proton decay operator
generated by quantum gravity at the TeV scale would have the form
\begin{equation}
\int d^5x \frac{1}{M_*^3}(Q^TC_5L)^\dagger (U^{c T}C_5 D^c).
\end{equation}
Again calculating the wave function overlap of the zero modes as before we get that the effective 4D
coupling is suppressed by the factor
\begin{equation}
\int dy (e^{-\mu^2 y^2})^3 e^{-\mu^2 (y-r)^2} \sim e^{-\frac{3}{4} \mu^2 r^2}.
\end{equation}
Here $r$ is the separation between quarks and leptons, and we can see that arbitrarily large suppression
of proton decay is possible in such models without invoking symmetries.

What would a model with large extra dimensions then look like in this scenario?
One would have the large extra dimensions of size TeV$^{-1} 10^{\frac{32}{n}}$, where the gravitational 
degrees of freedom propagate.
Within that large extra dimension there is a ``fat brane'', which is basically the domain wall that we 
have discussed above. The fat brane has the gauge fields and the Higgs scalar living along its world volume.
The width of this fat brane could be of the order TeV$^{-1}$, such that the KK modes of the gauge fields
arising from the ``fatness'' of the brane are sufficiently heavy. Within this fat brane fermions are localized
at different positions of this brane, which will eliminate the problem with proton decay and the geography
of the localized zero modes will generate the fermion mass hierarchy. The width of the localized fermions
is much smaller than the width of the fat brane, of order $0.1 - 0.01$ TeV$^{-1}$. For more on split fermions 
see~\cite{othersplit}.

\subsection{Mediation of supersymmetry breaking via extra dimensions
(gaugino mediation)}

Until now we have considered non-supersymmetric theories, and were trying to solve the hierarchy
problem using large extra dimensions. In the following two examples we will assume that the hierarchy problem
is solved by supersymmetry, and ask the question whether extra dimensions and branes could still play an  
important role in particle phenomenology. First we will discuss the issue of mediation of
supersymmetry breaking via an extra dimension, and then discuss how to break symmetries via the geometry
of the extra dimension. 

Assuming that the hierarchy problem is solved by supersymmetry, the most important issue would then
be how supersymmetry is broken. In the MSSM (the minimal supersymmetric standard model -- see the 
lectures by Carlos Wagner~\cite{Wagner}) the soft supersymmetry breaking parameters are usually inputted by hand, 
parameterizing our ignorance of what exactly the supersymmetry breaking sector is. However, most of the
arbitrary parameter space for the supersymmetry breaking parameters is excluded by experiments.
For example, if the susy breaking masses for the scalar partners of the fermions form an arbitrary mass 
matrix, then there would be large flavor changing neutral currents generated at the one loop level,
contributing for example to $K-\bar{K}$ mixing. In the SM the $K-\bar{K}$ is suppressed by the 
well-known GIM mechanism, namely the one loop diagram 

\vspace*{0.5cm}

\centerline{\includegraphics[width=0.35\hsize]{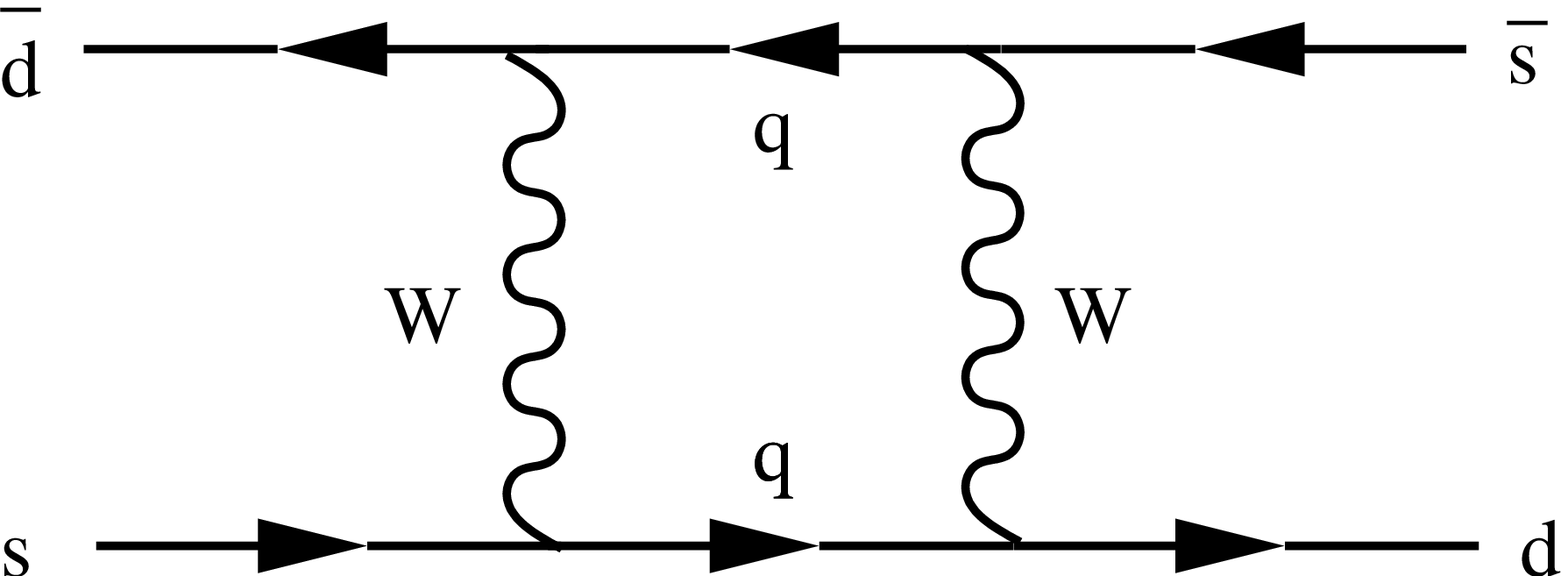}}
\noindent is proportional to leading order to the off-diagonal element of $(V^\dagger V)_{CKM}$, which vanishes due
to the unitarity of the CKM matrix. However, in the MSSM due to the presence of the superpartners of both 
the quarks and the gluino there is an additional vertex of the form 

\vspace*{0.5cm}

\centerline{\includegraphics[width=0.35\hsize]{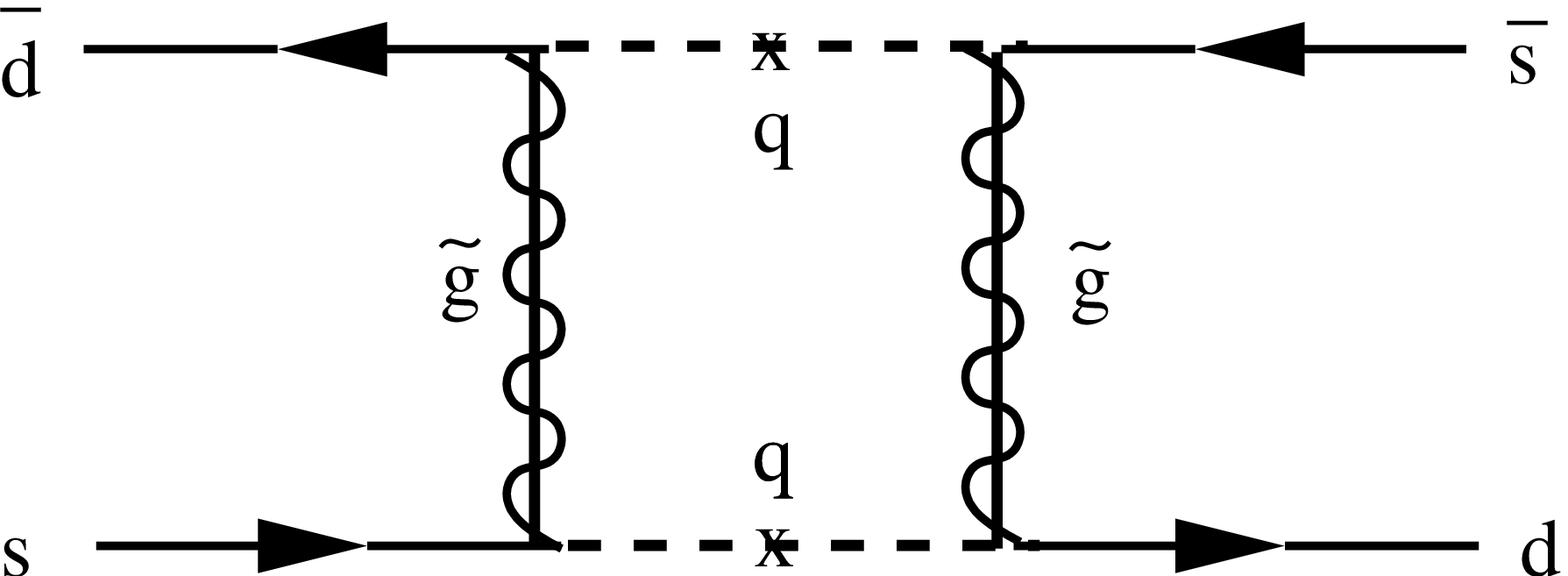}}
\noindent which is proportional to $V^\dagger_{CKM} M^2_{squarks} V_{CKM}$, and is therefore not suppressed, 
unless the mass matrix of the squarks themselves is close to the unit matrix, in which case the ordinary
GIM mechanism would operate here as well. The question of why should the squarks (at least for the first two
generations) be almost degenerate is called the susy flavor problem (see much more on this in the 
lectures by Ann Nelson~\cite{Nelson}). Here we will show that extra dimensions can be used to transmit supersymmetry
breaking to the SM in a way that this susy flavor problem could be resolved. There are two prominent
proposals for this: anomaly mediation~\cite{AMSB} (proposed by Randall and Sundrum, simultaneously to a 
similar proposal by Guidice, Luty, Murayama and Rattazzi), and gaugino mediation
\cite{GMSB} (proposed simultaneously by D.E.Kaplan, Kribs and Schmaltz and by Chacko, Luty, Nelson and Ponton). 
Anomaly mediation involves supergravity in the bulk of extra dimensions, and therefore is technically more 
involved. Here we will only consider proposal for gaugino mediation of supersymmetry breaking following the 
paper by Kaplan, Kribs and Schmaltz in~\cite{GMSB},
and for anomaly mediation we refer the reader to the original papers~\cite{AMSB} and~\cite{otheranomaly}. 

In both anomaly mediated and gaugino mediated 
scenarios the main assumption is that the MSSM matter fields are localized to a brane along an 
extra dimension, on which supersymmetry is unbroken. Supersymmetry is only broken on another brane in the 
extra dimension (it is not broken in the bulk either), and is transmitted to the MSSM matter fields
via fields that live in the bulk, and thus couple to both the susy breaking brane (the hidden brane)
and the visible MSSM brane. The difference between the two scenarios is what those bulk fields are that
transmit the supersymmetry breaking. In the anomaly mediated scenario one has pure supergravity in the bulk,
while in the gaugino mediated scenario the MSSM gauge fields live in the bulk, and thus the gauginos of the
MSSM will directly feel supersymmetry breaking, and will get the leading supersymmetry breaking terms.
The MSSM scalars, since they are localized on a different brane will only get susy breaking masses via
loops in the extra dimension, and will be therefore suppressed compared to the gaugino masses. 
The arrangement of fields for the gaugino mediated scenario is given in the figure below:

\vspace*{0.5cm}

\centerline{\includegraphics[width=0.5\hsize]{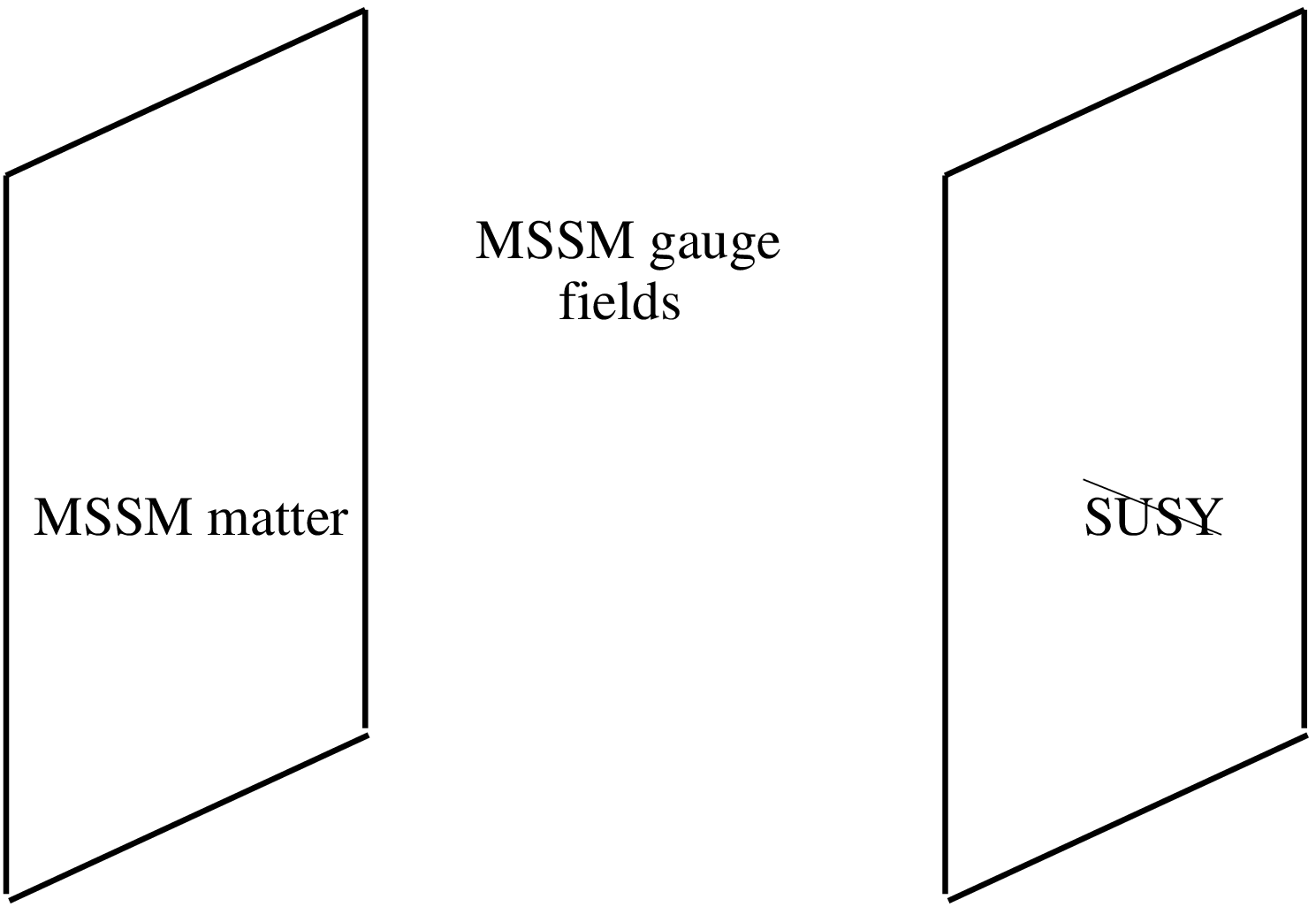}}

As mentioned before, these models have been formulated in the presence of a single extra dimension.
In order to be able to proceed with the discussion of the gaugino mediated model, we need to first formulate
supersymmetry in 5D. The issue is similar to the discussion of fermions in 5D when we were considering the
split fermion model: since in 5D the smallest spinor is the 4 component Dirac spinor, in 5D the smallest number
of supercharges must be 8. (As a reminder, in 4D the smallest supersymmetry algebra corresponds to the 
situation when there is a single complex two-component Weyl spinor supercharge, which means there 
would be four real components. Thus ${\cal N}=1$ susy in 4D corresponds to 4 supercharges.) Since in 5D the 
smallest spinor has 8 real components, ${\cal N}=1$ supersymmetry in 5D is twice as large as in 4D, and the 
dimensional reduction of the 5D theory would correspond to ${\cal N}=2$ in 4D. For example, the
5D vectormultiplet would have to contain a massless 5D vector, and also a {\it Dirac} spinor. Since the Dirac 
spinor has four components on-shell, and the 5D massless vector 3, there needs to be an additional massless
real scalar in the vectormultiplet, which is then:
\[ (A_M,\lambda ,\Phi ), \]
where $A_M$ is the 5D vector, $\lambda$ is a Dirac spinor, and $\Phi$ is a real scalar. 
When reduced to 4D the vector will go into a 4D vector plus $A_5$ which is a scalar, and the Dirac fermion
into two Weyl fermions. The $A_5$ together with $\Phi$ will form a complex scalar, and we can see that from the
4D point of view the 5D vector multiplet is a 4D vector multiplet plus a 4D chiral superfield in the adjoint
(or equivalently from the 4D point of view an ${\cal N}=2$ vector superfield, which is exactly
a vector plus a chiral superfield). This would mean that we would get two gauginos in the 4D theory, if 
5D Lorentz invariance is completely intact. However, when one has branes in the extra dimension, 5D Lorentz
invariance is anyways broken. 

One of the simplest ways of implementing the breaking of the 5D Lorentz invariance is to compactify the extra
dimension on an {\it orbifold} instead of a circle. The meaning of an orbifold is to geometrically identify certain
points along the extra dimension, and then require that the bulk fields have a definite transformation
property under this symmetry geometric symmetry. We will discuss such orbifolds in much more detail in the
next subsection, and there is also a whole lecture about them in these volumes by Mariano Quiros~\cite{Quiros}. 

For now we will simply compactify the extra dimension on an interval (an $S^1/Z_2$ orbifold) rather than
a circle, and require that the bulk fields are either even or odd under the $Z_2$ parities 
$y\to -y$, and $L-y\to L+y$. This is basically a fancy way of saying that we require that the bulk fields
satisfy either Dirichlet or Neumann boundary conditions (bc's). 
A convenient choice of these $Z_2$ parities
for the bulk 5D vector superfield discussed above is 
\begin{eqnarray}
&& (A_\mu ,\lambda_L)\to (A_\mu ,\lambda_L), \nonumber \\
&& (A_5 ,\Phi ,\lambda_R)\to -(A_5 ,\Phi ,\lambda_R). 
\end{eqnarray}
This means that $A_\mu$ and $\lambda_L$ satisfy Neumann boundary conditions at $y=0,L$ while
the other fields Dirichlet boundary conditions. Therefore, the KK expansions for these fields will differ from
the usual expansion on a circle. The modes for the various fields will be of the form:
\begin{eqnarray}
&& A_\mu ,\lambda_L\sim \cos \pi \frac{ny}{L}, \ \   \nonumber \\
&& A_5 ,\Phi ,\lambda_R\sim  \sin  \pi \frac{ny}{L}.
\end{eqnarray}
Here the mass of the $n$-th mode is as usual $m_n^2=n^2\pi^2/L^2$. We can see, that for $n=0$ the wave
functions of the odd modes are vanishing. Thus the orbifolding procedure effectively ``projects out''
the zero mode for the states that have odd parities (that is the ones that have a Dirichlet bc).

Therefore, the zero mode spectrum is no longer necessarily vectorlike, but orbifold compactifications
can lead to a chiral fermion zero mode spectrum. What we see is that with the above choice of parities
the zero modes will only appear for the fields that also appear in the MSSM. Thus the boundary conditions
take care of the unwanted modes as compared to the MSSM!

The Lagrangian of the gaugino mediated model will then schematically be of the form:
\begin{equation}
{\cal L} =\int dy \left[ {\cal L}_5 +\delta (y) {\cal L}_{MSSM matter} +\delta (y-L) {\cal L}_{susy breaking}
\right].
\end{equation}

We will not specify the dynamics that leads to supersymmetry breaking on the hidden brane, but rather assume
that there is a field $S$ which has a supersymmetry breaking $F$-term: 
\begin{equation}
\langle S \rangle = F \theta^2. 
\end{equation}
Since the gauginos live in the bulk, they can couple to this susy breaking field via the operator
\begin{equation}
\int d^2 \theta \frac{S}{M^2} W_\alpha W^\alpha +h.c.
\label{gauginomass}
\end{equation}
What should be the suppression scale $M$ in the above operator be? We have assumed, that the gauge fields 
(contrary to the large extra dimensional scenario) live in the bulk. This means that we have a 5D gauge
theory. However, the 5D gauge coupling is {\it not} dimensionless, and therefore the 5D gauge 
theory is not renormalizable. This simply means, that at some scale $M$ the 5D gauge interactions will
become strong, above which the perturbative definition of the theory should not be trusted.
However, below that scale the theory is a perfectly well-defined effective field theory, which however has 
to be completed in the ultra-violet (above the scale $M$). The real question is how the scale $M$ 
is related to the length of the extra dimension and the 4D gauge coupling. The scale at which the theory
becomes strongly coupled is where the effective dimensionless coupling $g_5^2 M$ becomes as large as a loop 
factor,
\begin{equation}
\frac{g_5^2 M}{16 \pi^2} \sim 1.
\end{equation}
However, as we have seen already the effective 4D gauge coupling, which is of order one since these are
the visible MSSM gauge couplings are given by
\begin{equation}
\frac{1}{g_4^2} =\frac{L}{g_5^2} = {\cal O}(1).
\end{equation}
Combining these two equations we get that $ML \sim 16 \pi^2 \sim {\cal O} (100)$, which implies that the
scale $M$ is a factor of 100 or so larger than $1/L$, the scale that sets the mass of the first KK mode.
This means, that there is a hierarchy between $M$ and $1/L$, and between these two scales it is
justified to use an effective 5D gauge theory picture. 

Coming back to the size of the supersymmetry breaking gaugino mass, substituting the vev of $S$ into
(\ref{gauginomass}) we get that the gaugino mass term is 
\begin{equation}
\frac{F}{M^2} \lambda \lambda |_{y=L}.
\end{equation}
This is a 4D mass term (since it is on the surface of the susy breaking brane only) for a 5D bulk field $\lambda$.
In order to get the effective 4D mass term for the zero mode, we have to rescale the 5D field to have a 
canonically normalized kinetic term for the zero mode. Since the 5D kinetic term is just the usual
\begin{equation}
{\cal L}_{kin} = \int \bar{\lambda} i \partial_M \gamma^M \lambda d^5x,
\end{equation}
after integrating over the extra dimension we get for the zero mode that $\lambda_{(5)}L^{\frac{1}{2}} =
\lambda_{(4)}$. Then the gaugino mass for the correctly normalized 4D field will be 
\begin{equation}
m_{gaugino} = \frac{F}{M} \frac{1}{ML}.
\label{gm2}
\end{equation}
Note, that the physical mass is suppressed by the size of the extra dimension, and this is the consequence of the
fact that the mass is localized on a brane while the field extends everywhere in the extra dimension.
(\ref{gm2}) implies that at the high scale $M$ all gauginos have a common mass term.

The most important question is what the magnitude of the masses of the scalar partners of the SM fermions will
be, since as we have seen at the beginning of this section, this is what leads to the susy flavor problem.
Qualitatively we can already see its features: since we have assumed that the SM matter fields are 
on the visible brane, they will not directly couple to the susy breaking sector. Thus they will feel
susy breaking effects only through loops that go across the extra dimension and connect the scalar fields on one
brane with the susy breaking vev on the other brane. This means, that this operator should be a loop factor 
smaller, than the gaugino mass term. However, even this would be too large, if arbitrary flavor structure
was allowed. The important point is that the loops in the extra dimensions involve the gaugino fields,
which couple universally (as determined by the gauge couplings) to the MSSM matter fields. Therefore,
just as in the so called gauge mediated models (which we have not discussed here since those are not relying 
on extra dimensions) the corrections to the soft scalar masses will be flavor universal, and thus this 
setup has a chance of solving the susy flavor problem. 

Even though we basically already roughly know the answer for what the size of the scalar masses is going to be,
it is still illustrative to go through the calculation in detail, since this should also teach us how to
handle diagrams that involve loops in the bulk. Before we go into the detail of the calculation, let us just 
make one comment: every extra dimensional loop which involves propagation from one brane to the other must 
be necessarily UV finite. The reason is that usually divergences appear when the size of the loops in a usual
Feynman diagram shrinks to zero. However here the size of the extra dimension provides a UV cutoff, and the 
integral will be finite (the loop can not shrink to zero in coordinate space since the fields have to
propagate from one brane to the other.)

The diagram that one needs to calculate is illustrated below:
\vspace*{0.5cm}

\centerline{\includegraphics[width=0.5\hsize]{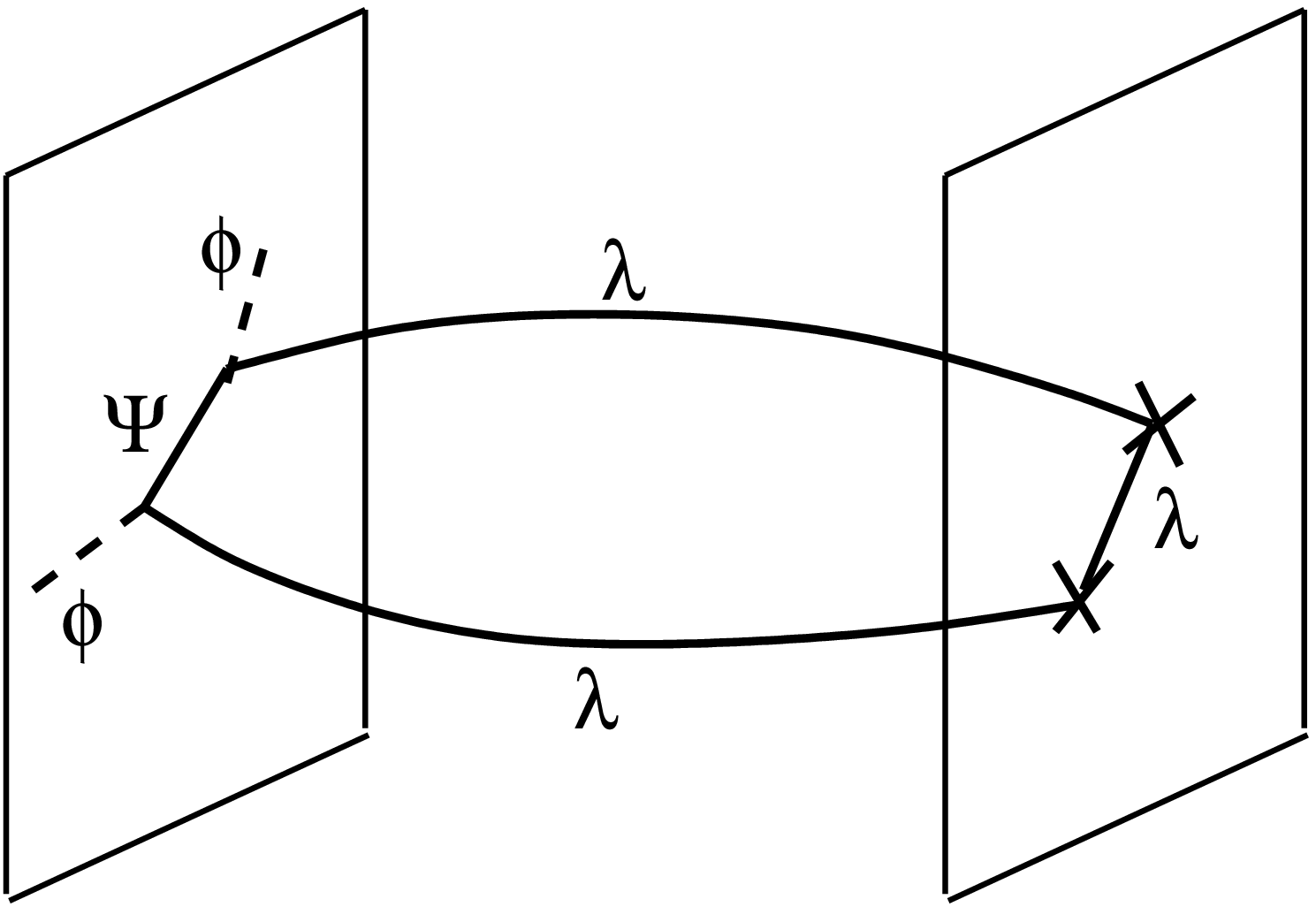}}

Since we know that the fields have to propagate from one brane to the other and back,
it is useful to use the propagators in the $y$ coordinate, however, for the 4D propagators we want to 
use the usual momentum representation. Therefore we need the propagator in a {\it mixed} $q,y_i,y_f$ 
representation. Even though what we really need is the propagator for fermions, since in the above loop
it is the gaugino that is propagating through the bulk, it is instructive to first calculate the
scalar propagator $P(q^2,a,b)$, which corresponds to a scalar with four momentum $q^2$ propagating from
$y=a$ to $y=b$. Since the bulk scalars are odd under the $Z_2$, their wavefunctions are 
\begin{equation}
\phi_n(y)= \sqrt{\frac{2}{L}}\sin \frac{\pi n y}{L}.
\end{equation}
The bulk propagator can then be written in terms of the wave functions as
\begin{equation}
P(q^2,a,b)=\sum_{m,n}  \phi_n^*(a)\phi_m(b) \frac{\delta_{mn}}{q^2+p_n^2}=
\frac{2}{L} \sum_n \sin \left( \frac{-\pi n a}{L}\right) \sin \left( \frac{\pi n b}{L}\right) \frac{1}{q^2+p_n^2}.
\end{equation}
Here $p_n=\frac{n\pi}{L}$, and one can convert the sum to an integral 
via $\Delta n = dp \ L/\pi$ to get
\begin{equation}
P(q^2,a,b)= -\frac{1}{\pi} \int_{-\infty}^\infty dp \frac{\sin (bp)\sin (ap)}{p^2+q^2} \sim
-\frac{1}{2q} e^{-|b-a|q}.
\end{equation}

A similar calculation for the fermions yields 
\begin{equation}
P(q^2,0,L)\sim
\frac{2 P_L q_\mu \gamma^\mu}{q} e^{-qL}.
\end{equation}
Evaluating the diagram above we get
\begin{equation}
g_5^2 \left( \frac{F_S}{M^2}\right)^2 \int \frac{d^4 q}{(2\pi)^4} {\rm Tr} \frac{1}{q_\mu \gamma^\mu}
P_L P(q,0,L) C P^T(q,L,L) C^{-1} P(q,L,0).
\end{equation}
Plugging in the propagator and evaluating the integral we get for the scalar mass
\begin{equation}
m_S^2 \sim \frac{g_5^2}{16\pi^2} \left( \frac{F_S}{M^2}\right)^2 \frac{1}{L^3} = \frac{g_4^2}{16\pi^2} m_g^2.
\end{equation}
Thus we find that as expected the scalar mass is flavor universal (it depends only on the gauge coupling)
and a loop-factor suppressed compared to the gaugino mass. 

In the end, the free parameters of the gaugino mediated model are the size of the extra dimension
$L$, the unified gaugino mass $M_{\frac{1}{2}}$, the ratio of Higgs VEVs $\tan \beta$, and the sign of the
$\mu$-term. With these input parameters one has to start running down the above susy breaking soft masses from
the scale $M$ to the electroweak scale, and find the appropriate physical mass spectrum. This gives a phenomenologically
acceptable mass spectrum, for details see \cite{gauginospectrum}.

\subsection{Symmetry breaking via orbifolds}

We have seen in the previous section, that compactification can break some
of the symmetries of the higher dimensional theory. The particular example
that we discussed was the breaking of supersymmetry from 8 supercharges 
(${\cal N}=2$ in 4D) to just 4 supercharges (${\cal N}=1$ in 4D).
In this section we will consider examples both of gauge symmetry breaking and
of supersymmetry breaking via boundary conditions in extra dimensions. 
Such symmetry breakings have been considered by the string theorists for a long time, see for example~\cite{stringorbifold}.
Here the first example we will discuss in detail  will be the breaking of
the grand unified gauge group as proposed by Hall and Nomura~\cite{HN}, see also~\cite{KawaAF,BHN2,HMN,otherorbiGUT}.
while the discussion of the breaking of supersymmetry is based on the
model by Barbieri, Hall and Nomura~\cite{BHN}, see also~\cite{otherBHN}.

We will use orbifold projections on a theory on a circle 
to obtain the symmetry breaking boundary conditions,
even though these can be also formulated very conveniently just by considering
a theory on an interval. 

Through out this section we will just consider a single extra dimension, and 
try to find what are the orbifold projections one can do to arrive at the 
final set of boundary conditions. By orbifold boundary projections we mean a 
set of identifications of a geometric manifold which will reduce the 
fundamental domain of the theory. One can give a different description of theories
with boundary conditions by starting with the variational principle for theories
on an interval. For this approach see~\cite{CGMPT}.

Let us first start with an infinite extra dimension, an infinite line $R$,
parametrized by $y$, $-\infty <y<\infty$. One
can obtain a circle $S^1$ from the line by the identification
$y\to y+2\pi R$, which we can denote as ``modding out the infinite line
by the translation $\tau$, $R\to S^1 = R^1/\tau$. This way we obtain the 
circle. 

Another discrete symmetry that we could use to mod out the line 
is a reflection $Z_2$ which takes $y\to -y$. Clearly, under this reflection
the line is modded out to the half-line $R^1\to R^1/Z_2$.

If we apply both discrete projections at the same time, we get 
$S^1/Z_2$, the orbifold that we have already used in the previous section
when we discussed gaugino mediation. This orbifold is nothing else but the 
line segment between $0$ and $\pi R$. 

Let us now see how the fields $\varphi (y)$ 
that are defined on the original infinite line 
$R^1$ will behave under these projections. The fields at the identified 
points have to be equal, except if there is a (global or local) symmetry 
of the Lagrangian. In that case, the fields at the identified points don't
have to be {\it exactly} equal, but merely equal {\it up to a symmetry}
transformation, since in that case the fields are still physically equal.
Thus, under translations and reflection the fields behave as
\begin{eqnarray}
&& \tau (2\pi R)\varphi (y) =T^{-1} \varphi (y+2\pi R), \\
&& {\cal Z} \varphi (y) = Z \varphi (-y),
\end{eqnarray}
where $T$ and $Z$ are matrices in the field space corresponding to some 
symmetry transformation of the action.
This means that we have made the field identifications
\begin{eqnarray}
&& \varphi (y+2\pi R)=T \varphi (y), \\
&& \varphi (-y)=Z \varphi(y).
\end{eqnarray}
Again, $Z$ and $T$ have to be symmetries of the action. However, $Z$ and 
$T$ are not completely arbitrary, but they have to satisfy a consistency
condition. We can easily find what this consistency condition is by
considering an arbitrary point at location $y$ within the fundamental 
domain $0$ and $2 \pi R$,  apply first a reflection around $0$ ${\cal Z}(0)$,
and then a translation by $2\pi R$, which will take 
$y$ to $2\pi R -y$. However, there is another way of connecting these two
points using the translations and the reflections: we can first translate $y$
backwards by $2\pi R$, which takes $y\to y-2\pi R$, and then reflect around
$y=0$, which will also take the point into $2\pi R -y$. This means that
the translation and reflection satisfy the relation:
\begin{equation}
{\cal Z} (0) \tau^{-1} (2\pi R)=\tau (2\pi R){\cal Z} (0).
\end{equation}
When implemented on the fields $\varphi$ this means that we need to have the 
relation
\begin{equation}
TZ=ZT^{-1}, {\rm or} \ \ ZTZ=T^{-1}
\label{consistency}
\end{equation}
which is the consistency condition that the field transformations $Z$ and $T$
have to satisfy.

As we have seen, the reflection ${\cal Z}$ is a $Z_2$ symmetry, and so
$Z^2=1$. $T$ is not a $Z_2$ transformation, so $T^2\neq 1$. 
However, for non-trivial $T$'s  $T\neq 1$ 
(T is sometimes called the Scherk-Schwarz-twist) one can always 
introduce a combination of $T$ and $Z$ which together act like another
$Z_2$ reflection. We can take the combined transformation
$\tau (2\pi R) {\cal Z}(0)$. This combined transformation takes any point 
$y$ into $2\pi R-y$. That means, that it is actually a reflection around
$\pi R$, since if $y=\pi R-x$, then the combined transformation takes it to 
$\pi R+x$, so $x\to -x$. So this is a $Z_2$ reflection. And using the 
consistency condition (\ref{consistency}) we see that for the 
combined field transformation $Z'=TZ$
\begin{equation}
{Z'}^2=(TZ)^2= (TZ)(ZT^{-1})=1,
\end{equation}
so indeed the action of the transformation on the fields is also acting 
as another $Z_2$ symmetry.  Thus we have seen that the description of a generic
$S^1/Z_2$ orbifold with non-trivial Scherk-Schwarz twists can be 
given as two non-trivial $Z_2$ reflections $Z$ and $Z'$, one which acts
around $y=0$ and the other around $y=\pi R$. These two reflections do not 
necessarily commute with each other. A simple geometric picture to visualize
the two reflections is to extend the domain to a circle of radius $4\pi R$, 
with the two reflections acting around $y=0,2\pi R$ for $Z$ and 
$\pi R, 3\pi R$ for $Z'$. One can either use this picture with the fields 
living over the full circle, or just living on the fundamental domain
between $y=0$ and $2\pi R$. The two pictures are equivalent. 
In the remainder of this section we will focus on two particular models that employ  an orbifold structure
of the form above, with the two reflections $Z$ and $Z'$ commuting with each other. 
 
\begin{figure}
\centerline{\includegraphics[width=0.5\hsize]{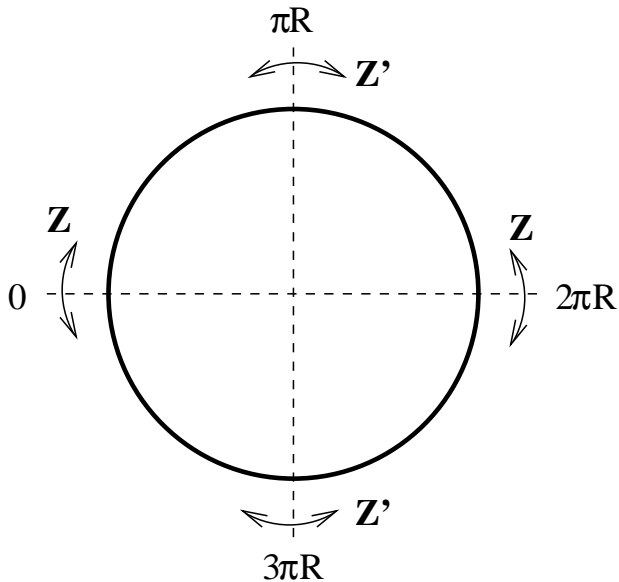}}
\caption{The action of the two $Z_2$ reflections in the extended circle picture. The fundamental domain of the
$S^1/Z_2$ orbifold is just the interval between $0$ and $\pi R$, and the theory can be equivalently
formulated on this line segment as well.}
\end{figure}

\subsubsection{Breaking of the grand unified gauge group via orbifolds in SUSY GUT's}

One of the main motivation for considering supersymmetric gauge theories (besides the resolution of the 
hierarchy problem) is the unification of gauge couplings. The SM matter fields seem to fall nicely
into SU(5) representations ${\bf 10 +\bar{5}}$. This has motivated people in the 70's to consider
theories based on an SU(5) gauge symmetry, which is broken down at high scales to SU(3)$\times$SU(2)$\times$ U(1).
If one calculates the running of the SM gauge couplings in the non-supersymmetric SM, they get close to each other
at a scale $\sim 10^{14}$ GeV, but don't really unify to the level required by the experimental precision of the 
three gauge couplings. However, in supersymmetric theories the precision of unification of the couplings
is much better than in the non-supersymmetric case and the scale of unification is pushed to somewhat larger 
values, of order $2 \cdot 10^{16}$ GeV. This is a consequence of the modification of the beta functions
that determine the running of the couplings due to the addition of the superpartners at some relatively low
scale. The other important consequence of the higher unification scale is that proton decay mediated
by the exchange of gauge bosons that transform both under SU(3) and SU(2) (the so-called $X,Y$ gauge 
bosons) will be more suppressed, since the decay rate depends on the fourth power of the $X,Y$ gauge boson mass,
which is proportional to the unification scale. Thus it looks like a complete supersymmetric GUT model 
could be completely realistic without fine tuning. There is however one important conceptual problem
in SUSY GUT's (and also in ordinary GUT's in general): the doublet-triplet splitting problem.

In order to get the set of beta functions in a SUSY GUT which leads to a sufficiently accurate prediction
of the strong coupling at low energies one needs to have the beta functions of the MSSM. As explained before,
the SM matter fields fall into complete SU(5) multiplets, however the two Higgs doublets of the MSSM do not.
Thus if one takes the GUT idea seriously one needs to embed the Higgses into some complete SU(5) multiplets,
and then somehow make the extra fields much heavier than the doublet. However, this seems very difficult:
one needs to give a mass to the extra fields that are $10^{14}$ times larger than the doublets. For example in the 
simplest case the two Higgs doublets are embedded as $H_u={\bf 2} \to {\bf 5}$, $H_d={\bf 2^*} \to {\bf \bar{5}}$.
In this case we had to add an extra triplet and antitriplet of SU(3), which have to be very heavy, while the 
doublets from the same multiplet light. This is another naturalness problem that is specific to 
SUSY GUT's. 
If the mass of the triplet was too low, the beta functions would change, and unification of couplings would not occur.
Even if one can somehow arrange naturally for the triplets to be heavy (there are some nice natural solutions
to the doublet-triplet splitting problem), they still contribute to proton decay at a rate that is usually too large.
The reason is that due to supersymmetry and grand unification the fermionic partners of 
heavy color triplet Higgses  necessarily couple to the SM fermions and the scalar partners via a Yukawa coupling
term. Since the triplets have to be massive there is necessarily a diagram of the form

\centerline{\includegraphics[width=0.5\hsize]{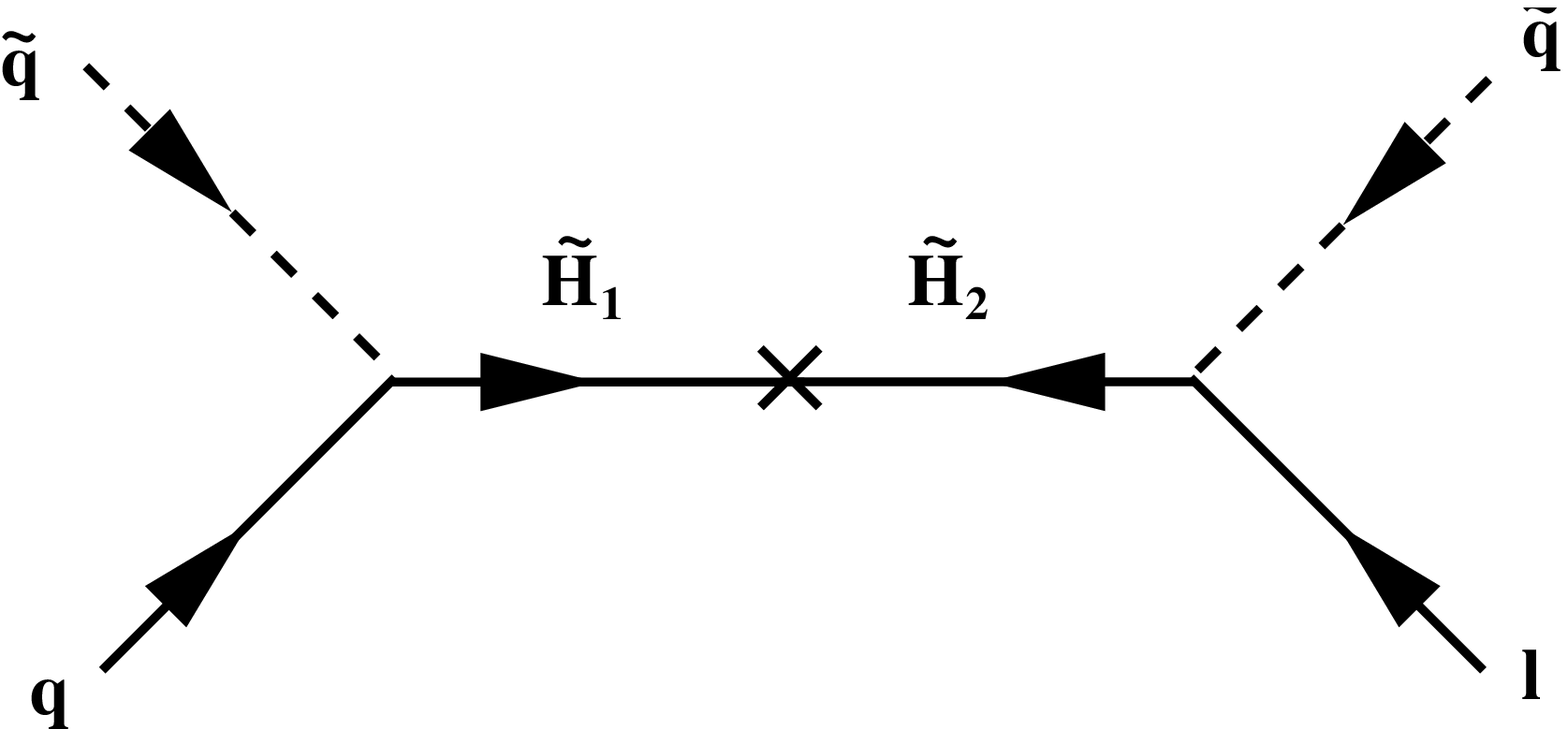}}
\noindent which is baryon number violating. Here the dotted line is a squark, and $\tilde{H}_{1,2}$ are the color triplet
Higgsinos. The squark-quark-color triplet Higgsino vertex must exist, since the ordinary Yukawa coupling must exist,
and this is the grand unified extension of the superpartner of the ordinary Yukawa vertex. The cross denotes the mass 
term between the two color triplet Higgsinos, which should also exist to lift the mass of these triplets to above the 
GUT scale. Now one can take this diagram and dress it up with some more couplings that turn the
squarks back to quarks to get a contribution to proton decay:

\centerline{\includegraphics[width=0.5\hsize]{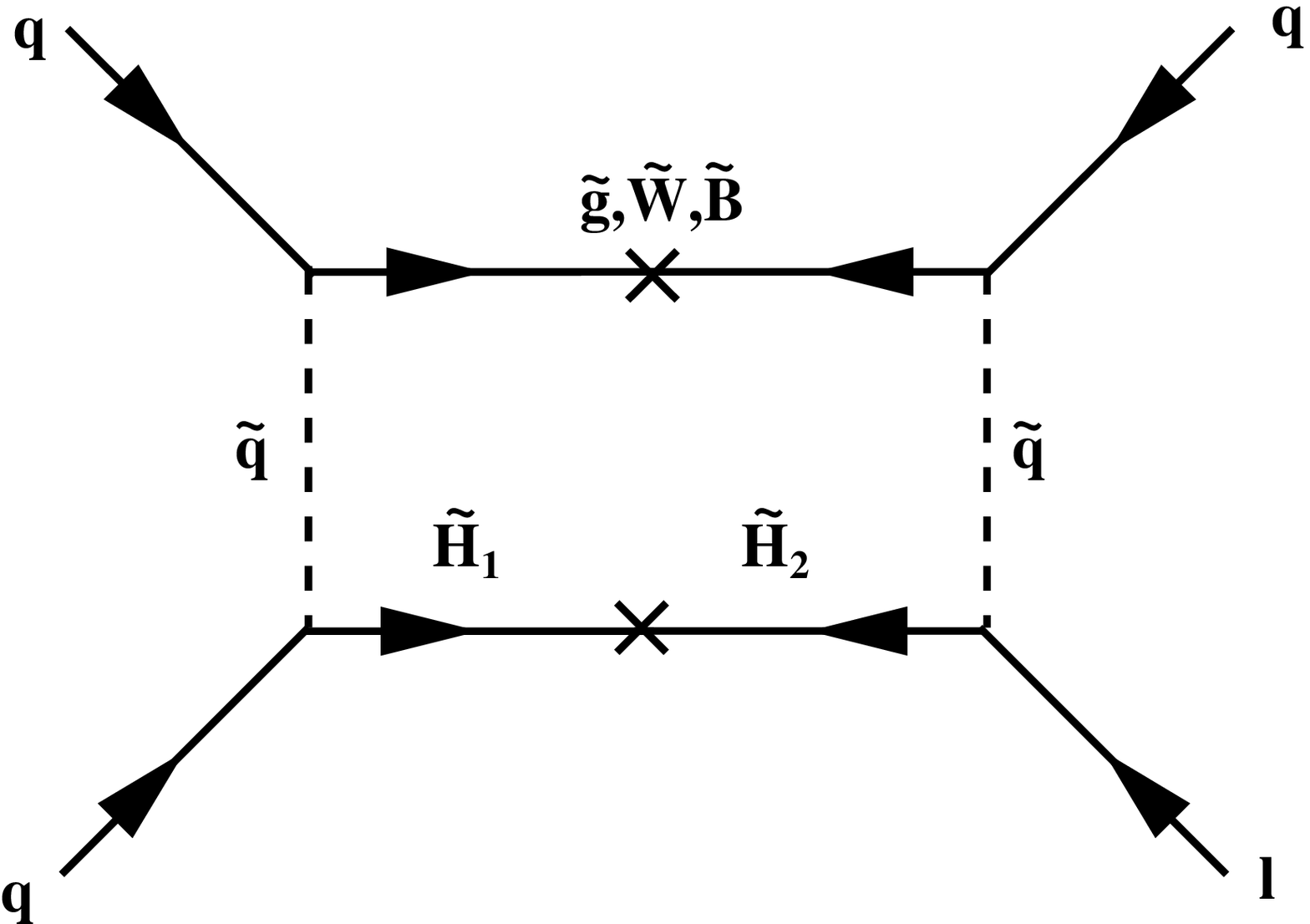}} 
\noindent The additional vertices used in this diagram must exist due to supersymmetry, as well as the gaugino mass insertions since
the gauginos have not been observed. 

In extra dimensional theories an $S^1/Z_2$ orbifold structure with a $Z_2\times Z_2'$ reflection structure could
give a nice explanation for the absence of the light triplets. In order to start building up an extra dimensional
GUT, we take a 5D SU(5) gauge theory with the minimal amount of supersymmetry ${\cal N}=1$ in 5D (meaning eight
superchages) in the bulk. Then by definition we need to have the 5D ${\cal N}=1$ vector superfields in the bulk,
while for the matter fields one can choose if they are brane fields or bulk fields. For now we will choose the 
fermionic matter fields to be brane fields, while the Higgs fields to be bulk fields. One of the two $Z_2$ 
reflections will act exactly as in the case of gaugino mediation: its role is simply to split the 
4D ${\cal N}=2$ multiplets into a 4D ${\cal N}=1$ vector superfield which has a zero mode, while the 
4D ${\cal N}=1$ chiral superfield which does not have a zero mode:

\vspace*{1cm}
\centerline{\includegraphics[width=0.5\hsize]{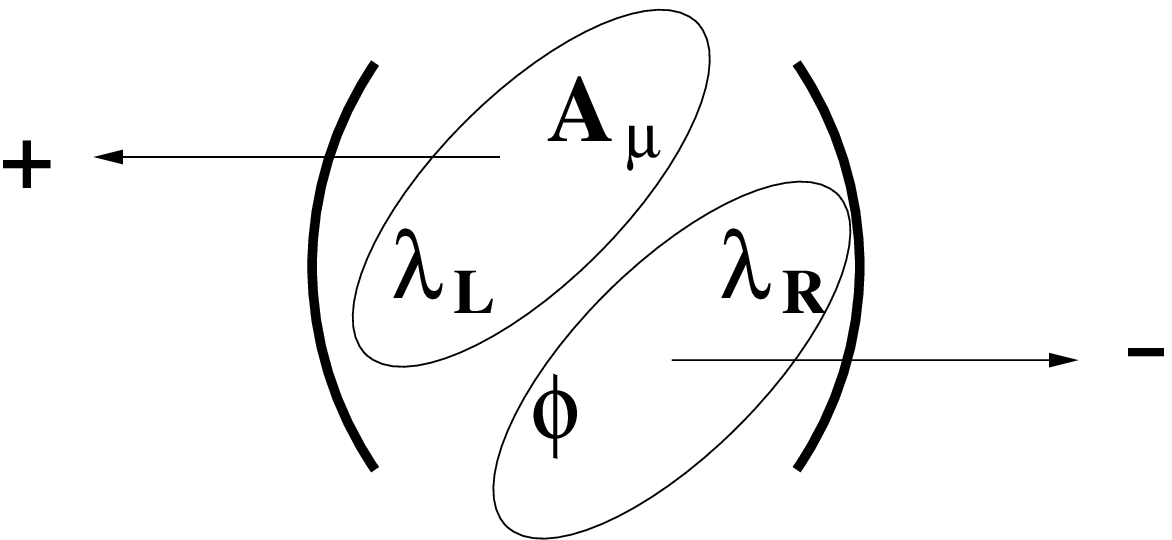}} 

In order to have a grand unified extra dimensional SUSY GUT, the Higgs doublets will live in 
5D chiral superfields in the ${\bf 5}$ representation of SU(5), 
which correspond to a 4D ${\cal N}=2$ hypermultiplet. Since we have two Higgses in the 
MSSM, we will effectively introduce {\it four} 4D chiral superfields in the ${\bf 5}$ of SU(5):
\begin{equation}
H= (5,\bar{5}), \ \ \ H'=(5',\bar{5}').
\end{equation}
We will use the second $Z_2$ reflection to break the SU(5) GUT symmetry down to the SU(3)$\times$SU(2)$\times$U(1)
of the MSSM. We will do this by picking the representation matrix for $Z_2'$ on the fundamental of SU(5) to be 
non-trivial:
\begin{equation}
Z_2': \ \ \ \left( \begin{array}{c} \\ \\ 5 \\ \\ \\ \end{array} \right) \to \left( \begin{array}{ccccc} -\\ & - \\ & & - \\& & & +  \\ & & & & +\\ \end{array} \right)\left( \begin{array}{c} \\ \\ 5 \\ \\ \\ \end{array} \right).
\end{equation}
Once the action of the second reflection is fixed on the fundamental ${\bf 5}$ of SU(5), we can also determine its
action on the adjoint ${\bf 24}$. Since an adjoint is a product of a fundamental and anti-fundamental, the action of 
$Z_2'$ will be such, that the various components of the adjoint matrix pick up the following signs:
\begin{equation}
Z_2': \ \ \ \left( \begin{array}{cccc} &&&\\&24&& \\ &&& \\ &&&\\ 
\end{array} \right) \to \left( \begin{array}{ccc|c} 
&&&\\ & +&& - \\ & &  &  \\ \hline & -&&  +  \\ \end{array} \right).
\end{equation}
This implies, that the fields at the position of the $X,Y$ gauge bosons will be odd under this second 
reflection, while the others will be even. The fields that are even under both $Z_2$ reflections will have a zero mode, 
while the others will not. We can now list the parities of all the bulk fields and find out what kind of 
KK towers these fields will have. The bulk fields are the following. The 5D vector superfield of SU(5) 
consists of the 4D vector superfield and the 4D chiral superfield in the adjoint:
\begin{equation}
\left( \begin{array}{ccc|c} 
&&&\\ & V_{SM}^a && X \\ & &  &  \\ \hline & Y&&  V_{SM}^a  \\ \end{array} \right), \ \ \ \left( \begin{array}{ccc|c} 
&&&\\ & \lambda^a && x  \\ & &  &  \\ \hline & y&&  \lambda^a  \\ \end{array} \right).
\end{equation}
The bulk Higgs fields decompose as 
\begin{equation}
H=(3+2,\bar{3}+\bar{2}), \ \ \ H'=(3'+2',\bar{3}'+\bar{2}').
\end{equation}
With this notation we can give the two $Z_2$ charges of all fields, and the corresponding KK masses:
\begin{equation}
\begin{array}{c|c|c|c} (Z_2,Z_2') & {}\rm mode& {\rm KK\ mass} & {\rm wave\ function}\\ 
\hline 
(+,+) & V^a_{SM}, 2,2' & \frac{2n}{R} & \cos \frac{2ny}{R}\\
(+,-) & \lambda^a,3, \bar{3}' & \frac{2n+1}{R} &  \sin \frac{(2n+1)y}{R}\\
(-,+) & x,y,\bar{3},3' & \frac{2n+1}{R} &  \cos \frac{(2n+1)y}{R}\\
(-,-) & X,Y, \bar{2},\bar{2}' & \frac{2n+2}{R} & \sin \frac{2ny}{R}\end{array}, \ \ n=0,1,2,\ldots 
\end{equation}
We can see from this table that with the charge assignments as above we would get a zero mode only for the fields
that are present in the MSSM as well, while the KK towers for the other fields would start at a non-zero value,
which is determined by  the radius of the extra dimension.

\begin{figure}
\vspace*{1cm}
\centerline{\includegraphics[width=0.5\hsize]{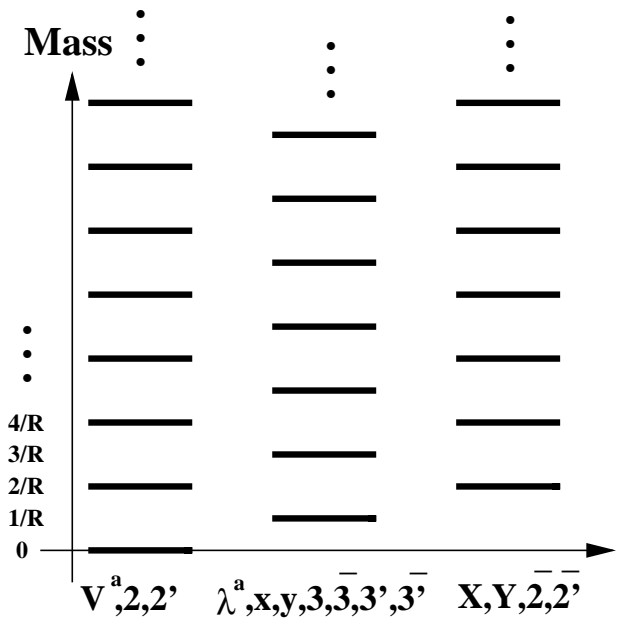}} 
\label{fig:KKGUT}
\caption{The KK towers for the various modes of the orbifold GUT theory
of Hall and Nomura.}
\end{figure}

The structure of the KK towers for the various fields is depicted in 
Fig.~\ref{fig:KKGUT}. From this figure we can see that by simply
assigning different $Z_2$ transformation properties to the doublet and 
the triplet in $5$ one can achieve the doublet-triplet splitting, due to the 
fact that the boundary condition for the triplet will not allow the existence of a triplet zero mode, while there is one for the doublet. Also note, that since
3 and $\bar{3}$ have different wave functions, there can be no mass term
of the form $3\bar{3}$, which implies that the dimension five proton decay 
operator discussed above is vanishing, since in that diagram the 
mass insertion (the cross) is vanishing. Instead the 3 and $\bar{3}$ get 
masses with $\bar{3}'$ and 3' respectively, which will not give 
any contribution to proton decay. The vanishing of the $3\bar{3}$ mass 
term is a consequence of ${\cal N}=2$ supersymmetry. However, at the brane 
where ${\cal N}=2$ supersymmetry is broken to ${\cal N}=1$, one could
in principle add an operator that would include a mass term for these fields. 
These mass terms can nevertheless be forbidden by requiring that a $U(1)_R$
global R-symmetry is obeyed, under which the charges of the fields are given by
\begin{equation}
\begin{array}{c|ccccccc}
&\Sigma & H & \bar{H} & H' & \bar{H}' & T_{10} & F_{\bar{5}} \\ \hline
U(1)_R & 0 & 0 & 0 & 2 & 2 & 1 & 1 \end{array}
\end{equation}

Since we are talking about a GUT model, it is important to make sure that the 
model would actually predict the unification of the gauge couplings. Since 
the GUT symmetry is broken at one of the endpoints of the interval, 
there can be a brane-induced kinetic term for the SM gauge fields on the 
GUT breaking brane which does not necessarily have to be SU(5) symmetric.
Therefore the tree-level 
matching between the 4D effective SM couplings and the 
5D gauge  coupling is given by
\begin{equation}
\frac{1}{g_{eff, \ i}^2}=\frac{\pi R}{g_5^2}+\frac{1}{g_{brane,\ i}^2},
\end{equation}
where the first term on the right is the SU(5) symmetric bulk gauge coupling,
while the second term is the (possibly) SU(5) violating brane coupling. How 
precisely the couplings will unify depends on the ratio of the brane induced 
term to bulk term. If the size of the extra dimension is reasonably large 
compared to the cutoff scale (for example $1/R\sim M_{GUT}$, $M_{cutoff}\sim 100 M_{GUT}$)
then the SU(5) violating brane term will be volume suppressed, and it
is reasonable to expect unification to a good precision.

Of course it is not enough to build an extra dimensional GUT model, it is also
important to check that the appearance of the new states charged under 
the SM gauge groups (the KK modes of the various gauge and Higgs fields) do 
not ruin the actual unification of couplings of the MSSM. This has been 
shown to be the case in~\cite{HN}, and the reader is referred for the details of 
that calculation to the original paper.

\subsubsection{Supersymmetry breaking via orbifolds}

We have seen in the previous two models that we can use a $Z_2$ projection 
under the $y\to -y$ reflection symmetry to break ${\cal N}=2$ supersymmetry
down to ${\cal N}=1$ SUSY. We have also seen that in the case of a single extra 
dimension there are {\it two separate} $Z_2$'s at our disposal at both ends 
of the interval. This yields the following nice possibility: one could use one
$Z_2$ to break one half of the supercharges, and the other $Z_2$ to break the 
other half:

\vspace*{1cm}
\centerline{\includegraphics[width=0.5\hsize]{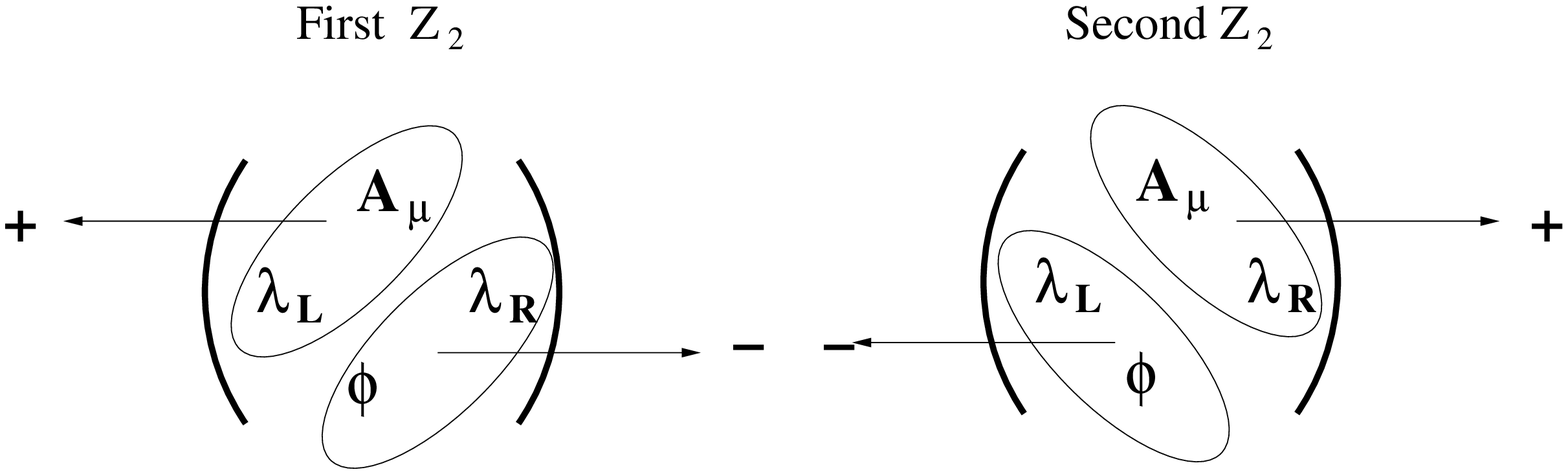}} 

The two $Z_2$'s leave a different combination of ${\cal N}=1$ supercharges
unbroken. The two projections together completely break supersymmetry, but {\it locally}
there is always some amount of supersymmetry unbroken. The full supersymmetry breaking
is only felt {\it globally}. This will have the important implication that the properties 
which follow from local supersymmetry will be maintained in the non-supersymmetric theory.

The next task is to also include the MSSM matter fields and Higgs fields into this theory.
A generic matter field of the SM will be an ${\cal N}=2$ hypermultiplet (two fermions and 
two scalars), and we can assign the $Z_2$ parities for the matter fields again such that 
only the one of the fermions will have a zero mode, while all the other fields from this 
hypermultiplet will have no zero modes.  For example for quarks this can be achieved by
the following choice of $Z_2$ projections:

\vspace*{1cm}
\centerline{\includegraphics[width=0.5\hsize]{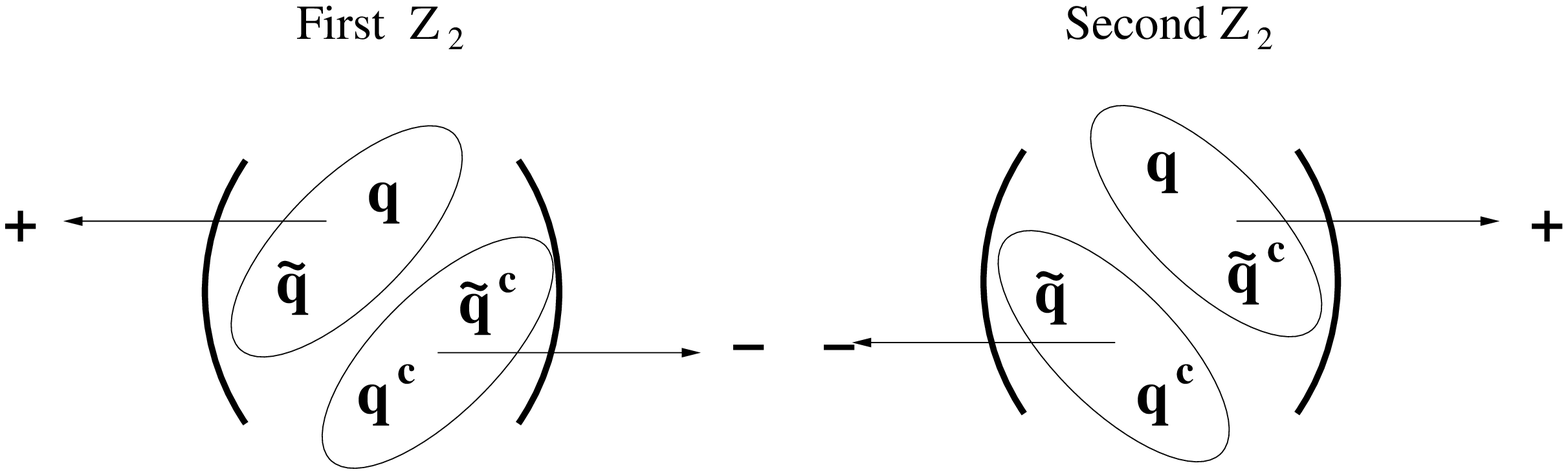}} 

For the Higgs, one has two possibilities: introducing two separate hypermultiplets for the 
two MSSM Higgs fields, which is the more straightforward but less exciting possibility,
or introduce just a {\it single} Higgs doublet hypermultiplet which contains both
of the MSSM Higgs doublets. We will consider the latter possibility, and then 
use the projections

\vspace*{1cm}
\centerline{\includegraphics[width=0.5\hsize]{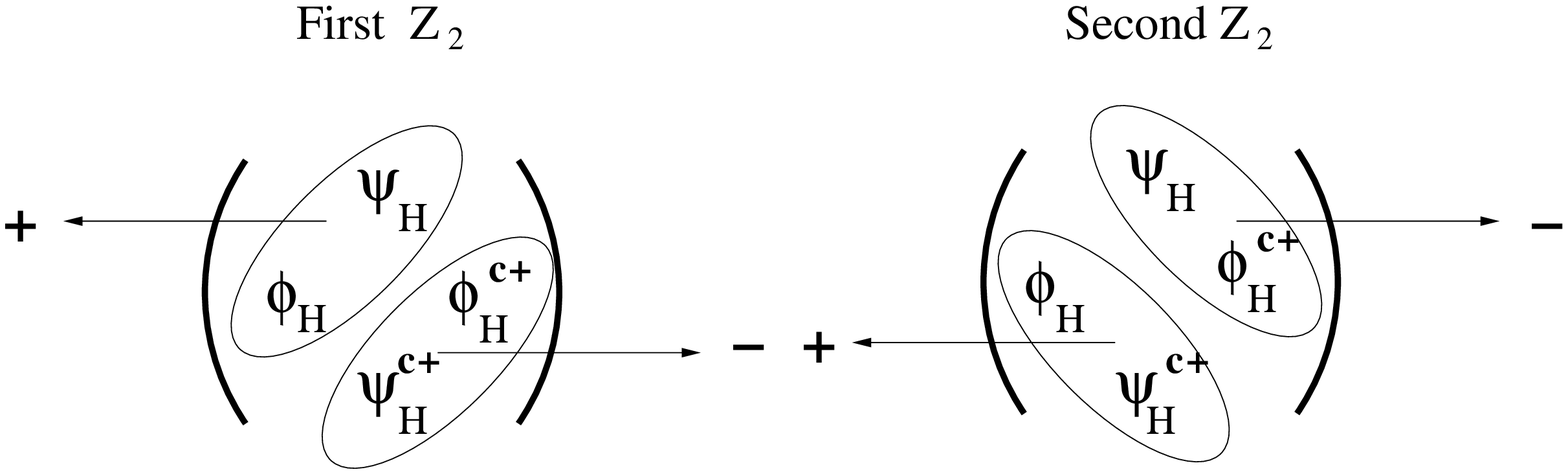}} 

In this case, there will only be a single Higgs zero mode, contrary to the MSSM which has
two Higgs doublets. So the zero mode spectrum really reproduces that of the SM, and 
{\it not} that of the ordinary particles of the MSSM. 

${\cal N}=2$ supersymmetry does not allow a Yukawa-type coupling between hypermultiplets. This means
that in the bulk one can not write down the appropriate couplings necessary for generating 
the fermion masses from the Yukawa couplings to the Higgs field. However, since 
on the fixed points ${\cal N}=2$ supersymmetry is broken to ${\cal N}=1$, these couplings 
can be added on the fixed points. One may wonder how this could be possibly done in a 
supersymmetric theory with just one Higgs field. The point is that since it is a 
{\it different} set of supercharges that remains unbroken at the two fixed point, one needs to use
a different splitting of the bulk Higgs hypermultiplet into ${\cal N}=1$ chiral multiplets
on the two fixed points. Thus from the full bulk Higgs hypermultiplet 
\begin{equation}
\left( \begin{array}{ccc} & \Psi_H & \\ H & & H^{c\dagger}\\ & \Psi_H^{c\dagger} & \end{array}
\right)
\end{equation}
we form the chiral multiplets
\begin{equation}
H_u \sim \left( \begin{array}{c} H \\ \Psi_H \end{array} \right), \ \ 
H_d \sim \left( \begin{array}{c} H^* \\ \Psi_H^c \end{array} \right).
\end{equation}
Note, that it is the same scalar component (the one with a zero mode) appearing in both
chiral multiplets, while one has a different fermionic partner corresponding to them
depending on which set of supercharges are unbroken. Thus we can write down the following
superpotential terms at the fixed point yielding the required Yukawa couplings:

\vspace*{1cm}
\centerline{\includegraphics[width=0.5\hsize]{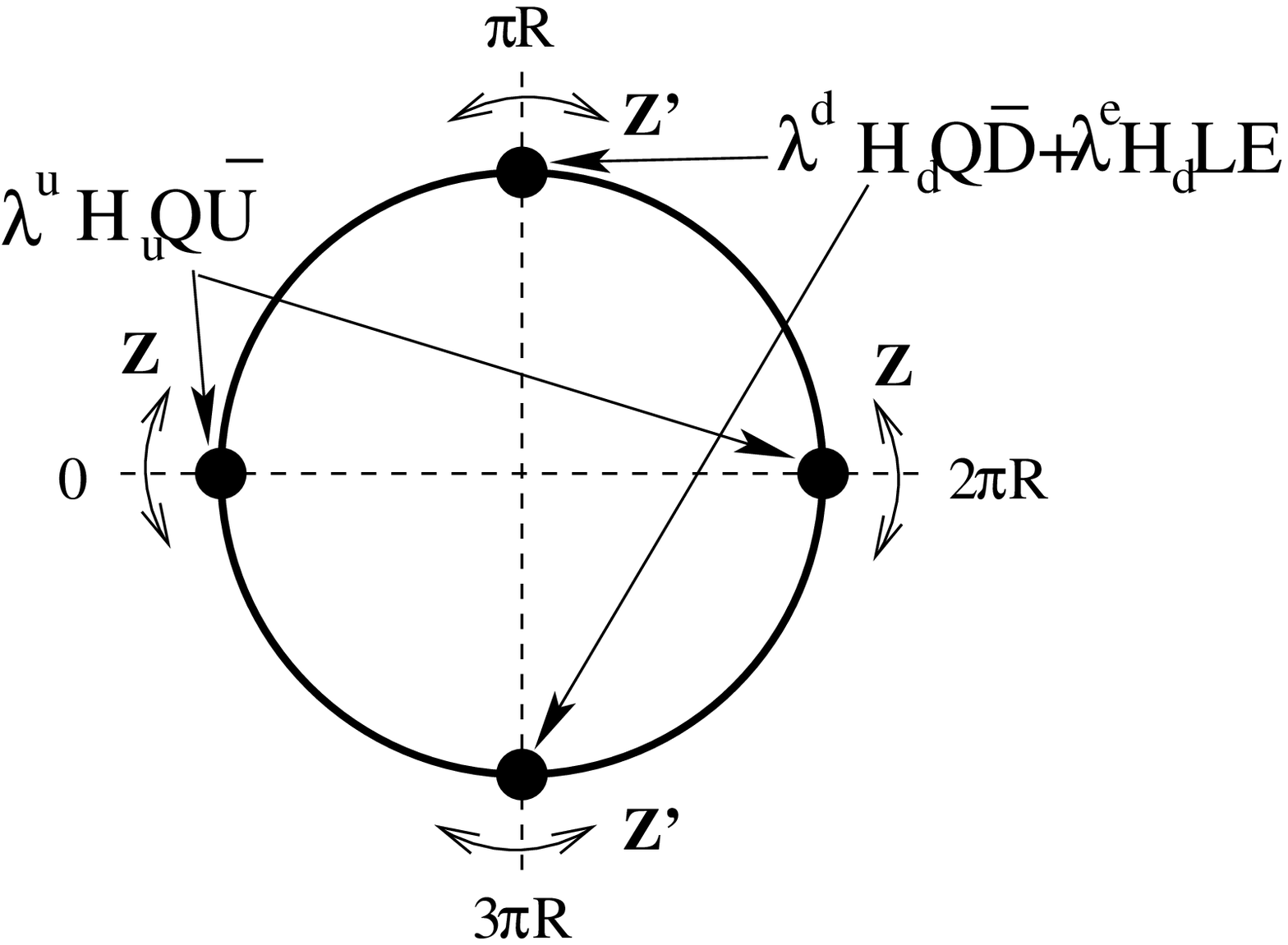}} 

The tree-level Higgs potential will be fixed just as in ordinary supersymmetric
theories. Without a bulk mass term the form of the potential is just determined from the 
D-terms, and is given by:
\begin{equation}
V_{tree}= \frac{g^2+g'^2}{8} |H|^4.
\end{equation}
This on its own determines that the Higgs mass has to be light, as in any supersymmetric 
theory.
Electroweak symmetry breaking can then happen radiatively, due to the top quark loops.
Since the top Yukawa coupling is localized at one of the fixed points, the resulting 
contribution to the Higgs potential can also only be a term at the fixed point. Since at least 
some of supersymmetry is locally unbroken everywhere, the contribution can not be quadratically
divergent. In fact, the contribution to the potential from the top quark loop is actually 
finite. However, there can be localized U(1) Fayet-Iliopoulos terms at the fixed points
which can be quadratically divergent, which will contribute to the Higgs potential. Also, one can
add a bulk mass term for the Higgs hypermultiplet, which also obviously contributes
to the Higgs mass. For further details on the prediction of the physical Higgs mass 
of this model see ~\cite{BHN,Nilles,BHN2}.

\section{Warped Extra Dimensions}
\label{sec:warped}
\setcounter{equation}{0}
Until now we have considered flat extra dimensions, that is assumed that whatever
gravitational sources there are present, their effects will be negligible for determining
the background metric, that is we ignored the {\it backreaction} of gravity to the 
presence of the branes themselves. If the tensions of the branes are small, then this is a 
good approximation. However, interesting situations may arise, if this is not the case. 

One can immediately see that taking the reaction of gravity to brane sources into account 
could be very important. The reason is that if one has a 4D theory with only 4D sources, these will
necessarily lead to an expanding universe. However, if one has 4D sources in 5D, one can {\it balance}
the effects of the 4D brane sources by a 5D bulk cosmological constant to get a theory where
the {\it effective} 4D cosmological constant would be vanishing, that is the 4D Universe
would still appear to be static and flat for an observer on a brane~\cite{RubShap}. The price to pay for this 
is that the 5D background itself will be curved, which is clear from the fact  that one had to introduce
a bulk cosmological constant. Thus one can ``off-load curvature'' from the brane into the 
bulk~\cite{RubShap,selftune},
and keep the brane to be still flat by curving the extra dimension. Such curved extra dimensioned
are commonly referred to as ``warped extra dimensions''. This possibility has been 
first pointed out by Rubakov and Shaposhnikov~\cite{RubShap}. This shows that for example the 
cosmological constant problem will be placed into a completely different light from an 
extra dimensional point of view: in 4D one had to explain why the vacuum energy is very small, while
in higher dimensions one has to explain why the vacuum energy of the SM fields is 
exactly canceled by a bulk vacuum energy~\cite{selftune}.

In the following we will discuss the best known and most concrete example of warped extra dimensions,
the Randall-Sundrum model~\cite{RS1,RS2}. First we discuss in detail how to solve the Einstein equations
to find the RS background, and discuss the mass scales in the model. Then we discuss the phenomenology
of various interesting models based on the Randall-Sundrum backgrounds.

\subsection{The Randall-Sundrum background}

We will show how to derive the Randall-Sundrum
solution (and the conditions under which a 4D flat background exists). We will begin again with a finite 
length~\cite{RS1} for the extra dimension, that is we consider an $S^1/Z_2$ orbifold, 
which by now is familiar to everyone. Since we are interested in warped solutions, we will
assume that there is a non-vanishing 5D cosmological constant $\Lambda$ in the bulk. We are also
interested in solutions that have the property mentioned above, that is even though the extra 
dimension is curved, the brane itself remains static and flat, that is it preserves 4D Lorentz
invariance. This means that the induced metric at every point along the extra dimension has to be 
the ordinary flat 4D Minkowski metric, and the components of the 5D metric depend only on the 
fifth coordinate $y$. The ansatz for the most general metric satisfying these properties is 
given by:
\begin{equation}
ds^2=e^{-A(y)} dx^\mu dx^\nu \eta_{\mu\nu} -dy^2.
\end{equation} 
The amount of curvature (warping) along the extra dimension depends on the function $e^{-A(y)}$,
which is therefore called the warp-factor. Our task is to find this function $A(y)$.
Note, that there are of course various ways 
of writing this metric, by doing various coordinate transformations. 
For example one could as well have gone into the coordinate system where there is an overall
prefactor in front of all coordinates, which is called the conformally flat metric. This is the 
simplest coordinate system, and this is what we will use to find $A$. To go into the conformally
flat frame, we need to make a coordinate transformation of the form
$z=z(y)$. The coordinate transformation should not depend on the 4D coordinates $x$, since 
those would induce off-diagonal terms in the metric. In order to ensure that the metric be conformally 
flat in the new frame, $dy$ and $dz$ have to be related by
\begin{equation}
e^{-A(z)/2}dz=dy,
\end{equation}
such that the full metric in the $z$ coordinates will be:
\begin{equation}
ds^2=e^{-A(z)} (dx^\mu dx^\nu \eta_{\mu\nu} -dz^2). 
\end{equation}
This metric is conformally flat, so a conformal transformation (that is an overall rescaling of the
metric) connects it to the flat metric:
\begin{equation}
g_{MN}=e^{-A(z)} \tilde{g}_{MN}, \ \ \tilde{g}_{MN} =\eta_{MN}.
\end{equation}
A very useful relation connects the Einstein tensor $G_{MN}=R_{MN}-\frac{1}{2}g_{MN}R$
calculated from $\tilde{g}$ and 
$g$ in any number of dimensions and for arbitrary function $A$ (see for example~\cite{Wald}):
\begin{equation}
G_{MN}=\tilde{G}_{MN} +\frac{d-2}{2} \left[
\frac{1}{2} \tilde{\nabla}_M A\tilde{\nabla}_N A +\tilde{\nabla}_M \tilde{\nabla}_N A
-\tilde{g}_{MN} (\tilde{\nabla}_K \tilde{\nabla}^K A-\frac{d-3}{4}
\tilde{\nabla}_K A\tilde{\nabla}^K A)\right].
\label{conformal}
\end{equation}
Here the covariant derivatives $\tilde{\nabla}$ are with respect to the metric
$\tilde{g}$, while on the left the Einstein tensor $G_{MN}$ is calculated from
$g_{MN}$. For us $\tilde{g}_{MN}=\eta_{MN}$, and therefore the covariant derivative
$\tilde{\nabla}_M \to \partial_M$, and $d=5$.
Using these expression we can now easily evaluate the non-vanishing components of the Einstein tensor:
\begin{eqnarray}
G_{55}=\frac{3}{2} A'^2 \nonumber \\
G_{\mu\nu}=-\frac{3}{2} \eta_{\mu\nu} (-A''+\frac{1}{2}A'^2).
\end{eqnarray}
This takes care of the left hand side of the bulk Einstein equation $G_{MN}=\kappa^2 T_{MN}$,
where $\kappa^2$ is the higher dimensional Newton constant, in 5D it is related to the 
5D Planck scale by 
\begin{equation}
\kappa^2=\frac{1}{M_*^3}.
\end{equation}
The 5D Einstein action for gravity with a bulk cosmological constant $\Lambda$ is 
\begin{equation}
S=-\int d^5x \sqrt{g} (M_*^3 R+\Lambda ).
\end{equation}
One can then use the definition of the stress-energy tensor to find the Einstein equation:
\begin{equation}
G_{MN}=\kappa^2 T_{MN}=\frac{1}{2 M_*^3}\Lambda g_{MN}.
\end{equation}
The 55 component of the Einstein equation will then be:
\begin{equation}
\frac{3}{2} A'^2 =\frac{1}{2M_*^3} \Lambda e^{-A}.
\end{equation}
The first thing that we can see is that a solution can only exist if the bulk cosmological
constant is negative $\Lambda <0$. This means that the important case for us will be 
considering anti-de Sitter spaces, that is spaces with a negative cosmological constant. Once 
a negative cosmological constant is fixed, we can take the root of the above equation
\begin{equation}
A'=\sqrt{-\frac{\Lambda}{3M_*^3}} e^{-A/2}.
\end{equation}
Introducing the new function $e^{-A/2}\equiv f$ we get the equation
\begin{equation}
-f'/f^2=\frac{1}{2}\sqrt{-\frac{\Lambda}{3M_*^3}},
\end{equation}
from which we can read off the solution for the metric
\begin{equation} 
\label{AdS1}
e^{-A(z)}=\frac{1}{(kz+{\rm const.})^2},
\end{equation}
where we have introduced 
\begin{equation}
k^2=-\frac{\Lambda}{12 M_*^3}.
\end{equation}
The constant in (\ref{AdS1}) can be fixed by fixing the value of the metric at some point 
(this is an irrelevant rescaling of the units), we choose $e^{-A(0)}=1$, from which 
the constant is set to 1.
So a simple form for writing the metric is 
\begin{equation}
e^{-A(z)}=\frac{1}{(kz+1)^2}.
\end{equation}
However since we are on a $S^1/Z_2$ orbifold, the solution must be symmetric under 
$z\to -z$ reflection, and therefore the final form of the RS solution can be written in the form
\begin{equation}
ds^2=\frac{1}{(k|z|+1)^2} (\eta_{\mu\nu}dx^\mu dx^\nu-dz^2).
\label{RS}
\end{equation}
With this the 55 component of the Einstein equations is solved everywhere. However one needs to 
check whether the 4D components are also satisfied. We have seen above that these components 
are given by $-3/2 \eta_{\mu\nu} (-A''+A'^2/2)$. This means that (unlike in the 55 component)
$A''$ will appear, which will imply that there are delta function contributions to the 
Einstein tensor at the fixed points. Thus (\ref{RS}) can be a solution only, if 
there are also localized energy densities on the brane that compensate these delta functions. 
This is exactly as we have expected at the beginning of this section: there can be a non-vanishing 
bulk cosmological constant and still have flat induced 4D space if there is some energy density on the 
brane that compensates. Note however, that the 55 component of the Einstein equation already determined 
(together with the $S^1/Z_2$ assumption) the full solution. So the energy density on the brane necessary
to have a static solution would be fine-tuned against the bulk cosmological constant. Evaluating
$A''$ we get
\begin{equation}
A''=-\frac{2k^2}{(k|z|+1)^2 }+\frac{4k}{k|z|+1}(\delta (z)-\delta(z-z_1)),
\end{equation}
where $z_1$ is the location of the negative tension brane, 
and so the $\mu\nu$ components of the Einstein tensor are
\begin{equation}
G_{\mu\nu} = -\frac{3}{2} \eta_{\mu\nu}\left[ \frac{4k^2}{(k|z|+1)^2}
- \frac{4k(\delta (z)-\delta(z-z_1))}{k|z|+1}\right]
\end{equation}
The first term (the term without the delta functions) exactly matches the bulk contribution to the energy momentum
tensor from the bulk cosmological constant. Therefore the remaining delta function terms are the ones that need to
be compensated by adding additional sources onto the branes. To match these sources we need to find out what the
energy-momentum tensor of a brane tension term $V$ (an energy density localized on the brane) would be. The action 
for this is given by
\begin{equation}
\int d^4 x V \sqrt{g^{induced}}=\int d^5 x V \frac{\sqrt{g}}{\sqrt{g_{55}}} \delta (y)
\end{equation}
for a flat brane at $y=0$. This implies that the energy momentum tensor is
\begin{equation}
T_{\mu\nu}=\frac{1}{\sqrt{g}} \frac{\delta S}{\delta g^{\mu\nu}}=\frac{1}{2 \sqrt{g_{55}}} g_{\mu\nu} V \delta (y).
\end{equation}
Thus the final form of the energy-momentum tensor for a brane tension is:
\begin{equation}
T_{\mu\nu}^{tension}=\frac{1}{2}{\rm diag} (V,-V,-V,-V,0) e^{-A/2} \delta (y).
\end{equation}
Therefore to satisfy the Einstein equations at the brane we need to have two brane tensions, one at each end
of the interval (at the two fixed points), and so we need the equality
\begin{equation}
 -\frac{3}{2} \eta_{\mu\nu}\left[-\frac{4k(\delta (z)-\delta(z-z_1))}{k|z|+1}\right]=
\frac{\eta_{\mu\nu}}{2M_*^3} \left[\frac{V_0 \delta (z)+V_1\delta(z-z_1))}{k|z|+1}\right].
\end{equation}
This implies that the two brane tensions at the two fixed points will have to be opposite, and given by
\begin{equation}
V_0=-V_1=12 k M_*^3.
\end{equation}
Plugging in the expression for $k$ we obtain that the bulk cosmological constant and the brane 
tensions have to be related by
\begin{equation}
\Lambda=-\frac{V_0^2}{12M_*^3}, \ \ V_1=-V_0.
\label{RSfinetune}
\end{equation}
Thus there is a static flat solution only, if the above {\it two} fine tuning conditions are satisfied.
At this point it is not clear why we ended up with two fine tuning conditions. We could have expected at least 
one fine tuning, related to the vanishing of the 4D cosmological constant. However, at this point the 
meaning of the second fine tuning condition is obscure, but we will find out the reason behind the second fine tuning
shortly in Sec.~\ref{radionstabilize}. 

Now that we have found that the RS solution in the $z$ coordinates where the metric is conformally flat is
\begin{equation}
ds^2=\frac{1}{(k|z|+1)^2} (\eta_{\mu\nu}dx^\mu dx^\nu -dz^2),
\end{equation}
we can ask what the metric in the original $y$ coordinates would be. The reason why this is 
interesting is that $y$ is the physical distance along the extra dimension, since in that metric 
there is no warp factor in front of the $dy^2$ term. Since the relation between $z$ and $y$ was given 
by 
\begin{equation}
e^{-A(z)/2} dz = \frac{dz}{k |z|+1}=dy,
\end{equation} 
we get that (by choosing $y=0$ to correspond to $z=0$):
\begin{equation}
\frac{1}{(k|z|+1)^2}=e^{-2k |y|},
\end{equation}
and so the RS metric in its more well-known form is finally given by:
\begin{equation}
ds^2=e^{-2k|y|} dx^\mu dx^\nu \eta_{\mu\nu}-dy^2.
\end{equation}

Let us now discuss what the physical scales in a theory like this would be, if the matter fields were localized
on one of the fixed points. First we assume this to be the brane with the negative tension (the brane where the induced 
metric is exponentially small compared to the other fixed point). Consider thus a scalar field (for example the 
Higgs scalar) on the negative tension brane. Its action would be given by
\begin{equation}
S^{Higgs}=\int d^4 x \sqrt{g^{ind}} [g_{\mu\nu}D^\mu HD^\nu H -V(H)], \ \ \ V(H)=\lambda [(H^\dagger H)-v^2)^2].
\end{equation}
If the size of the extra dimension is $b$, then the induced metric at the negative tension is given by
\begin{equation}
g_{\mu\nu}^{ind}|_{y=b}=e^{-2kb} \eta_{\mu\nu}.
\end{equation}
Plugging this in for the above action we get that the action for the Higgs is given by
\begin{equation}
S^{Higgs}=\int d^4x e^{-4kb}[e^{2kb} 
\eta_{\mu\nu} \partial^\mu H\partial^\nu H -\lambda (H^\dagger H-v^2)^2].
\end{equation}
We can see that due to the non-trivial value of the induced metric on the negative tension brane the
Higgs field will not be canonically normalized. To get the action for the canonically normalized field one 
needs a field redefinition $\tilde{H}=e^{-kb} H$. In terms of this field the action is
\begin{equation}
S^{Higgs}=\int d^4x [ 
\eta_{\mu\nu} \partial^\mu \tilde{H}\partial^\nu \tilde{H} -\lambda [(\tilde{H}^\dagger \tilde{H})-
(e^{-kb}v)^2]^2].
\end{equation}
This is exactly the action for a normal Higgs scalar, but with the VEV (which sets all the mass parameters)
``warped down'' to 
\begin{equation}
\tilde{v}_{Higgs}=e^{-kb} v.
\end{equation}
This shows, that all mass scales are exponentially suppressed on the negative tension brane, but not on the positive tension
brane. Therefore, the positive tension brane is often also referred to as the {\it Planck-brane}, since the 
fundamental mass scale there would be unsuppressed of the order of the Planck scale, while the 
negative tension brane is referred to as the {\it TeV-brane} since the relevant mass scale there is TeV. 
The curvature in the bulk leads
to a redshifting of all energy scales away from the positive tension brane. This means that if one introduces a 
{\it bare} Higgs VEV (and thus also Higgs mass) of order the 5D Planck scale $M_*$ into the Lagrangian, the 
{\it physical} Higgs mass and VEV will be exponentially suppressed. 

In order to find out, whether this exponential suppression is interesting we need to also find out what the 
effective scale of gravity (the 4D Planck scale) would be in this model. For this we need to find out how 
the effective 4D gravitational action depends on the radius $b$ of the extra dimension. To find this 
out quickly, for now we will just assume that the 4D graviton $h_{\mu\nu}$ is embedded into the full 5D metric as
\begin{equation}
\label{gravitonansatz}
ds^2=e^{-2k|y|} [\eta_{\mu\nu}+h_{\mu\nu}(x)] dx^\mu dx^\nu-dy^2.
\end{equation}
The subject of the next section will be mostly to verify this and to analyze the consequences of this equation in 
more detail, but for now we just accept this result. In this case the 5D Ricci tensor $R_{\mu\nu}^{(5)}$ 
will contain the Ricci tensor calculated from $h_{\mu\nu}$:
$R_{\mu\nu}^{(4)}\subset R_{\mu\nu}^{(5)}$. The reason behind this is that $R_{\mu\nu}$ is invariant under a 
constant rescaling of the metric. Plugging in the explicit metric factors that appear in the action we get 
that the relevant piece of the 5D Einstein-Hilbert action containing the 4D metric is 
\begin{equation}
S=-M_*^3 \int d^5x\sqrt{g}R^{(5)} \supset  -M_*^3 \int e^{-4k|y|} \sqrt{g^{(4)}}e^{2k|y|}
R^{(4)} d^5x.
\end{equation}
From this we can read off the value of the 4D effective Planck scale:
\begin{equation}
M_{Pl}^2=M_*^3 \int_{y=-b}^{y=b} e^{-2k|y|} dy= \frac{M_*^3}{k} (1-e^{-2kb}).
\end{equation}
For moderately large values of $b$ this expression {\it barely depends} on the size of the extra dimension,
and it is completely different from the analogous expression (\ref{grmatching}) for large extra dimensions.
It also shows, that if all bare parameters $M_*,\Lambda ,V_0, v_{Higgs}$ are of the same order and of the 
order of the Planck scale, then while the scale of the Higgs VEV (the weak scale) gets exponentially suppressed,
the scale of gravity itself will remain of the order of the Planck scale. This means that with a moderately large
$b$ one can naturally introduce an exponential hierarchy between the weak and the Planck scales! This is exactly
what one would like to achieve when solving the hierarchy problem. So we can see that the Randall-Sundrum 
model has the possibility of giving a completely new explanation to the hierarchy problem. We will have a much 
better understanding of how the RS model solves the hierarchy problem after the next section, where
we consider the behavior of gravity at different locations along the extra dimension in this model.

\subsection{Gravity in the RS model}
To find out more why the RS model solves the hierarchy problem, we need to study the properties of gravity in 
this AdS background in more detail~\cite{RS2}. Thus we really need to find the KK decomposition of the graviton 
in this curved background. Normally, when one has a flat extra dimension on a circle, one expects that there would be a
graviton zero mode, a scalar zero mode and a vector zero mode, making up for the five degree of freedom
in the 5D massless graviton, while at the massive level there would just be massive 4D gravitons, which also have 
5 degrees of freedom. However, due to the $y\to -y$ orbifold projection the zero mode of the vector will be 
eliminated. The reason is that from the generic form of the metric
\begin{equation}
ds^2=e^{-2k|y|} g_{\mu\nu} dx^\mu dx^\nu +A_\mu dx^\mu dy -b^2 dy^2,
\end{equation}
where $g_{\mu\nu}$ parametrizes the graviton, $A_\mu$ the vector and $b$ the scalar fluctuations, one can see 
that since $ds^2$ is symmetric under $y\to -y$, $A_\mu$ has to be flipping sign, and thus can not have a zero
mode. Therefore at the zero mode level one expects only a graviton and a scalar, while at the massive level only
gravitons. First we will focus exclusively on the graviton modes setting the scalar fluctuation to zero at first,
and then later we will discuss the relevance of the scalar mode. 

In order to find the KK expansion of the graviton modes, we will go to the conformal frame for the metric,
and parametrize the graviton fluctuation by
\begin{equation}
ds^2=e^{-A(z)} \left[ (\eta_{\mu\nu}+h_{\mu\nu} (x,z)) dx^\mu dx^\nu-dz^2\right].
\label{gravipert}
\end{equation}
Clearly the most general graviton fluctuations can be parametrized like this. Now we use our formula (\ref{conformal})
that relates the Einstein tensor of conformally related metrics. We take as the metric $g$ the metric following
from (\ref{gravipert}), while the conformally related metric as $g=e^{-A(z)} \tilde{g}$. Then the Einstein 
tensors are related by
\begin{equation}
G_{MN}=\tilde{G}_{MN}+\frac{d-2}{2}\left[ \frac{1}{2} \tilde{\nabla}_M A \tilde{\nabla}_N A
+\tilde{\nabla}_M \tilde{\nabla}_N A -\tilde{g}_{MN}(\tilde{\nabla}_R \tilde{\nabla}^R A-
\frac{d-3}{4} \tilde{\nabla}_R A \tilde{\nabla}^R A)\right],
\label{Einstpert}
\end{equation}
in general $d$ dimensions. Next we impose the gauge choice for the perturbations
\begin{equation}
h^\mu_\mu= \partial_\mu h^\mu_\nu=0,
\end{equation}
which is usually called the RS gauge choice. We will get back to the possibility of always choosing such a gauge 
later.

Since $\tilde{g}$ is the fluctuation of a flat metric, the linearized Einstein tensor for this case $\tilde{G}_{\mu\nu}$
is written in every book on general relativity. This is however not everything, since the covariant derivatives
$\tilde{\nabla}$ is evaluated with respect to the {\it perturbed} metric $\eta_{\mu\nu}+h_{\mu\nu}$, and so 
the Christoffel symbols are {\it not} all vanishing (as they would be for the flat background). Taking these 
extra terms in the derivatives in (\ref{Einstpert}) into account  is a tedious but doable work, which we will leave
to the reader as an exercise. Finally, we need the perturbation of the right hand side of the Einstein 
equation (the perturbation of the energy-momentum tensor $T_{MN}$),
which is quite trivial since the only change is due to $g_{MN}\to g_{MN}+\delta g_{MN}$. The 
perturbed Einstein equation is of course $\delta G_{MN}=\kappa^2 \delta T_{MN}$, which after putting all
terms together yields the linearized Einstein equation in a warped background to be
\begin{equation}
-\frac{1}{2} \partial^R\partial_R h_{\mu\nu} +\frac{d-2}{4} \partial^R A \partial_R h_{\mu\nu}=0.
\end{equation}
In order to get to an equation that we have more intuition about, we can transform it into the form of a 
conventional one-dimensional Schr\"odinger equation, by rescaling the perturbation as
\begin{equation}
h_{\mu\nu} = e^{\frac{d-2}{4} A} \tilde{h}_{\mu\nu}.
\end{equation}
This field redefinition was chosen such, that the first derivative terms in the differential equations cancel,
so one is really left with a second derivative (kinetic energy) term and a no derivative potential term.
The form of the equation in the new basis is
\begin{equation}
-\frac{1}{2} \partial^R\partial_R \tilde{h}_{\mu\nu} +\left[ \frac{(d-2)^2}{32} \partial^R A \partial_R A
-\frac{d-2}{8} \partial^R\partial_R A\right] \tilde{h}_{\mu\nu}=0.
\end{equation}
The second derivative is given by $\partial^R\partial_R=-\Yfund_x-\nabla_z^2$, where the 4D box operator is defined
as $\Yfund=-\eta^{\mu\nu}\partial_\mu\partial_\nu$. Finally, we separate the variables as 
$\tilde{h}_{\mu\nu}(x,z)= \hat{h}_{\mu\nu} (x) \Psi (z)$, and require that the $\hat{h}$ be a four dimensional
mass eigenstate mode $\Yfund \hat{h}_{\mu\nu}=m^2 \hat{h}_{\mu\nu}$. Then the final Schr\"odinger type equation that 
we get for the KK modes is given by~\cite{RS2,CEHS,dWGFK,sugraRS}
\begin{equation}
-\partial_z^2\Psi +\left( \frac{9}{16} A'^2-\frac{3}{4}A''\right) \Psi=m^2 \Psi.
\end{equation}
Thus the Schr\"odinger potential is given by
\begin{equation}
V(z)=\frac{9}{16} A'^2-\frac{3}{4}A''.
\end{equation}
In our case $e^{-A(z)}=1/(1+k|z|)^2, A=2 \log (k|z|+1)$, and so the potential is
\begin{equation}
V(z)=\frac{15}{4} \frac{k^2}{(1+k|z|)^2}-\frac{3k \delta (z)}{1+k|z|}.
\end{equation}
This potential has the shape of a volcano, since there is a peak in it due to the first term, but then there is a
delta function which is like the crater of the volcano. This is why it is commonly referred to as the 
``volcano potential''. Note, that for any warp factor $A(z)$ one can always define the operators
$Q=\partial_z +\frac{3}{4} A'$ and $Q^\dagger =-\partial_z+\frac{3}{4} A'$, such that the equation is written
as $Q^\dagger Q \Psi =m^2 \Psi$ \cite{CEHS,sugraRS,dWGFK}. This means, that we are again facing a SUSY quantum 
mechanics problem,
which can be solved as in the case when discussing the localization of fermions. Therefore, as we know there is 
{\it always} a zero mode solution to this equation given by
\begin{equation}
\Psi_0= e^{-\frac{3}{4} A(z)}= \frac{1}{(1+k|z|)^{\frac{3}{4}}}.
\end{equation}
The existence of a zero mode is not very surprising, since we have not broken 4D Lorentz invariance, so 
we expect a massless 4D graviton to exist. The interesting question to ask is under what circumstances will
this graviton zero mode actually be normalizable. For this we have to evaluate the usual quantum mechanical
norm 
\begin{equation}
\int_0^{z_0} dz |\Psi_0|^2 =\int_0^{z_0} dz e^{-3/2 A(z)}=\int_0^{z_0} dz \frac{1}{(1+k|z|)^\frac{3}{2}}.
\end{equation}
If one is wondering whether this is really the relevant normalization, one can simply follow all the 
field and coordinate redefinitions backwards and find that the coefficient of the 4D kinetic term of the 
graviton zero mode is indeed proportional to this factor. Above $z_0$ is the size of the extra dimension
in the $z$ coordinates. 

The most important comment about the above integral is that it is converging even in the limit when 
the size of the extra dimension becomes infinitely large, $z_0\to \infty$. This usually does not happen in extra 
dimensional theories: when there is a compact extra dimension there is usually a graviton zero mode and KK tower.
As the size of the extra dimension gets infinitely large, the zero mode becomes more and more decoupled due to its
huge spread in the extra dimension, however the KK modes become lighter and lighter, and eventually form a continuum
that reproduces higher dimensional gravity. Here however one does not find that the zero mode would decouple in 
the infinitely large extra dimension limit, and the reason for that is that contrary to the flat extra dimensional
case, the zero mode of the graviton has a non-trivial wave function that is peaked around the positive tension
brane. This implies that in the RS model {\it gravity itself becomes localized} around the positive tension 
brane, and far away one only has a small tail for the graviton wave function. In the original $y$ coordinates the
wave function of the graviton zero mode is given by
\begin{equation}
ds^2=e^{-2k|y|} (\eta_{\mu\nu}+h_{\mu\nu})-dy^2.
\end{equation}
This confirms the ansatz that we have used in (\ref{gravitonansatz}) to calculate the effective 4D Planck scale, and also
sheds much more light on the nature of the solution to the hierarchy problem in the RS model. Since gravity is 
confined to the positive tension brane, observers living far away from the positive tension brane (for example
on the other end of the interval) will only feel the tail end of the graviton wave function, which is 
exponentially suppressed  compared to the Planck brane. On the other hand particle physics interactions will
have equal strength irrespectively of where one is along the extra dimension. So the solution of the hierarchy 
problem can be summarized by saying that gravity is weak compared to particle physics because we happen to live
at a point in extra dimension that is far away from where gravity is localized. As a comparison, in large extra 
dimensions the way that the hierarchy problem is resolved is by saying that gravity is much weaker than particle
physics because its fundamental strength does get diluted by fluxes spreading out into the large dimensions,
while since particle physics is localized on a brane it does not get diluted and we feel the fundamental strength
of interactions. 

\begin{figure}
\vspace*{1cm}
\centerline{\includegraphics[width=0.5\hsize]{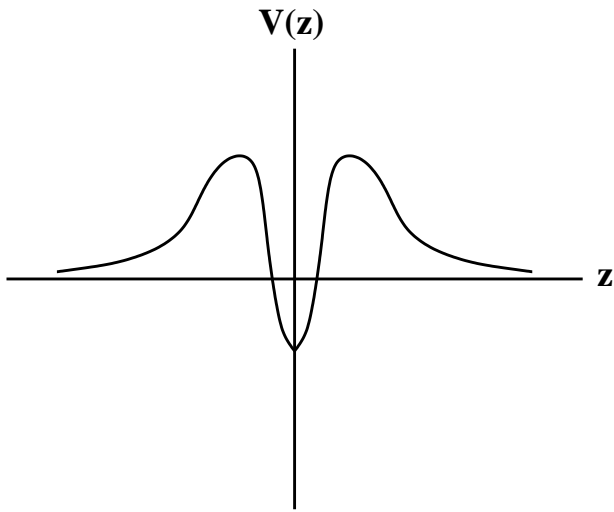}}
\caption{The shape of the volcano potential that localizes gravity in the RS model.}
\label{volcpot} 
\end{figure}

Getting back to the issue of the wave function of the graviton zero mode: since it is localized around the 
Planck brane, and remains normalizable even in the limit of infinite extra dimension, it opens the possibility
of recovering 4D gravitational interactions even when the size of the extra dimension is very large, 
since the zero mode remains localized around the Planck brane. However, this on its own is not sufficient to really
have a 4D gravity theory without compactification. As explained above in the infinite extra dimension limit there
will be a {\it continuum} of KK modes for the graviton, and usually these continuum modes are responsible for 
turning gravity into higher dimensional gravity. There will be a continuum of KK modes in the RS model as well,
however there is hope that these would not turn gravity over into a higher dimensional theory of gravity,
since these continuum modes also feel the volcano potential of Fig.~\ref{volcpot}, 
and this means that in order to get to the
TeV brane they have to tunnel through a large barrier, and thus their wave functions will be strongly suppressed
at the Planck brane~\cite{RS2}. 

One can easily estimate the suppression of the wave function by a the WKB approximation for the tunneling 
rate~\cite{CEHS}. 
Consider a continuum mode with mass $m$, then the semiclassical tunneling amplitude is
\begin{equation}
T(m)\sim e^{-2\int_{z_0(m)}^{z_1(m)} dz \sqrt{V(z)-m^2}},
\end{equation}
where $z_0(m)$ and $z_1(m)$ are the classical turning points for a particle with mass $m$ in the 
volcano potential. In order for the KK modes to not give a large contribution to the 4D gravitational potential 
one would need that $\lim_{m\to 0} T(m)=0$. The reason is that the potential between two test charges located at the 
Planck brane are given by the exchange of the zero mode and the KK modes:
\begin{equation}
U(r)\sim \frac{G_N M_1M_2}{r} +\frac{1}{M_*^3} \int_0^\infty dm \frac{M_1M_2 e^{-mr}}{r} \Psi_m^2 (0),
\end{equation}
where the first term comes from the exchange of the zero mode which would cause 4D gravitational interactions,
but the second term is due to the exchange of the KK modes. If for small $m\to 0$ we would have 
$\Psi_m(0)\to $ const. we would get unsuppressed $1/r^2$ corrections due to the integral around $m=0$. Thus we need
that $\Psi_m(0)\to 0$ for small $m$, and this can only be due to $T(m)\to 0$.  In the RS case 
$V(z)\sim 1/z^2$ for large $z$, and so as $m\to 0$ we get that 
\begin{equation}
T(0) \sim e^{-\int_0^\infty \frac{1}{z} dz}.
\end{equation}
The integral in the exponent diverges, so that we expect the KK modes to indeed decouple in the RS model and
to recover 4D gravity even in the infinitely large extra dimension limit. In order to really establish this rigorously,
a slightly more precise analysis is needed. Assume that the volcano potential for large $z$ behaves like
$\alpha (\alpha +1)/z^2$ (for the RS case $\alpha =3/2$). Then the asymptotic form of the Schr\"odinger equation
\begin{equation}
-\Psi_m''+\frac{\alpha (\alpha +1)}{z^2}\Psi_m =m^2 \Psi_m
\end{equation}
has solutions in terms of the Bessel functions
\begin{equation}
\Psi_m=a_m z^{\frac{1}{2}} Y_{\alpha +\frac{1}{2}}(m(z+1/k))+ b_m z^{\frac{1}{2}} J_{\alpha +\frac{1}{2}}(m(z+1/k))
\end{equation}
If one does a careful matching of the wave function inside and outside the potential we get that in the RS case
\begin{equation}
\Psi_m (0)\sim \left( \frac{m}{k}\right)^\frac{1}{2}.
\end{equation}
Therefore the correction to Newton's law is of the form
\begin{equation}
\int_0^\infty dm \frac{M_1M_2}{r} e^{-mr} \left(\frac{m}{k}\right).
\end{equation}
Since $\int_0^\infty e^{-mr}m\sim 1/r^2$ we get the final form of the corrected Newton potential to be
\begin{equation}
\label{Newton}
V(r)=G_N \frac{M_1M_2}{r} \left( 1+\frac{C}{(kr)^2}\right),
\end{equation}
where $C$ is a calculable constant of order one. This shows, that the full correction due to the KK modes is extremely
small for distances larger than the AdS curvature $1/k$, which in our case is of the order of the Planck length.
This means that for distances above the Planck length one would really get a 4D gravitational potential irrespectively
of the compactification. Thus in the presence of localized gravity there is no need to compactify~\cite{RS2}
 the extra dimension! Also, due to the wave function suppression of the KK modes the production of the continuum KK modes would
be suppressed on the Planck brane by $p^2/k^2$.

Besides reproducing the ordinary 4D Newton potential, one should however go somewhat further before claiming that 
4D gravity as we know it (that is 4D {\it Einstein} gravity, not just Newtonian gravity) is in fact reproduced on the 
brane. The reason is that as we mentioned several times before, there could be an extra massless scalar in the graviton.
That would also give a $1/r$ potential, so would also lead to Newtonian gravity, however the tensor structure
of the graviton propagator could be modified, which would lead to a so called scalar-tensor theory of gravity.
It is somewhat subtle to show that this does not indeed happen, and one needs to take so called brane bending 
effects into account for that. The analysis that shows that RS reproduces 4D Einstein gravity completely on the 
Planck brane has been performed in~\cite{GT,GKR}, see also \cite{RSGR}. 

\subsection{Intersecting branes, hierarchies with infinite extra dimensions}
We close this section by discussing two pressing issues. The first is what happens if we have more than a single 
extra dimensions: can we still localize gravity somehow? The second issue is whether if one has an infinite extra 
dimension, can one still solve the hierarchy problem (since now there is no negative tension brane)?

\subsubsection{Localization of gravity to brane intersections}
The easiest way to localize 4D gravity in more than one extra dimension is to consider the intersection
of co-dimension one branes (that is intersection of $d-1$ spatial dimensional branes in $d$ spatial dimensions).
The simplest possibility for intersecting branes is for two 4-branes in 6 space-time dimensions to intersect
orthogonally~\cite{ADDK}. 
To find the solution we should first think about what the RS solution with just the Planck brane
is (this situation is usually referred to as the RS2 model): it is nothing else but two copies of slices of AdS$_5$
spaces glued together at the Planck brane. This is what we need to do in the case of orthogonally intersecting
4-branes as well: we need to take four ``quarters'' (or quadrangles) of the AdS$_6$ space and glue them 
together at the branes (of course in the general case of intersecting $2+n$ branes one needs to glue 
$2^n$ copies). One quadrangle can just be given by the usual AdS$_6$ metric. An AdS$_6$ metric is exactly the same
form as the AdS$_5$ metric in the conformally flat coordinates:
\begin{equation}
ds^2=\frac{1}{(kz+1)^2} \left[ dx^\mu dx^\nu \eta_{\mu\nu} -dy^2 -dz^2\right],
\end{equation}
where we can bring this metric by the coordinate redefinition $z=(z_1+z_2)/\sqrt{2}$, $y=(z_1-z_2)/\sqrt{2}$
to
\begin{equation}
ds^2=\frac{1}{(k(z_1+z_2)+1)^2} \left[ dx^\mu dx^\nu \eta_{\mu\nu} -dz_1^2 -dz_2^2\right].
\end{equation}
A simple way of patching these solutions together at two four-branes is by the metric
\begin{equation}
ds^2=\frac{1}{(k(|z_1|+|z_2|)+1)^2} \left[ dx^\mu dx^\nu \eta_{\mu\nu} -dz_1^2 -dz_2^2\right].
\end{equation}
This metric will be the same AdS$_6$ metric in all four quadrangles, and continuous at the intersections. 
Generically for the intersection of $n$ branes in $4+n$ space-time dimensions the metric can be written as
\begin{equation}
ds^2=\Omega^2 (z) (dx^Mdx^N\eta_{MN}), \ \ \Omega (z)=\frac{1}{(k\sum_{i=1}^n|z_i|+1)}.
\end{equation}
To find out what the relation of the tensions of the intersecting branes to the bulk cosmological constant is 
we need to find the Einstein tensor for the above metric. One can again use the formula for a conformally flat 
metric that gives the Einstein tensor:
\begin{eqnarray}
G_{MN}=&& (n+2)\left[ \tilde{\nabla}_M \log \Omega \tilde{\nabla}_N \log \Omega
-\tilde{\nabla}_M \tilde{\nabla}_N \log \Omega +\tilde{g}_{MN}(\tilde{\nabla}_R \tilde{\nabla}^R \log \Omega +
\right. \nonumber \\
&&\left. \frac{n+1}{2} \tilde{\nabla}_R \log \Omega \tilde{\nabla}^R \log \Omega)\right],
\label{Einstpert2}
\end{eqnarray}
and we find that 
\begin{eqnarray}
G^M_N &=&\frac{n(n+2)(n+3)}{2} k^2 \delta^M_N 
-\frac{2(n+2)k}{\Omega} \delta (z_1) \left( \begin{array}{ccccccccc} 1&&&&&&&&\\ &1\\ &&1\\ &&&1\\ &&&&0\\&&&&&1\\
&&&&&&1\\ &&&&&&&\ddots \\ &&&&&&&&1 \end{array} \right)-\ldots \nonumber \\ 
&&-\frac{2(n+2)k}{\Omega} \delta (z_n) \left( 
\begin{array}{ccccccccc} 1&&&&&&&&\\ &1\\ &&1\\ &&&1\\ &&&&1\\&&&&&1\\
&&&&&&1\\ &&&&&&&\ddots \\ &&&&&&&&0 \end{array} \right).
\end{eqnarray}
This needs to be equated with the energy momentum tensor from the bulk cosmological constant, which exactly
matches the first term, and the brane tensions on the intersecting branes exactly have to match the individual 
subleading terms. From these we get matching relations analogous to that of the 5D RS model:
\begin{equation}
\frac{n(n+2)(n+3)}{2} k^2=\Lambda \kappa_{4+n}^2, \ \ 2(n+2) k =V\kappa_{4+n}^2,
\end{equation}
where $\Lambda$ is the $4+n$ dimensional bulk cosmological constant, while $V$ is the (common) brane tension.
One can again find the perturbed Einstein equation for the graviton, the result is given by:
\begin{equation}
\left[ -\frac{1}{2} \nabla_z^2 +\frac{n(n+2)(n+4)k^2}{8} \Omega^2-\frac{(n+2)k\Omega}{2} \sum_j \delta (z_j)
\right] \Psi =\frac{1}{2}m^2 \Psi.
\end{equation}
One can again show~\cite{ADDK}
that there is a single bound state zero mode solution to this equation, and its wave function
is proportional to $\Omega^\frac{n+2}{2}$. The zero mode is localized to the intersection of the 
branes. The relevant length scale for gravity is again the AdS curvature scale $L\sim k^{-1}$, where this is 
given by $k^2 \sim \Lambda \kappa_{4+n}^2 \sim \Lambda /M_*^{2+n}$, where $\kappa_{4+n}$ is the $4+n$ 
dimensional Newton constant, while $M_*$ as always is the $4+n$ dimensional Planck scale. Therefore
\begin{equation}
L\sim \left( \frac{M_*^{2+n}}{\Lambda}\right)^\frac{1}{2},
\end{equation}
and so we get a relation between the AdS length, the $4+n$ D Planck scale and the 4D Planck scale of the 
form~\cite{ADDK}
\begin{equation}
M_*^{2+n} L^n =M_{Pl}^2.
\label{AdSmatch}
\end{equation}
For scales much larger than $L$ gravity will behave like ordinary 4D gravity, while for distances smaller than that
scale gravity will behave as $4+n$ dimensional gravity. (\ref{AdSmatch}) implies that the scale $L$ is analogous
to the real radius of compact {\it flat} extra dimensions. Thus one can try to solve the hierarchy problem in the 
presence of infinite extra dimensions by lowering the fundamental Planck scale $M_*$ down to a TeV, which would
of course also imply that the brane tensions and the bulk cosmological constants would be themselves much smaller than
$M_*$. 

\begin{figure}
\vspace*{1cm}
\centerline{\includegraphics[width=0.5\hsize]{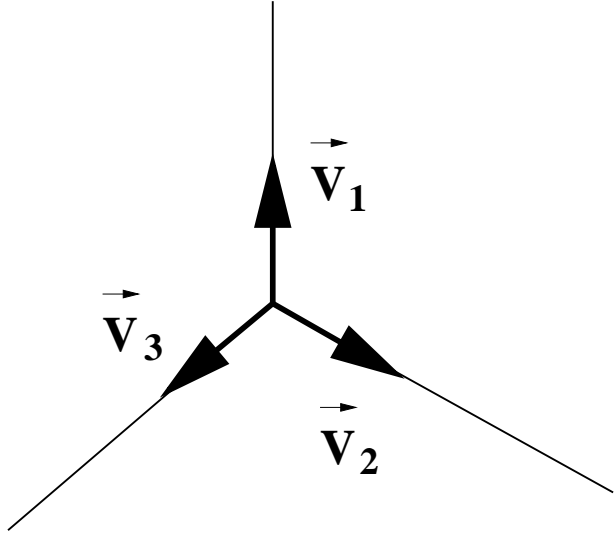}}
\caption{Intersecting branes in the RS model.}
\label{junction} 
\end{figure}

One can go further with the intersecting brane picture, and consider brane junctions~\cite{CS} -- 
that is semi-infinite 
co-dimension one branes ending at the same point, the brane junction (see Fig.~\ref{junction}). One can show,
that such static junctions may exist only, if the total force from the tensions acting on the junction vanishes:
\begin{equation}
\sum_i \vec{V}_i=0.
\end{equation}
In that case there will be a single remaining fine tuning relation between the bulk cosmological constant and the 
brane tensions, equivalent to the vanishing of the 4D cosmological constant. For more details see~\cite{CS}.

\subsubsection{The hierarchy problem in the infinite RS case}

Finally, one may ask whether a solution to the hierarchy problem really necessitates the compactification
as discussed at the beginning of this lecture. One possible way to avoid this we have seen above, where
we simply lower the fundamental Planck scale $M_*$ down to the TeV as in the large extra dimension scenarios. 
However, this would then raise the issue of why the AdS curvature $L$ is so large compared to its natural value.
However, one could just live in the infinite extra dimensional scenario on a brane (that has a very small tension)
away from the Planck brane~\cite{LR}. 
Due to the small curvature gravity would not be much affected, and due to the large 
warp factor one would still get that the particle physics scales on this brane would be warped down to an 
exponentially small scale. But would we still get 4D gravity on this brane? The corrections to Newton's law would 
be given by the form
\begin{equation}
V(r)=G_N \frac{M_1M_2}{r} \left(1+\int_0^\infty dm \frac{e^{-mr}}{r} \left( \frac{\Psi_m (z_0)}{\Psi_0 (z_0)}
\right)^2\right).
\end{equation}
The important point is that it is the {\it local} ratio of the KK amplitudes vs. the zero mode that will 
appear in this expression. One can show that this ratio is still scaling the same way as at the Planck brane
\begin{equation}
\frac{\Psi_m (z_0)}{\Psi_0 (z_0)}= \frac{\Psi_m (0)}{\Psi_0 (0)}\sim m^\frac{1}{2},
\end{equation}
for small values of $m$. Thus the corrections to Newton's potential will be similarly negligible far away 
from the Planck brane as on the Planck brane. Thus the conclusion is that one really does not need to 
compactify, one can recover 4D gravity {\it and} solve the hierarchy problem. In fact, one can show without too 
much calculation that this indeed needs to be the case: the zero mode contribution gives a potential of the form
$k/(M_*^3 r)$, while the KK modes at worst would give $1/(M_*^3 r^2)$. For relatively large values of $r$ the second
term is always suppressed compared to the first and gravity will have to look four dimensional.

\section{Phenomenology of Warped Extra Dimensions}
\label{sec:RS1}
\setcounter{equation}{0}
\setcounter{footnote}{0}

This lecture will cover some important topics about further developments in the RS model, and warped 
spaces in general.

\subsection{The graviton spectrum and coupling in RS1}
The first issue that we would like to discuss is what the phenomenology of the graviton KK modes would be
in the RS model with two branes, which is often also referred to as the RS1 model. We have already seen that the
graviton fluctuations obey the Schr\"odinger-type equation
\begin{equation}
-\partial_z^2\Psi +(\frac{9}{16} A'^2 -\frac{3}{4} A'')\Psi =m^2 \Psi,
\end{equation}
where $h_{\mu\nu}=e^{3A/4} \hat{h}_{\mu\nu}(x) \Psi (z)$, where the $\hat{h}_{\mu\nu}(x)$ are the 4D modes
that satisfy $\Yfund \hat{h}_{\mu\nu} =m^2\hat{h}_{\mu\nu}$. For the RS background
$A=2\log (k|z|+1)$. Until now we have not discussed much what the BC of these equations should be, since we were 
imagining them in the infinite RS2 case. For the case with finite branes we need to find the appropriate set of
BC's, which can be read off from the fact that we  imposed an orbifold projection $y\to -y$, under which
the graviton was symmetric. Therefore, in $y$ coordinates $\partial_y h_{\mu\nu} =0$ at the two fixed 
points, from which it follows that also $\partial_z h_{\mu\nu} =0$. Since the graviton can be recovered from
the Schr\"odinger frame wave function $\Psi (z)$ as $h_{\mu\nu}=e^{3A/4} \hat{h}_{\mu\nu}(x) \Psi (z)$, and
$A=2\log (k|z|+1)$ we can translate the BC on $h_{\mu\nu}$ to a BC on $\Psi$:
\begin{eqnarray}
&& \partial_z \Psi =-\frac{3}{2} k \Psi|_{z=0} \nonumber \\
&& \partial_z \Psi =-\frac{3}{2} \frac{k}{k|z_1|+1} \Psi|_{z=z_1},
\end{eqnarray}
where $z_1$ is the location of the second fixed point (with negative tension) in the $z$ coordinates,
$z_1=e^{k b}/k$. The equation between the two boundaries for the RS case is
\begin{equation}
-\partial_z^2 \Psi+\frac{15}{4} \frac{k^2}{(kz+1)^2}\Psi=m^2 \Psi.
\end{equation}
Generically, the solution of this equation is given in terms of Bessel functions:
\begin{equation}
\Psi(z)=a_m (kz+1)^\frac{1}{2} Y_2(m(z+1/k))+b_m (kz+1)^\frac{1}{2} J_2(m(z+1/k)).
\end{equation}
The two boundary conditions will completely determine the masses of the KK tower $m$'s, where the $n$th solution
is given by
\begin{equation}
m^{(n)}=k x_n e^{-kb} ={\cal O} (TeV), \ \ J_1(x_n)=0,
\end{equation}
where $b$ is the distance between the two branes. 
Thus the spacing of the graviton zero modes is of order TeV, and {\it not} of order $M_{Pl}$ as one would have
naively suspected from the fact that the proper length of the extra dimension is of order $1/M_{Pl}$. 
The fact that the KK modes are spaced by TeV suggests that it is better to think of this model with extra dimension
of size $1/$TeV, that is a relatively large extra dimension, rather than a really small one. The answer to any 
physical observable will convey this impression. For example, if one were to ask how long it takes for light to 
bounce back and fourth between the two branes, the time we get will be of order $1/$TeV due to the warping,
much longer than the naively expected $1/M_{Pl}$. 

Let us now ask how one would detect these KK graviton resonances if one were to live on the TeV brane (the negative
tension brane, where the hierarchy problem is resolved). Ordinarily one would think that these states will 
be completely unobservable, since they are relatively heavy (of order TeV), and only gravitationally 
coupled. However, our naive intuition fails us again! The point is that even though these KK modes are indeed 
gravitationally coupled, since they are repelled from the Planck brane by the tunneling through the volcano
potential, their wave function must be exponentially peaked on the TeV brane. Therefore their couplings
will also be exponentially {\it enhanced} to matter on the TeV brane (compared to gravitational strength coupling).
Thus the picture that emerges is the following: on the TeV brane one has the graviton zero mode, 
that is peaked on the other brane therefore its coupling to TeV brane matter is weak (this is what sets the 
scale of gravitational coupling), and one also has the KK modes of the graviton whose masses are of order
TeV, but their couplings are enhanced to $1/$TeV rather than $1/M_{Pl}$ on the TeV brane:
\begin{equation}
{\cal L}_{TeV}=-\frac{1}{M_{Pl}} T^{\alpha \beta}h_{\alpha\beta}^{(0)}-\frac{1}{M_{Pl}e^{-kb}}  T^{\alpha \beta}
\sum_{n=1}^{\infty} h_{\alpha\beta}^{(n)}.
\end{equation}
The detailed coefficients in the equation above have been worked out in \cite{DHR1}, but by now none of the
features of this coupling are surprising: the graviton modes couple to the stress-energy tensor of matter
on the TeV brane, and the scales of coupling come from the consideration of the wave functions of the 
zero and KK modes as discussed above.

\begin{figure}
\vspace*{1cm}
\centerline{\psfig{figure=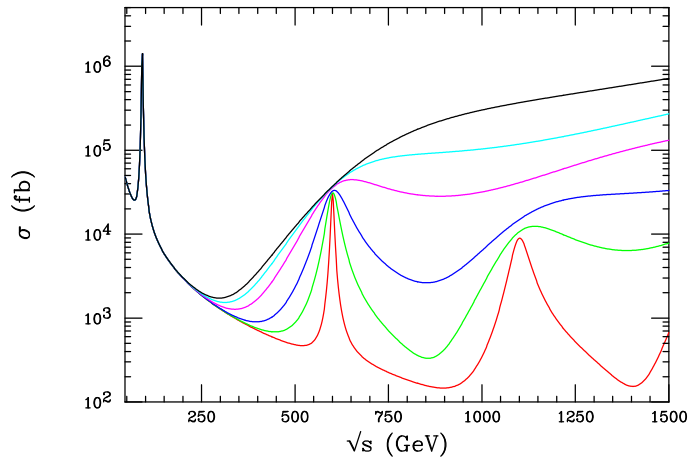,height=6.cm,width=9cm,angle=90}}
\caption{Scattering cross section for $e^+e^-\to \mu^+\mu^-$ at a linear collider in the presence of the tower
of  Randall-Sundrum KK gravitons, as calculated and plotted by Davoudiasl, Hewett and Rizzo~\cite{DHR1}.
The mass of the first KK resonance is fixed to be 600 GeV, and the various curves correspond to different 
values of $k/M_{Pl}=1,0.7,0.5,0.3,0.2,0.1$ from top to bottom. For details and LHC processes see the 
original paper~\cite{DHR1}.
\label{DHRfig}} 
\end{figure}

This implies that in TeV scale colliders such as the LHC will be (and already at the Tevatron) one could study
these KK modes {\it as individual resonances} in the scattering cross section. Some of these cross-sections have been
worked out in detail by Davoudiasl, Hewett and Rizzo in~\cite{DHR1}. Some sample cross sections in the presence
of the KK resonances for the LHC and a linear collider calculated in~\cite{DHR1}
are given in Fig.~\ref{DHRfig}.  Note, that the phenomenology
of the RS model is {\it very different} from that of large extra dimensions. In the case of large extra dimensions
the spacing of the KK modes is very small, in the extreme case of 2 large extra dimensions of order
$10^{-3}$ eV. This is tiny enough to allow energetics to produce it in all colliders, however since the 
couplings of the individual modes is extremely small, of gravitational strength $1/M_{Pl}$ one can 
certainly not see the individual modes. Their {\it collective} effects will be visible as a raise in cross sections
at the LHC or the Tevatron. However, for RS as we discussed the individual KK modes are much heavier and much 
more strongly coupled, leading to well-defined resonances in high energy scatterings. 

\subsection{Radius stabilization\label{radionstabilize}}

Until now we have treated the radius of the RS1 model as a fixed given constant, and found that radius $r$ 
(which we have denoted $b$ until now) has to be $r\sim 30/k$ in order for the hierarchy problem to be resolved
(since $M_{weak}\sim 10^{-16} M_{Pl}$, from which $rk\sim \log 10^{16}\sim 30$).
This raises several important questions, that need to be addressed:
\begin{itemize}
\item Since the radius is not {\it dynamically} fixed at the moment, but rather just set to its 
desirable value, there will be a corresponding {\it massless} scalar field in the effective theory,
which corresponds to the fluctuations of the radius of the extra dimensions, called the 
{\it radion}~\cite{GW1,CGRT,GW2,CGR}.
One can understand the masslessness of this field by realizing, that the RS solution discussed until now
did not make any reference to the size of the extra dimension, it was a solution for arbitrary values. This means
that in the effective theory this parameter is also arbitrary, and thus has no potential, and so is a flat direction.
Thus it can have no mass. This massless
radion would contribute to Newton's law and result in violations of the equivalence principle (would cause
a fifth force), which is phenomenologically unacceptable, and therefore it does need to obtain a mass -- the radius
{\it has} to be stabilized.

\item The radius has to be stabilized at values somewhat larger than its natural value (we need $kr\sim 30$,
while one would expect $r\sim 1/k$). Does this
reintroduce the hierarchy problem?

\item We have seen that one needed two fine tunings to obtain the static RS solution, one of which was 
equivalent to the vanishing of the 4D cosmological constant, and is thus expected. But can we shed light on 
what the nature of the second fine tuning is and whether we can eliminate it somehow?
\end{itemize}

A mechanism for radius stabilization will address all the above mentioned issues. The simplest and most elegant 
solution for stabilization of the size of the extra dimension was proposed by Goldberger and Wise~\cite{GW1}, and is 
known as the Goldberger-Wise (GW) mechanism. Here we will discuss the details of this mechanism and its effect on 
the radion mass and radion physics, however before plunging into the details and the formalism let us first 
summarize the main idea behind the GW mechanism. Radius stabilization at non-trivial values of the radius
usually occurs dynamically, if there are different forces, some of which would like to drive the extra 
dimension very large, and some very small. Then there is a hope that these forces may balance each other at some 
value and a stable non-trivial minimum for the radius could be found. A possible way to find such a tension between
large and small radii is if there is a tension between a kinetic and a potential term, one which would want
derivatives to be small (and thus large radii) and the other which would want small radii to minimize the potential.
The proposal of Goldberger and Wise was to use exactly this scenario. Introduce a bulk scalar field into the RS
model, and add a bulk mass term to this field. This will result in a non-trivial potential for the radius,
since due to the bulk mass the radius wants to be as small as possible. However, if we somehow achieve that there is 
also a non-trivial profile (a VEV that is changing with the extra dimensional coordinate) for this scalar, then
the kinetic term would want the radius to be as large as possible, so as to minimize the kinetic energy in the 
5th direction. Then there would be a non-trivial minimum. To achieve the non-trivial profile for this scalar
field GW suggested to add {\it brane potentials} for this scalar on both fixed points, which have minima at different
values from each other. Then one is guaranteed the non-trivial profile, and thus radius stabilization as well. 
The main question is not whether this mechanism works, it is intuitively clear that it should, but rather whether
one can naturally achieve the somewhat large radius using this mechanism without fine tuning, and if yes what will
be the characteristic mass of the radion.

To describe the GW mechanism in some detail one needs to set up some formalism for theories with extra scalar fields
in the bulk. The reason is that as we will see the {\it back-reaction} of the metric to the presence of the scalar
field in the bulk will be important, therefore it would be very nice to simultaneously solve the Einstein and the
bulk scalar equations, to have the back-reaction exactly under control. One can get away without such an exact 
solution either by just calculating a 4D effective potential for the radion~\cite{GW1,CGRT},
or by calculating the back-reaction order-by-order. However, with some formalism we will be able to solve
the coupled equations exactly for certain scalar potentials, and therefore we will discuss the issue of radius 
stabilization using this approach~\cite{dWGFK,CEHS,CEGH,CGK}. 
Denote the scalar field in the bulk by $\Phi$, and consider the action
\begin{equation}
\int d^5x\sqrt{g} \left[-M_*^3 R++\frac{1}{2}(\nabla\Phi)^2 -V(\Phi )\right]-
\int d^4x \sqrt{g_4} \lambda_P (\Phi )-
\int d^4x \sqrt{g_4} \lambda_T (\Phi ),
\end{equation}
where the first term is the usual 5D Einstein-Hilbert action and the bulk action for the scalar field,
while the next two terms are the brane induced potentials for the scalar field on the Planck (P) and on the 
TeV (T) branes. We will denote the 5D Newton constant as always by $\kappa^2=1/2M_*^3$, and look for an ansatz
of the background metric again of the generic form as in the RS case to maintain 4D Lorentz invariance:
\begin{equation}
ds^2 =e^{-2 A(y)} \eta_{\mu\nu} dx^\mu dx^\nu -dy^2. 
\end{equation}
The Einstein equations are {\it exactly} as we have derived for the RS case, except here we have an energy-momentum
tensor that is derived from the action of the scalar field. So everyone should be able to derive the following 
equations on their own using the tools presented until now:
\begin{eqnarray}
\label{bulkEinst}
&& 4 A'^2-A''=-\frac{2\kappa^2}{3} V(\Phi_0) -\frac{\kappa^2}{3} \sum_{i=P,T} \lambda_i (\Phi_0) \delta (y-y_i)
\nonumber \\
&& A'^2=\frac{\kappa^2}{12} \Phi_0'^2-\frac{\kappa^2}{6} V(\Phi_0).
\end{eqnarray}
The first equation is the $i,j$ component of the Einstein equations, while the second one is the 
$5,5$ component. $\Phi_0$ denotes the solution of the scalar field, which (by the requirement of Lorentz invariance)
is assumed to be only a function of $y$: $\Phi = \Phi_0 (y)$. In addition to these two equations that we are already 
used there will be another one determining the shape of the scalar field in the bulk, and it is simply the 
bulk scalar equation of motion in the warped space, derived from the generic scalar equation
\begin{equation}
\partial_\mu \sqrt{g} g^{\mu\nu} \partial_\nu \Phi =\frac{\partial V}{\partial\Phi}\sqrt{g}.
\end{equation}
By the substituting the scalar and metric ansatz into this equation we get
\begin{equation}
\Phi_0''-4A'\Phi_0'=\frac{\partial V}{\partial\Phi_0}+\sum_i\frac{\partial\lambda_i (\Phi_0)}{\partial\Phi}
\delta (y-y_i).
\label{bulkscalar}
\end{equation}
We can separate these equations into the bulk equations that do not contain the delta functions, and the boundary 
conditions which will be obtained by matching the coefficients of the delta functions at the fixed points. For example,
the first Einstein equation contains the explicit delta function term $-\frac{\kappa^2}{3} \sum_{i=P,T} \lambda_i 
(\Phi_0) \delta (y-y_i)$ at the branes which naively does not seem to be balanced by anything. However, there is 
no requirement for the derivative of the metric to be continuous (the metric itself is a physical quantity which 
should be continuous), and so there could be a jump in the derivative $A'$ at the branes, which would 
imply that $A''$ also contains a term proportional to a delta function. If the derivative $A'$ jumps from 
$A'(0-\epsilon)$ to $A'(0+\epsilon)$, this implies that locally $A'$ contains a term of the form
$(A'(0+\epsilon)-A'(0-\epsilon))\theta (y)$, where $\theta (y)$ is the step function. Therefore, $A''$ will
contain the term $(A'(0+\epsilon)-A'(0-\epsilon))\delta (y)$, thus the delta function is proportional
to the {\it jump} of the derivative of A, sometimes denoted by $[A']_0\equiv A'(0+\epsilon)-A'(0-\epsilon)$.
Therefore, the boundary conditions derived this way are sometimes also called the {\it jump equations},
which in our case will be given by
\begin{eqnarray}
&& [A']_i=\frac{\kappa^2}{3}\lambda_i(\Phi_0), \nonumber \\
&& [\Phi_0']_i=\frac{\partial\lambda_i (\Phi_0)}{\partial\Phi}.
\end{eqnarray}
The bulk equations (\ref{bulkEinst}-\ref{bulkscalar}) together with these boundary conditions form the 
equations of the coupled gravity-scalar system. These are coupled second order differential equations, which are
quite hard to solve. However, for specific potentials it is possible to simplify the solution. Let us assume
that we have found a solution to the system of equations above given by
$A(y), \Phi_0(y)$. We can {\it define} the function $W(\Phi )$ via the equations
\begin{eqnarray}
&&A'\equiv \frac{\kappa^2}{6} W (\Phi_0), \nonumber \\
&&\Phi_0'\equiv \frac{1}{2} \frac{\partial W}{\partial \Phi}.
\end{eqnarray}
If we plug in these expressions for $A'$ and $\Phi_0'$ into the Einstein and scalar equations 
we will find that {\it all} equations are satisfied, if the following consistency condition holds:
\begin{equation}
\label{Weq}
V (\Phi )=\frac{1}{8} \left( \frac{\partial W}{\partial \Phi}\right)^2-\frac{\kappa^2}{6} W(\Phi )^2,
\end{equation}
and the jump conditions are given by
\begin{eqnarray}
\label{jump}
&& \frac{1}{2}[W(\Phi_0)]_i=\lambda_i(\Phi_0), \nonumber \\
&& \frac{1}{2}[\frac{\partial W}{\partial\Phi}]_i=\frac{\partial\lambda_i(\Phi_0)}{\partial\Phi}.
\end{eqnarray} 
If $W$ were given we would reduce the equations from coupled second order differential equations to 
ordinary first order equations that are quite easy to solve. The price to pay is that it is very hard to find
$W$ for a given $V(\Phi )$: one needs to solve a non-linear second order differential equation (\ref{Weq})
to find the ``superpotential'' $W$. However, if our goal is not to solve the equations for a very 
particular bulk potential $V(\Phi )$, but rather a bulk potential which has some properties, we can simply
start with a superpotential $W$ that will produce a $V$ with the required properties, and we will be able to easily
get the full solution of the equations. In our case we would like the bulk potential to include a cosmological
constant term (independent of $\Phi$) and a mass term (quadratic in $\Phi$), but we will not care if there are
some other terms as well if those make solving the equations simpler. So we choose~\cite{dWGFK}
\begin{equation}
W(\Phi )=\frac{6k}{\kappa^2}-u\Phi^2.
\end{equation}
The first term is just what one needs for a cosmological constant, while the second term will provide the mass term
when taking the derivative. Let us first discuss the jump conditions. Since the metric is an even function
of $y$, if we try to solve the equations on the $S^1/Z_2$ orbifold again as in the RS case $A(y)$ must also be an even
function, and therefore $A'$ is odd. However, one the equations we have above is that $A'\sim W$, which means that 
$W$ must be an odd function at the branes, it must change signs. To satisfy the jump conditions (\ref{jump})
I must choose the brane potentials to be of the form
\begin{equation}
\lambda (\Phi )_\pm=\pm W(\Phi_\pm)\pm W'(\Phi_\pm) (\Phi-\Phi_\pm)+
\gamma_\pm (\Phi-\Phi_\pm)^2,
\end{equation}
where $\Phi_\pm$ are the values of the scalar field at the two branes, which we will also denote 
by $\Phi_+=\Phi_P$ at the Planck brane, and $\Phi_-=\Phi_T$ at the TeV brane. Then the solution will be given by
the solution of the equation 
\begin{equation}
\Phi'=\frac{1}{2} \frac{\partial W}{\partial \Phi}=-u \Phi,
\end{equation}
which is simply
\begin{equation}
\Phi_0 (y) =\Phi_P e^{-uy}.
\end{equation}
From this the value of the scalar field at the TeV brane is determined to be
\begin{equation}
\Phi_T =\Phi_P e^{-vr}.
\end{equation}
This means that the radius is no longer arbitrary, but given by
\begin{equation}
r=\frac{1}{u} \ln \frac{\Phi_P}{\Phi_T}.
\end{equation}
The value of the radius is determined by the equations of motion, which is exactly what we were after. This is the 
GW mechanism. The metric background will then be obtained from  the equation
\begin{equation}
A'=\frac{\kappa^2}{6}W(\Phi_0)= k-\frac{u\kappa^2}{6} \Phi_P^2 e^{-2uy}
\end{equation}
given by the solution
\begin{equation}
A(y)=k y +\frac{\kappa^2\Phi_P^2}{12}e^{-2uy}.
\end{equation}
The first term is the usual RS warp factor (remember that $A$ has to be exponentiated to obtain the metric),
while the second term is the back-reaction of the metric to the non-vanishing scalar field in the bulk.
We will assume that the back-reaction is small, and thus that $\kappa^2\Phi_P^2, \kappa^2\Phi_T^2 \ll 1$, and that
$v>0$. The values of $\Phi_P$ and $\Phi_T$ are determined by the bulk and brane potentials, so $\Phi_P/\Phi_T$ is a fixed
value. Since we want to generate the right hierarchy between the Planck and weak scales we need to ensure that
\begin{equation}
kr\sim 30,
\end{equation}
from which we get that 
\begin{equation}
\frac{k}{u} \ln \left( \frac{\Phi_P}{\Phi_T}\right) \sim 30,
\end{equation}
which implies that $u/k$ is somewhat small (but not exponentially small). This is the ratio that will set the hierarchy
in the RS model, and we can see that indeed one can generate this hierarchy using the GW stabilization mechanism by a very 
modest (${\cal O}(50)$) tuning of the input parameters of the theory. 

Once we have established the mechanism for radius stabilization, we know that the radion is no longer
massless. The next obvious question then is what will be the value of the radion mass: is it naturally sufficiently
large to avoid the problems mentioned at the beginning of this section or not? We will answer this question by 
finding what the radion mode is in the GW stabilized RS model, and explicitly find its mass~\cite{CGK,TM},
see also~\cite{CGRT,GW2,KKOP}. For this, we need to find the 
scalar fluctuations of the coupled gravity-scalar system. This can be parametrized in the following way:
\begin{eqnarray}
&& ds^2=e^{-2A-2F(x,y)}\eta_{\mu\nu}dx^\mu dx^\nu-(1+G(x,y))^2 dy^2, \nonumber \\
&&\Phi (x,y)=\Phi_0(y)+\varphi (x,y).
\end{eqnarray}
At this moment it looks like there would be three different scalar fluctuations, $F,G$ and $\varphi$. However,
if we plug this ansatz into the Einstein equation the 4D off-diagonal $\mu\nu$ components are satisfied only if
\begin{equation}
G=2F,
\end{equation} 
while the $\mu 5$ components imply the following further relation among the fluctuations:
\begin{equation}
\varphi = \frac{1}{\Phi_0'} \frac{3}{\kappa^2}(F'-2A'F).
\end{equation}
This means, that in the end there is just a single independent scalar fluctuation in the coupled equation,
which we can choose to be $F$. Requiring the above two relations we find that the rest of the Einstein equations are
{\it all} satisfied if the following equation holds:
\begin{equation}
F''-2A'F'-4A''F-2\frac{\Phi_0''}{\Phi_0'}F'+4A'\frac{\Phi_0''}{\Phi_0'}F=e^{2A}\Box F
\label{radeq}
\end{equation}
in the bulk and the following boundary condition:
\begin{equation}
(F'-2A'F)_i=0.
\end{equation}
Let us first assume that there is no stabilization mechanism, that is the background is {\it exactly} the RS 
background given by $A=k|y|$, and $\Phi_0=0$. In this case most of the terms in the above equation disappear, and 
we are left with 
\begin{equation}
F'-2kF= e^{2ky} m^2 F, \ \ (F'-2kF)_i=0.
\end{equation}
We can see that the only solution is for $m^2=0$, and the wave function of the un-stabilized radion will be 
given by
\begin{equation}
F(y)=e^{2k|y|}.
\end{equation}
This wave-function has been first found by Charmousis, Gregory and Rubakov, and thus the metric corresponding to 
radion fluctuations in the RS model corresponds to
\begin{equation}
ds^2=e^{-2k|y|-2e^{k|y|}f(x)}\eta_{\mu\nu} dx^\mu dx^\nu-(1+2e^{2k|y|}f(x))dy^2.
\end{equation}
This is a single scalar mode, that is exponentially 
peaked at the TeV brane, just like all the graviton KK modes. Since it is peaked on the 
TeV brane, it means that its coupling on the Planck brane will be strongly suppressed, while it will be of its natural 
size (suppressed by 1/TeV) on the TeV brane. The exponential peaking on the TeV brane also implies, that if we move the 
TeV brane to infinity (that is consider the RS2 model) then the radion will no longer be a normalizable mode and completely
decouples from the theory. This implies that in that case one is really recovering 4D gravity with just the tensor 
couplings on the Planck brane in RS2. 

Generically, for the RS1 case we now have a complete set of modes (the radion mode is the one that is missing if we are 
imposing the RS gauge choice): without stabilization
at the zero mode level there is a graviton and a radion mode (which does not have a KK tower), while at the massive
level there are the graviton KK modes only. In the presence of the GW stabilization the scalar mode {\it will} acquire a 
KK tower: the lowest mass mode of this tower will behave like the radion, while the higher mass modes will be identified with
what used to be the KK tower of the GW bulk scalar. 

To find the radion mass for the case with stabilization, we simply need to plug into (\ref{radeq}) the full background
for $A$ and $\Phi_0$ with stabilization:
\begin{equation}
F''-2A'F'-4A''F+2uF'-4uA'F+m^2e^{2A}F=0.
\end{equation}
We can see that if the back-reaction of the metric was neglected ($A''=0$) we would still not get a mass for the radion 
zero mode. Thus to find the leading term for the radion mass we expand in terms of the back-reaction of the metric
in the parameter $l=\kappa \Phi_P/\sqrt{2}$, and obtain the mass of the radion
\begin{equation}
m_{radion}^2=\frac{4l^2(2k+u)u^2}{3k} e^{-2(u+k)r}.
\end{equation}
As a reminder, the mass of the graviton KK modes is of the order
\begin{equation}
m_{graviton}\sim k e^{-kr},
\end{equation}
and so the radion/graviton mass ratio is
\begin{equation}
\frac{m_{radion}^2}{m_{graviton}^2} \sim l^2 \frac{u^2}{k^2} e^{-2ur}.
\end{equation}
Thus we can see that the radion mass is smaller than the graviton mass, and the radion would be likely the lightest
new particle in the RS1 model. The other scalar KK modes that originate mostly from the GW bulk scalar 
will have masses of the same order as the graviton. 

In order to find the coupling of the radion to SM matter fields on the TeV brane we need to find the canonically 
normalized radion field. The wave function (not yet normalized) is of the form
\begin{equation}
F(x,y)=e^{2k|y|}(1+{\rm corrections}) R(x),
\end{equation}
where the correction terms above are suppressed by the back-reaction, and since these are not the leading terms we can
neglect them. $R(x)$ is the 4D radion field. The induced metric will be 
\begin{equation}
g_{\mu\nu}^{ind}=e^{-2A}\eta_{\mu\nu}(1-2F(x,y)).
\end{equation}
The coupling to matter on the TeV brane will be as usually given by $T_{\mu\nu}g^{\mu\nu}$, which will result in a 
coupling of the form $T^\mu_\mu R(x)$. The question is what will be the coefficient in front of this coupling be?
Once we go to the canonically normalized radion field $r(x)$ we find (after calculating the normalization of the kinetic term 
for the radion) that the radion coupling is given by
\begin{equation}
\frac{1}{\sqrt{6} M_{Pl}e^{-kr}} r(x) {\rm Tr}\, T.
\end{equation}
Thus the coupling is suppressed by the TeV scale just like for the graviton KK modes, except the masses are somewhat smaller.
One can also write this coupling in the form
\begin{equation}
\frac{v}{\sqrt{6} \Lambda} \frac{r(x)}{v} {\rm Tr}\, T \equiv \gamma \frac{r(x)}{v} {\rm Tr}\, T,
\end{equation}
where we have denoted $\Lambda = M_{Pl}e^{-kr}$ as the ``TeV scale'' that one obtains from the RS hierarchy,
$\gamma = v/\sqrt{6} \Lambda$, and $v$ is the usual Higgs VEV $v=246$ GeV. This coupling is exactly equivalent to the 
coupling of the SM Higgs boson. The reason is that both the radion and the Higgs will couple to the mass terms in the 
Lagrangian, the Higgs via $H/v$, the radion via $r/v$, the only difference is the extra suppression factor $\gamma$ in the
strength of the radion coupling. Thus the phenomenology of the radion would be very similar to that of the SM Higgs 
boson. 

There is one more important possibility for the physics of the radion. There could be a brane induced operator on the 
TeV brane mixing the radion and the SM Higgs field $H$~\cite{GRW2}:
\begin{equation}
\int d^4 x \xi H^\dagger H R(g^{ind}) \sqrt{g^{ind}}.
\end{equation}
Using our master-formula (\ref{conformal}) for calculating curvature tensors for conformally flat metrics we get 
\begin{equation}
R(\Omega^2(r)\eta_{\mu\nu})=-6\Omega^{-2} (\Box\ln\Omega+(\nabla\ln\Omega )^2).
\end{equation}
The interaction term will then be given by 
\begin{equation}
{\cal L}_\xi = -6\xi \Omega^2 (\Box\ln\Omega+(\nabla\ln\Omega )^2) H^\dagger H.
\end{equation}
Expanding both the Higgs $H$ and the warp factor $\Omega$ in the uneaten physical Higgs and the radion respectively
\begin{equation}
H=\left( \begin{array}{c} 0 \\ \frac{v+h}{\sqrt{2}} \end{array}\right), \ \ \Omega (r)=1-\gamma \frac{r}{v}+\ldots
\end{equation}
we get 
\begin{equation}
{\cal L}_\xi = 6\xi \gamma h\Box r+3\xi\gamma^2 (\partial r)^2.
\end{equation}
Thus there will be kinetic mixing induced between the Higgs and the radion fields. One needs to diagonalize the full 
quadratic Lagrangian
\begin{equation}
{\cal L}=-\frac{1}{2} h\Box h-\frac{1}{2} m_h^2 h^2 -\frac{1}{2} (1+6\xi\gamma^2) r\Box r -\frac{1}{2} m_r^2 r^2
+6\xi\gamma h\Box r.
\end{equation}
After the diagonalization, we find that not only is the coupling of the radions affected, but the couplings of the 
SM Higgs will also be modified from their SM expressions, and thus the radion-Higgs mixing could significantly affect
ordinary Higgs physics as well~\cite{CGK,Gunion}.

\subsection{Localization of scalars and quasi-localization}
We have seen in previous sections that the zero mode graviton is strongly peaked at the Planck brane, that is 
gravity is localized by the RS background.  An important question to discuss is whether fields with spins other than 2 would 
also be localized or not. In the following we will discuss two examples of that: a bulk scalar field and a bulk gauge field.
We will not discuss the issue of bulk fermions, the reader is referred to 
references~\cite{GN,RSsusy,BG,HS1,nomurasmith,CGHST} for that case.

We will start by (again) considering a bulk scalar field (see for example~\cite{GW0,BG}. 
We have already discussed the equation of motion for the bulk
scalar, in the absence of a bulk potential it is given by
\begin{equation}
\partial_M (\sqrt{g}g^{MN}\partial_N \Phi )=0.
\end{equation}
If we are interested in whether or not a zero mode exists, we simply have to assume that the 4D derivative on $\Phi$ vanishes,
so the equation that a zero mode has to satisfy is
\begin{equation}
\partial_y e^{-4ky} \partial_y \Phi =0.
\end{equation}
The solution to this is simply $\Phi =const.=\Phi_0$. However, this is not enough. We need to figure out whether or not
this mode is normalizable in the limit when the extra dimension becomes infinitely large (that is we want to find out if 
the effective 4D kinetic term remains normalizable or not). We write the zero mode as $\Phi =\Phi_0 \varphi (x)$,
where $\varphi $ is the 4D wave function for a massless 4D mode $\Box_4 \varphi =0$. The normalization of the kinetic term
then comes from
\begin{equation}
\int \sqrt{g}\partial_\mu\Phi\partial_\nu\Phi g^{\mu\nu} dy d^4x = \int e^{-4ky}e^{2ky}\Phi_0^2 dy \int \partial_\mu\varphi
\partial_\nu\varphi \eta^{\mu\nu} d^4x.
\end{equation}
We can see that the $y$-integral converges even when the extra dimension is infinitely large, so this means that the 
scalar field is localized in the RS background, even though in $y$ coordinates its wave function is constant (if we went 
to the $z$ coordinates we would of course find a wave function peaked at the Planck brane). However, this analysis was done
in the absence of a mass term (or bulk potential in general), which is quite unnatural for a scalar field. Let us ask
how the solutions will be modified if there is a bulk potential (and still keep the extra dimension infinitely large).
The equation of motion will now be
\begin{equation}
\partial_M (\sqrt{g}g^{MN}\partial_N \Phi )=\sqrt{g}\frac{\partial V}{\partial \Phi}.
\end{equation}
Clearly, there will be no more zero mode solution. In order to find out what happened to the zero mode,
we need to consider the full continuum spectrum~\cite{DRT}. The scalar equation for a 4D mode with 4D momentum $p$ in the presence
of a bulk mass term $m^2\Phi^2$ in the bulk potential will be:
\begin{equation}
\partial_y^2\Phi -4k\Phi'+p^2 e^{2ky} \Phi -m^2\Phi =0.
\end{equation}
As we mentioned before, $p^2=0$ is {\it not} a solution to this equation, there is no zero mode. Naively we would 
have thought that the zero mode just picks up a mass from the bulk mass term and will sit at $p^2=m^2$, 
but $p^2=m^2$ is not a solution to the bulk equations either. So now we might be really puzzled what happened to our zero mode.
The continuum spectrum can be determined at large $y$, where the extra $m^2$ term from the bulk mass is unimportant. 
Thus the continuum spectrum will be unchanged compared to the $m^2=0$ spectrum, except the wave functions will be distorted 
for small $y$. The general solution to the equation above will be in terms of Hankel functions,
\begin{equation}
\Phi (y)={\rm const} \cdot e^{2ky} H_\nu^{(1)} (\frac{p}{k} e^{ky}), \ \ \nu=\sqrt{4+\frac{m^2}{k^2}}.
\end{equation}
Since we have not included a source on the Planck brane the $Z_2$ symmetry will imply that the BC (jump equation)
at the Planck brane will just be $\partial_y \Phi |_{y=0}=0$.
This implies that 
\begin{equation}
\frac{p H_{\nu-1}^{(1)} (\frac{p}{k})}{k H_{\nu}^{(1)} (\frac{p}{k})}+2-\nu =0.
\end{equation}
The solution to this equation for small values of the bulk mass $m$ is 
\begin{equation}
p=m_0-i\Gamma, \ \ m_0^2=\frac{m^2}{2}, \ \ \frac{\Gamma}{m_0}=\frac{\pi}{16}\left( \frac{m_0}{k}\right)^2.
\end{equation}
Thus what we find is that rather than having a single localized mode with mass $m$, the mass will be shifted to $m_0$,
and pick up an {\it imaginary} part. This implies that a discrete mode with a finite lifetime will exist. We call this 
mode a quasi-localized mode for the following reason: if the lifetime is sufficiently long, then this mode for relatively 
short times will behave like an ordinary localized 4D mode. However, after some long times it will decay into some continuum
bulk KK modes due to its finite lifetime. This mode can also be interpreted as a resonance in the continuum KK spectrum,
rather than an individual mode with complex mass. This interpretation could be understood by calculating the 
effective volcano-type potential for this system. What we find is that the tunneling out amplitude for the mode trapped in the 
volcano will now be finite, and therefore it will have a finite width and lifetime.

Either way the result is the same: for short times one has an effectively
localized 4D particles, which however decays into the bulk after a long time. If this field were to carry bulk gauge charges,
then it would lead to apparent non-conservation of this charge from the 4D observers point of view, even though of course
there is no real violation of charge conservation in the full 5D theory once the continuum KK modes are also taken into account
(which are however suppressed on the brane). This case is merely the simplest example of quasi-localization in an 
infinite volume space-time. Several attempts have been made to construct a viable quasi-localized model for gravity,
which would be very exciting, since that would tell that the graviton would not strictly be massless, and have a finite 
width. The examples include the GRS model~\cite{GRS}, and the DGP modes~\cite{DGP}. However, none of them are quite
satisfactory, either since there is an inherent instability in the spectrum, or because 4D gravity is not reproduced at 
large distances (the latter issue is still subject to debate). 

\subsection{SM Gauge fields in the bulk  of RS1}
The next example we will discuss is the theory with gauge fields in the bulk~\cite{RSbulk,BG}. In this case one has to solve
Maxwell's equation in a curved background, which is given by
\begin{equation}
\partial_\mu(\sqrt{g}g^{\mu\nu}g^{\alpha\beta} F_{\nu\beta})=0.
\end{equation}
We will choose a gauge where $\partial^\mu A_\mu =A_5=0$. The equation for the zero mode (similar to the considerations
for the scalar) is just given by
\begin{equation}
A_\mu''=0.
\end{equation}
The solution would again be a constant wave-function along the fifth dimension, $A_\mu =A_0 a_\mu (x)$, however in this
case (contrary to the case with the bulk scalar) the wave function is not normalizable (the kinetic term for the 4D gauge 
zero mode diverges), as one can see from
\begin{equation}
\int d^5x \sqrt{g} g^{\mu\nu} g^{\alpha\beta} F_{\mu\alpha}F_{\nu\beta}=
\int dy e^{-4ky} e^{2ky} e^{2ky} A_0^2 \int d^4x F^{(4)}_{\mu\nu}F^{(4) \mu\nu}.
\end{equation}
Due to the different power of the warp factor (which is simply due to the more indices in the gauge field) the 
$y$ integral is no longer normalizable, and thus there will be no localized zero mode. In fact, as we will see below the 
zero mode wave function is flat also in the $z$ coordinates, and is thus not localized. One can show similarly, that 
bulk fermions would also not be localized in RS2. 

Nevertheless, we can consider the RS1 model with the two branes in the presence of bulk gauge fields (and also possibly 
bulk fermions). However, we will assume that the Higgs is still localized on the negative tension brane (there is good 
reason to do that -- it turns out that if it was in the bulk one would need extreme fine tuning in the bulk mass parameter to 
lower the Higgs mass from TeV to $\sim$ 100 GeV). One should ask, how in this case with Higgs on the brane and gauge fields
in the bulk would electroweak symmetry breaking work. The Lagrangian for the gauge fields would be given by
(after the Higgs on the TeV brane gets a VEV)
\begin{eqnarray}
&& \int d^4x\int_R^{R'}dz \sqrt{G} \left[ -\frac{1}{4}G^{MP}G^{NQ}\left( \frac{1}{g_5^2} W_{MN}^aW_{PQ}^a+
\frac{1}{{g_5'}^2} B_{MN}B_{PQ}\right)+\nonumber \right.\\
&& \left. \frac{v^2}{8} \frac{\delta (z-R')}{\sqrt{G_{55}}} G^{MP}
(W_M^1W_P^1+W_M^2W_P^2+(W_M^3-B_M^3)(W_P^3-B_P^3))\right].
\end{eqnarray}
Here we have used the perhaps simplest form of the RS metric:
\begin{equation}
ds^2=\left(\frac{R}{z}\right)^2 (\eta_{\mu\nu}dx^\mu dx^\nu-dz^2),
\label{AdSmetric}
\end{equation}
where $R=1/k$ (do not confuse this with the size of the extra dimension in proper $y$ coordinates), and the variable
$z$ runs between $R$ and $R'$, $R'/R=e^{kr}=10^{16}$. $g_5$ and $g_5'$ are the 5D bulk SU(2)$_L\times$U(1)$_Y$ gauge 
couplings. 

One would actually like to calculate the gauge boson masses and check if the prediction agrees with that of the SM~\cite{CET}.
If there was no Higgs VEV, we have seen that the gauge boson would have zero modes and completely flat wave functions.
However, the Higgs VEV on the TeV brane will deform the gauge boson wave functions, and the expression for the gauge boson masses 
will not be as simple as in the SM. To find the $M_W, M_Z$ masses one needs to actually solve the bulk equations in the presence
of a non-trivial boundary condition set by the brane Higgs field, and find the lowest lying eigenstates which should be identified 
with the bulk gauge bosons. The bulk equation for an eigenmode is
\begin{equation}
(\partial_z^2-\frac{1}{z}\partial_z+m^2-\frac{1}{4} v^2 g_5^2 \delta(z-R')\frac{R}{R'})\Psi (z)=0.
\end{equation}
The solutions are as always Bessel functions $zJ_1(mz)$ and $zY_1(mz)$, and the BC's are that $\Psi$ is flat at the Planck brane
and satisfies a mixed BC depending on the Higgs VEV
 on the TeV brane. The equation determining the $W$ mass is then
\begin{eqnarray}
&&J_0(M_WR)(4M_WR'Y_0(M_WR')+g_5^2v^2RY_1(M_WR'))=\nonumber \\
&&Y_0(M_WR)(4M_WR'J_0(M_WR')+g_5^2v^2RJ_1(M_WR')).
\end{eqnarray}
Expanding the solution of this equation in powers of $R^2v^2$ we get that
\begin{equation}
M_W^2=\frac{g_5^2}{R\ln \frac{R'}{R}}\frac{R^2v^2}{4R'^2}-\frac{g_5^4R^2v^4R'^2}{32R'^4\ln \frac{R'}{R}}+\ldots 
\end{equation}
The term $\frac{g_5^2}{R\ln \frac{R'}{R}}$ can be identified with the tree-level 4D gauge coupling (see next section
on AdS/CFT), and the term $\frac{R^2v^2}{R'^2}=v^2e^{-2kr}$ is just the warped down (physical) Higgs VEV. Thus the first term
is just the usual SM expression for the $W$-mass, however there is a relatively large correction to this expression which is of 
order
\begin{equation}
\frac{\Delta M_W^2}{M_W^2} \sim \frac{(v^{SM}R')^2}{16} \ln \frac{R'}{R}.
\end{equation}
This on its own is not yet very meaningful, since a shift in a single mass can always be absorbed by redefining the value 
of the Higgs VEV $v$ slightly. What one needs to do is calculate the full set of electroweak precision observables (for example
the $M_W/M_Z$ mass ratio, etc) and compare those to the experimental values. This has been done in~\cite{CET}, where it was 
found that generically all electroweak precision observables will get these type of corrections, and can be brought in agreement 
with experiment only if the value of the TeV scale $1/R'$ is raised to above 10 TeV. This would be quite bad for these
models, since in a case like this both the graviton and the KK gauge boson masses would be raised to values above those 
observable at the LHC. However, it has been pointed out recently in~\cite{ADMS} that the reason for these corrections is that 
custodial SU(2) of the SM is violated. They have proposed a model based on $SU(2)_L\times SU(2)_R$ bulk gauge symmetry 
where the corrections to electroweak precision observables can be greatly reduced and the bound on $R'$ relaxed (see also the 
discussion in the next section).

\subsection{AdS/CFT}

A big advance of string theory has been recently the realization, that certain string theories in an AdS background are
completely equivalent to some 4D supersymmetric gauge theories. In particular, Maldacena conjectured~\cite{Maldacena}, that 
type IIB string theory on AdS$_5\times S^5$ space is equivalent to ${\cal N}=4$ supersymmetric SU(N) Yang-Mills theory for large
$N$. AdS$_5\times S^5$ is a ten dimensional space (as superstring theories live in 10 space-time dimensions), and the AdS$_5$ space
that appears here is the full, un-truncated AdS space, for example given by the metric (\ref{AdSmetric}) for $0<z<\infty$. 
This supersymmetric gauge theory is also a conformal field theory in four dimensions.
The meaning of this correspondence is that if you take any operator ${\cal O}$ in the ${\cal N}=4$ SU(N) theory, there will be 
a field $\Phi$ in the string theory that couples to this operator as ${\cal O}\Phi$ on the boundary of the AdS space. Then all
correlators in the field theory will be related to the boundary-to-boundary propagator of this field
$\Phi$ in the AdS space. The important piece for us is the notion that the bulk of AdS$_5$ space is equivalent to a conformal
field theory in 4D. By bulk here we really mean the KK modes of gravity in the AdS background as discussed in the RS model.
That it is the AdS and not the 5-sphere that really gives the CFT can be seen from the fact that the 
conformal symmetry group is exactly equivalent to the isometries of the AdS space. 

However, in the above correspondence gravity does not appear on the 4D CFT side. Why is that? The reason is that we have 
considered the full AdS space, without the Planck and the TeV branes as in the RS model. Without these branes the graviton zero mode
will not be normalizable, and decouples from the theory. So it is not so surprising that the 5D gravity theory on AdS space
corresponds to a 4D theory without gravity. However, once the Planck brane is introduced, and AdS space is cut off, the 
zero mode will become normalizable~\cite{Gubser,Verlinde}. 
This means that gravity will be coupled to the CFT, and also due to the appearance of 
a brane at high energies the CFT will have a UV cutoff determined by the position of the Planck brane. If we were to also 
introduce matter on the Planck brane, this matter would couple with order one strength to the zero mode graviton, however as we 
have seen the KK modes are exponentially suppressed on the Planck brane, and therefore will basically not couple to matter 
on the Planck brane. Since the KK modes represent correspond to the CFT, this means that in the dual CFT picture the matter 
fields on the Planck brane do not couple directly to the CFT, only through the fact that the graviton zero mode couples both 
to the matter fields and the CFT. This picture can be checked via a non-trivial calculation. If the above proposal is right,
one should be able to predict the form of the corrections to Newton's law on the Planck brane due to the fact that these
corrections come from the KK modes, which should also have a prediction in terms of a CFT. This has been done by 
Gubser~\cite{Gubser},
and below we will briefly summarize his argument. In AdS/CFT language the correction to the graviton propagator 
is due to the interaction with the CFT. A graviton always couples to the stress-energy tensor, so in this case
it would be $g_{\mu\nu}T_{CFT}^{\mu\nu}$. The correction will be of the form

\vspace*{1cm}
\centerline{\includegraphics[width=0.5\hsize]{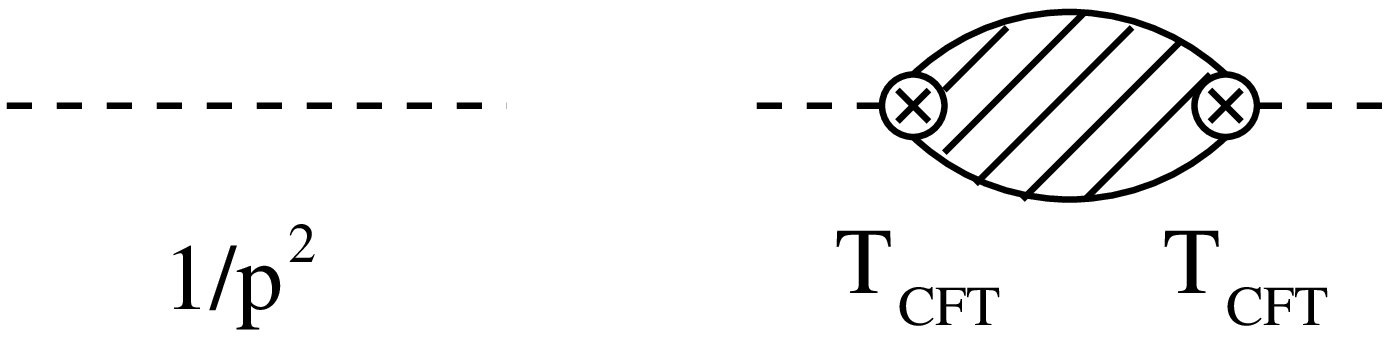}}
Thus we can see that the corrections to Newton's potential generically boil down to calculating the two-point function
of the stress-energy tensor in a CFT, $\langle TT\rangle_{CFT}$. In conformal field theories this correlator is 
well-know (and determined by conformal invariance).
The result is given by $\langle T(x) T(0)\rangle \sim c/x^8$, since the dimension of 
the stress-energy tensor in 4D is 4. Converting this to momentum space we get $\langle T(p)T(-p)\rangle =p^4\log p$.
Thus the Newton potential from the figure above will pick up the terms
\begin{equation}
\frac{1}{p^2}+\frac{1}{p^2} p^4\log p \frac{1}{p^2}.
\end{equation}
Fourier-transforming back to coordinate space we get that the first term is of course $1/x^2$, while the second is the 
Fourier transform of $1/x^4$, thus AdS/CFT predicts the form of the gravitational potential on the Planck-brane to be
\begin{equation}
F\sim \frac{Gm_1m_2}{r^2} (1+C\frac{1}{(kr)^2}).
\end{equation}
This is in agreement with what we have seen in (\ref{Newton}). 

What would be the interpretation of the negative tension (TeV) brane on the CFT side~\cite{APR,RZ}? 
Once the TeV brane is introduced, 
instead of the continuum spectrum of KK modes we will get the discrete spectrum of KK gravitons strongly peaked on the brane.
These have a well-defined mass scale, which implies that conformal invariance must have been broken on the CFT side in the IR.
 Thus we may
think of the RS1 scenario as an almost conformal field theory, that very slowly runs, but suddenly becomes strongly interacting 
at some scale around the TeV, spontaneously breaks the conformal invariance, confines and produces the bound state resonances 
that correspond to the KK gravitons. All the modes localized on the TeV brane 
should correspond to such CFT bound states, including 
the SM Higgs and SM matter and gauge fields. So the RS1 model can be simply interpreted as a CFT that becomes strongly interacting 
and produces the composite SM matter and Higgs fields. The hierarchy problem is simply solved due to the compositeness of the 
Higgs field!

What happens if we put gauge fields in the bulk of AdS~\cite{APR,RZ}? We have mentioned at the beginning of this section, that 
${\cal N}=4$ SYM corresponds to the full AdS$_5\times S^5$ type IIB string theory. The gauge theory has an SU(4) global
symmetry, called the R-symmetry. This global symmetry is reflected by the isometries of the $S^5$ internal space. Since an 
isometry implies gauge boson zero modes, which we can indeed find in the string theory side. We thus conclude, that the 
global symmetry of the CFT corresponds to a gauge symmetry in the bulk~\cite{APR,RZ}. 
This global symmetry may be weakly gauged, if 
there is a normalizable zero mode for the bulk gauge field (as in the case of RS1). But then the CFT modes couple directly 
to the gauge fields (since they are no longer on the Planck brane), and will give direct contribution to the 
running of the gauge coupling. In fact, we have seen before that the matching relation between the 4D and the 5D gauge 
couplings at tree level is
\begin{equation}
g_4^2=\frac{g_5^2}{R\log R'/R}.
\end{equation}
The extra log appearing in this matching relation exactly reflects this running due to the CFT modes first noted by Pomarol, 
which is a tree-level effect in the gravity side, however a quantum loop effect on the CFT side. 

The nicest interpretation of the RS1 model is when the gauge fields and fermions are in the bulk, while the Higgs is on the brane.
In this case there is a CFT that is slowly running, the gauge and matter fields are coupled to the CFT, but are not
composites. At some scale 
the slow running of the CFT will suddenly turn into strong interactions breaking the conformal symmetry,
and producing a composite Higgs particle, which then breaks the electroweak symmetry. Thus this should be interpreted as a 
walking technicolor model with a composite Higgs~\cite{APR}. From the AdS/CFT point of view we now also understand why we got the 
large corrections to the $W$ and $Z$ masses: in the SM the relation of these masses is protected by a global symmetry 
called custodial SU(2). However, we have seen that if there was a global symmetry like this, we would need to see 
bulk gauge fields. So the real problem is that the strong interaction (the CFT) does not obey the custodial SU(2) symmetry.
The resolution is simple~\cite{ADMS}: simply impose this additional global SU(2) by putting SU(2)$_L\times SU(2)_R$ gauge 
bosons in the bulk, and break the SU(2)$_R$ on the Planck brane to eliminate its zero mode. This way custodial SU(2) will
be restored and the corrections to electroweak observables will be smaller. However, this model would still correspond
to a walking technicolor with a composite Higgs. One may ask, what would real technicolor (without a Higgs) then correspond to?
In that case it is the strong interactions themselves (the appearance of the negative tension brane) which should break
electroweak symmetry, rather than a composite Higgs on the TeV brane. This means that to get real technicolor, one needs to break
electroweak symmetries by BC's on the TeV brane. A model for this has been recently proposed in~\cite{CGPT}, (see also
~\cite{CGMPT}).

\section{Epilogue}

In these lectures I was trying to summarize some of the important/interesting topics in theories with extra dimensions. These notes should
be sufficient for an advanced graduate  student to get started on research in this area. However, the topics covered
here are by far not complete. Some examples of entire fields left out are brane cosmology~\cite{branecosm,branecosmformal},
higher dimensional warped space~\cite{6D}, the self-tuning approach to the cosmological constant problem~\cite{selftune},
1/TeV size extra dimensions~\cite{universal}, dark matter from Kaluza-Klein modes~\cite{KKdm}, unification in warped
extra dimensions~\cite{AdSunification}, black holes in the RS models~\cite{RSBH},
electroweak symmetry breaking using the fifth component of the gauge field as a 
Higgs~\cite{manton,A5Higgs,ST}, supersymmetric Randall-Sundrum models~\cite{RSsusy,RSsusy2},
deconstruction of extra dimensions~\cite{Andy,deconstruction,otherdeconstruction},
and the list could go on and on. That the list of omitted topics is so extensive shows how much interest there is 
currently in this subject. Whether any of these ideas will become reality or remain speculation forever can only be decided
by experiment. All of those in the field hope that in ten years some of the topics listed will no longer be speculation about the
behavior of nature at high energies, but undeniable facts established by experiments at the LHC. Until then we will happily keep on
speculating...

\section*{Acknowledgments}
I thank Howie Haber, Ann Nelson and K.T. Mahanthappa for inviting me and arranging TASI 2002, and the 
students for a great atmosphere. I am also grateful
to Howie Haber for his regular friendly reminders: without them these lecture notes would have never been 
finished. Finally I thank Christophe Grojean,
Jay Hubisz, Patrick Meade and Maxim Perelstein
for carefully reviewing the manuscript and pointing out many errors and typos
in it. This research has been support in part by the NSF grants PHY-0139738 and PHY-0098631, and by 
the DOE OJI grant DE-FG02-01ER41206.


\end{document}